# PARAXIAL LIGHT BEAMS WITH ANGULAR MOMENTUM

A. Bekshaev[1], M. Soskin[2], M. Vasnetsov[2]

[1]I.I. Mechnikov National University, Dvoryanska, 2, Odessa 65026, Ukraine,

[2]Institute of Physics of the NAS of Ukraine, Prosp. Nauki, 46, Kiev 03028, Ukraine

**ABSTRACT**

Fundamental and applied concepts concerning the ability of light beams to carry a certain mechanical angular momentum (AM) with respect to the propagation axis are reviewed and discussed. In paraxial beams, the total beam AM can be represented as a sum of the spin (SAM) and orbital (OAM) angular momenta. SAM is an attribute of beams with elliptic (circular) polarization and is related to the spin of photons. OAM is conditioned by the macroscopic transverse energy circulation and does not depend on the beam polarization state. In turn, the OAM can be divided in two components which reflect different forms of this energy circulation. The "vortex" OAM describes the "hidden" internal circulation that occurs, e.g., in beams with the screw wavefront dislocation, the "asymmetry" OAM characterizes rotation of the beam as a whole. In beam transformations with symmetry breakdown both forms can mutually convert.

An important class of beams with OAM are vortex beams with helical geometric structure. They constitute a full set of azimuthal harmonics characterized by integer index $l$ each possessing AM $l\hbar$ per photon. Arbitrary paraxial beam can be represented as a superposition of helical beams with different $l$. Models of helical beams and methods of their practical generation are discussed.

Transverse energy flows in light beams can be described on the basis of a mechanical model assimilating them to fluid bodies; remarkably, in a helical beam the transverse flow distribution exactly corresponds to the laws of the vortex behavior in other fields of physics (fluid dynamics, electricity). Experiments on transmission of the beam AM to other bodies (optical elements and to suspended microparticles) are discussed.

The specific spatial structure of a vortex light beam is manifested in so called rotational Doppler effect: the beam observable frequency depends on relative rotation of the beam and observer around the beam axis. As a result, any rotation of an optical image appears to be equivalent to certain coordinated frequency shifts of its azimuthal harmonics; reciprocally, a variable signal generated in a fixed photodetector by a rotating inhomogeneous beam is caused by interference between different azimuthal harmonics. Mechanical properties of non-vortex beams that are subject to forced rotation are also analyzed. Such beams always possess complicated 3D configuration and are characterized by AM which is proportional but opposite to the rotation angular velocity.

Research prospects and ways of practical utilization of optical beams with AM are discussed.

**LIST OF ABBREVIATIONS**

| | |
|---|---|
| AM | angular momentum |
| CP | circular polarization |
| CS | circular-spiral |
| LG | Laguerre-Gaussian |
| OAM | orbital angular momentum |
| OV | optical vortex |
| RDE | rotational Doppler effect |
| SAM | spin angular momentum |
| SZP | spiral zone plate |
| WF | wavefront |

## 1. HISTORICAL REFERENCE

In accord with common experience [1], even approximate determination of the moment when a certain scientific doctrine had emerged for the first time often appears to be a thankless task, especially if the question relates to such a venerable branch of physics as optics. Almost all principal concepts were "invented" several times, and nobody can be sure that possible connections between the rotational motion and the light propagation had not been mentioned in the lost notebooks of Leonardo or come up in the Aristotle's conversations. Nevertheless, in order to follow the tradition, we dare to relate the origins of scientific ideas on the "rotational" properties of light with the name of Descartes [2, 3]. In his metaphysical system, the rotational motion generally occupies an outstanding position; he conceives the light as compressions of an absolutely elastic medium and explains the difference of colors by vortex motions of particles of this medium performed with different velocities.

Following steps in understanding the role of rotation in light phenomena were associated with development of the idea of light polarization, which is, first of all, a Fresnel's merit [4]. It had resulted in formation of the concept of elliptically polarized waves and to discovery of the optical activity effects consisting in the polarization plane rotation (Arago, 1811; Faraday, 1846 [5]). As for the mechanical aspect of such rotational phenomena, at that time, as a rule, it had been out of question, since all the light theories were essentially mechanistic; more likely, one had to look for explanations, sometimes rather artificial, why the mechanical behavior of light is not observable in the every day activity.

However, in spite of a number of remarkable achievements, the pre-Maxwell era should be recognized as an age of "prehistory"; the real history of the problem begins from the famous Maxwell's treatise [6], where ideas of the electromagnetic field and its mechanical properties were first expressed in clear and consistent form (1873). Grounding on the Maxwell equations, in 1898 A.I. Sadovsky who worked in Yuryev (now Tartu, Estonia) had predicted that the light with circular and elliptic polarization exerts the rotatory action upon material objects [7]. The same facts became commonly known after their repeated discovery by Poynting in 1909 [8], who had also introduced the transparent analogy between a light beam with circular polarization (CP beam) and a rotating mechanical body.

These ideas had found the experimental confirmation in 1936 in careful experiments by Beth [9], who managed to register the mechanical torque exerted by a CP wave and thereby to determine the value of its angular momentum (AM) and, consequently, the AM of a single circularly polarized photon.

Simultaneously, the electromagnetic theory had provided consistent means for analyzing energy flows associated with light waves. Probably, first ideas on the possibility of a circulatory flow of the light energy were formulated as early as in 1919 during the study of optical field focused by a collecting lens [10]. Later on, the problems of vortex-like and, generally, mechanical properties of light beams remained among the topics of physical literature but somewhere at the periphery of the main stream, being chiefly a subject of philosophical contemplations about the universal character of physical laws and the unity of physical world. They were indispensable in respectable monographs on classic and quantum electrodynamics [11–15]. The most important results of this period were expressed in the consistent theoretical analysis of vortex properties of general light fields and, first of all, development of the concept of orbital angular momentum (OAM) [12, 14, 15]. In the "proper" optical area, the main progress was associated with further elaboration of the traditional problems associated with the focused field structure. In particular, the detailed calculations of a light wave in the near-focal region of a limited-aperture lens (see, e.g., [16]) have demonstrated a circulatory flow of electromagnetic energy near each Airy ring.

Such an unhurried development continued until the beams with helical spatial structure were discovered (more exactly, presence of this helical structure was recognized in the well known Laguerre-Gaussian modes of laser resonators) [17]. But the real "boom" had begun in 1992 when L. Allen and co-workers [18] calculated AM of those modes and showed their remarkable similarity to CP beams. Since then, "optical vortices" (OV) [19] have become perhaps the most popular objects of the light beam optics, first of all due to characteristic peculiarities of the wavefront (WF) structure [20–22]. This circumstance has laid a bridge to the extensive field of the wavefield topology, which had just been basically developed by efforts of Nye, Berry and their co-workers [23–25]. In this context, mechanical characteristics of light beams acquire the role of effective instruments for studying their spatial configuration. In particular, the AM turns out to be an indispensable attribute of an OV, describing its morphology in terms of circular energy flows, and becomes one of the main notions of the so-called singular optics [26] – a new scientific area dealing with various singular points and catastrophes [27] of optical fields.

## 2. ANGULAR MOMENTUM OF A PARAXIAL BEAM AND ITS CONSTITUENTS

As is known [28], in an electromagnetic field with electric and magnetic strengths $\boldsymbol{E}$ and $\boldsymbol{H}$, an energy flow exists whose density is determined by the Poynting vector

$$\mathbf{S} = \frac{c}{4\pi}\left[\boldsymbol{E}\times\boldsymbol{H}\right]. \tag{1}$$

This expression holds in a homogeneous medium with light velocity $c$, to which case we will restrict our further consideration. The energy transfer expressed by Eq. (1) is related to the electromagnetic momentum of the field with density [29, 30]

$$\mathbf{\Pi} = \frac{1}{c^2}\mathbf{S}. \tag{2}$$

Operating formally, one always can choose a reference point or axis, define an arm $\mathbf{R}$ (see Fig. 1) and with the help of Eqs. (1) and (2), in a standard manner, "build" an expression for the local AM density of the electromagnetic field with respect to this point or axis. Then, integration over the whole space gives the total AM of the field [12]

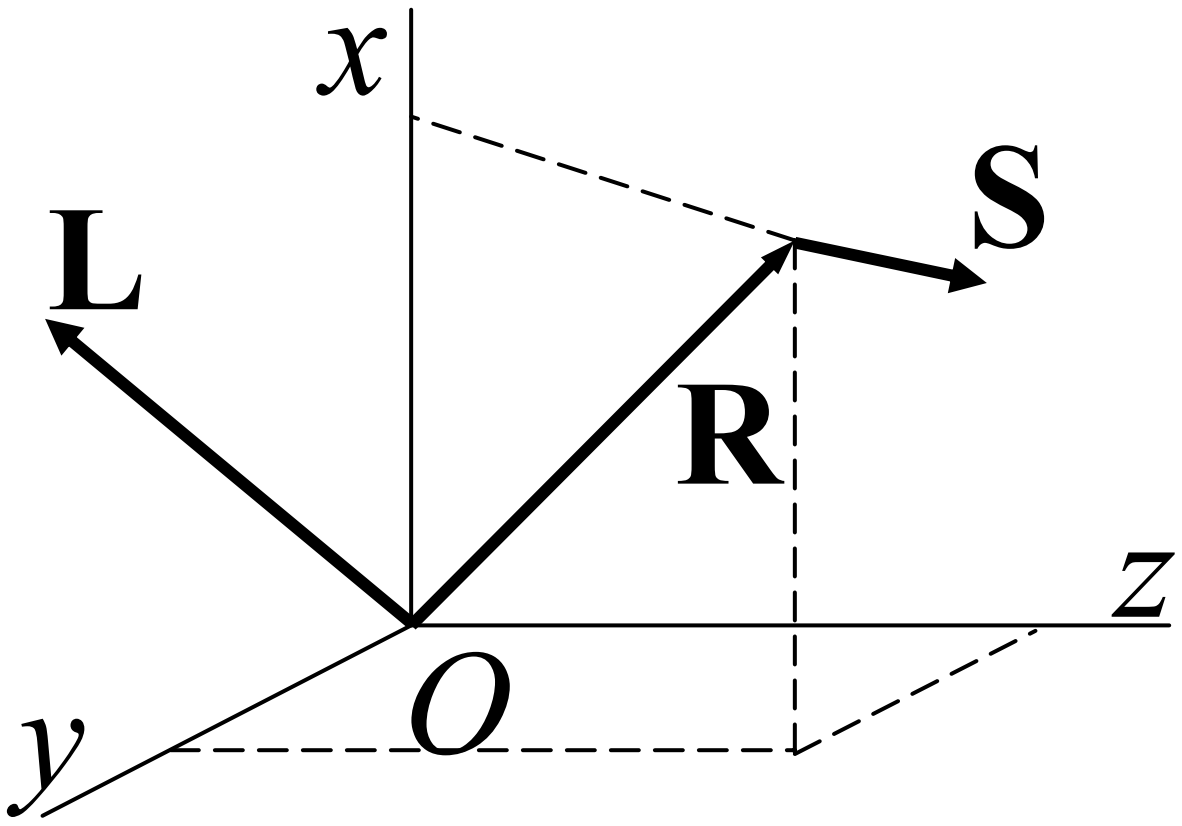

Fig. 1. Mutual disposition of the Poynting vector **S** and the angular momentum **L** in the Cartesian frame; *O* is the reference point, **R** is the angular momentum arm.

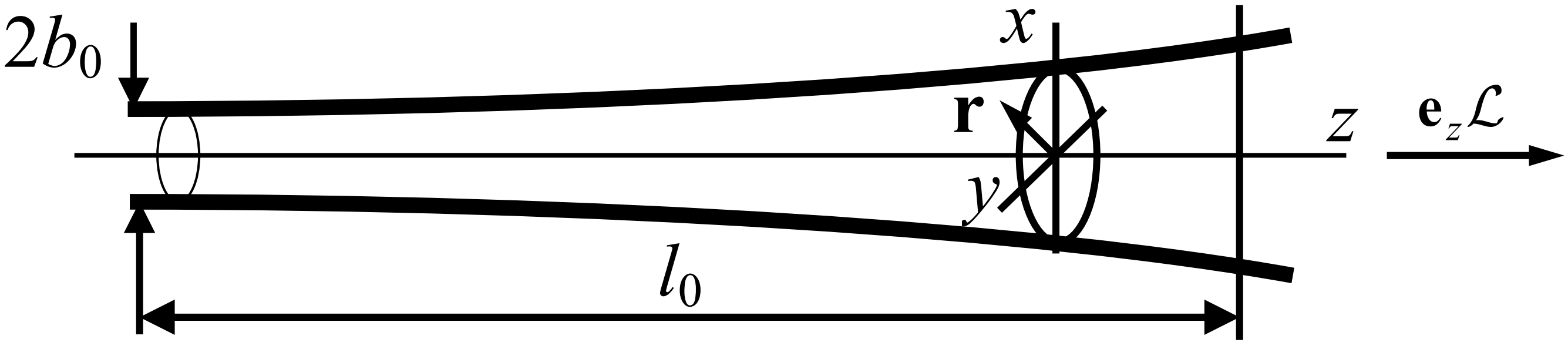

Fig. 2. Paraxial light beam: $b_0$ and $l_0$ are the longitudinal and transverse scales of the field spatial variations, ($x$, $y$, $z$) is the Cartesian frame, $\mathbf{r} = (x, y)$ is transverse radius-vector, $\mathbf{e}_z$ is the unit vector of axis $z$, $\mathbf{e}_z\mathcal{L}$ is vector of the beam AM with respect to axis $z$.

$$\mathbf{L} = \int [\mathbf{R} \times \mathbf{\Pi}](dR) = \frac{1}{c^2} \int [\mathbf{R} \times \mathbf{S}](dR), \tag{3}$$

where $(dR)$ is the elementary volume (from here on, no indication of the integration limits means the integration over the whole domain of the integrand).

Of course, in order to acquire a physical significance, this construction should be applied to certain physically preferential points or directions of the field. In this review, we deal with paraxial light beams [31, 32], where this preferential direction is dictated by the longitudinal direction of propagation (axis $z$ with unit vector $\mathbf{e}_z$), and the transverse coordinates form a radius-vector $\mathbf{r} = \begin{pmatrix} x \\ y \end{pmatrix}$ (Fig. 2). In such fields, a strong difference exists between characteristic scales for longitudinal $l_0$ and transverse $b_0$ spatial variations [33], so the ratio $\gamma = b_0/l_0 \ll 1$ (diffraction angle) is a small parameter of the paraxial approximation (it also determines relative value of the transverse energy flows within the beam [33–35] in respect of the longitudinal flow).

So, for paraxial beams, the AM with respect to the propagation axis $z$ constitutes the main interest. By using Eqs. (1), (2), we find the value of this AM belonging to a unit length of the beam [18]:

$$\mathbf{L} = \mathbf{e}_z \mathcal{L} = \int [\mathbf{r} \times \mathbf{\Pi}](dr) = \frac{1}{c^2} \int [\mathbf{r} \times \mathbf{S}](dr), \tag{4}$$

where integration is performed over the beam cross section and $(dr) \equiv dx \cdot dy$ is a symbol of the elementary area. Although quantity (4) has a meaning of the AM linear density, for the brevity we will often call it merely "AM of the beam", if this does not cause misunderstandings. A scalar measure of this AM is

$$\mathcal{L} = \frac{1}{c^2} \int \left( x S_y - y S_x \right) dx dy = \frac{1}{c^2} \int r S_\phi \, r dr d\phi \tag{5}$$

where polar coordinates $r = |\mathbf{r}|$, $\phi = \mathrm{arctg}(y/x)$ are introduced in the beam cross section, $S_\phi$ is the azimuthal component of $\mathbf{S}$. As is seen, AM with respect to axis $z$ is determined by transverse components of the energy flow, i.e. it is an amount of the first order of smallness in $\gamma$ relative to the main (longitudinal) field momentum. The form of the integrand in Eq. (5) emphasizes the determinative role of the azimuthal component of the transverse energy flow that expresses the energy circulation near axis $z$ (the radial component $S_r$ describing the beam divergence or convergence does not affect its AM). That is why the beam AM can be considered as a consequence and a characteristic of the transverse energy circulation in the beam.

In general case, the electromagnetic field of the beam can be presented as a superposition of orthogonally polarized components

$$\boldsymbol{E} = \boldsymbol{E}_X + \boldsymbol{E}_Y\,, \quad \boldsymbol{H} = \boldsymbol{H}_X + \boldsymbol{H}_Y\,, \tag{6}$$

where indices $X$ and $Y$ denote parameters of waves in which the electric vector oscillates within planes ($xz$) and ($yz$), correspondingly, and should not be confused with indices of $x$- and $y$-components of a vector (so, e.g., for a plane wave, vector $\boldsymbol{H}_X$ lies in plane ($yz$)). As usual [28], we employ the complex representation of monochromatic fields with cyclic frequency ω in the form

$$\boldsymbol{E}_j\left(\mathbf{r},z,t\right) = \mathrm{Re}\left[\mathbf{E}_j\left(\mathbf{r},z\right)\exp\left(-i\omega t\right)\right], \quad \boldsymbol{H}_j\left(\mathbf{r},z,t\right) = \mathrm{Re}\left[\mathbf{H}_j\left(\mathbf{r},z\right)\exp\left(-i\omega t\right)\right] \tag{7}$$

($j$ = $X$, $Y$). In their turn, vectors $\mathbf{E}_j$ and $\mathbf{H}_j$ in the first order of the paraxial approximation are expressed through slowly varying complex amplitudes $u_j$ [33–35]:

$$\begin{Bmatrix}\mathbf{E}_X\\ \mathbf{H}_X\end{Bmatrix} = \exp\left(ikz\right)\left(\begin{Bmatrix}\mathbf{e}_x\\ \mathbf{e}_y\end{Bmatrix}u_X + \frac{i}{k}\mathbf{e}_z\begin{Bmatrix}\partial/\partial x\\ \partial/\partial y\end{Bmatrix}u_X\right), \quad \begin{Bmatrix}\mathbf{E}_Y\\ \mathbf{H}_Y\end{Bmatrix} = \exp\left(ikz\right)\left(\begin{Bmatrix}\mathbf{e}_y\\ -\mathbf{e}_x\end{Bmatrix}u_Y + \frac{i}{k}\mathbf{e}_z\begin{Bmatrix}\partial/\partial y\\ -\partial/\partial x\end{Bmatrix}u_Y\right) \tag{8}$$

(here $\mathbf{e}_x$, $\mathbf{e}_y$ are the unit vectors of transverse coordinates, $k = \omega c$ is the radiation wave number), that satisfy the parabolic equation of paraxial optics which for a homogeneous medium has the form

$$i\frac{\partial u_j}{\partial z} = -\frac{1}{2k}\nabla^2 u_j\,, \tag{9}$$

$\nabla \equiv \begin{pmatrix}\partial/\partial x\\ \partial/\partial y\end{pmatrix}$ is the transverse gradient. Sometimes it is expedient to introduce the module $A$ and phase φ of the complex amplitude according to equation

$$u_j\left(\mathbf{r},z\right) = A_j\left(\mathbf{r},z\right)\exp\left[ik\varphi_j\left(\mathbf{r},z\right)\right]. \tag{10}$$

In view of relations (1) and (6), (7), the Poynting vector averaged over the light oscillation period acquires the form

$$\mathbf{S} = \frac{c}{8\pi}\mathrm{Re}\left(\left[\mathbf{E}_X \times \mathbf{H}_Y^*\right] + \left[\mathbf{E}_Y \times \mathbf{H}_X^*\right] + \left[\mathbf{E}_X \times \mathbf{H}_X^*\right] + \left[\mathbf{E}_Y \times \mathbf{H}_Y^*\right]\right). \tag{11}$$

Hence, in particular, it follows that the energy flow can be divided into two parts [36–38]. First of them contains terms with "mixtures" of different orthogonal polarizations (the first two summands of (11))

$$\mathbf{S}_C = -\frac{ic}{16\pi k}\left[\mathbf{e}_z \times \nabla\left(u_X u_Y^* - u_X^* u_Y\right)\right]. \tag{12}$$

It is of purely transverse character (vector $\mathbf{S}_C$ is always orthogonal to axis $z$) and, by substituting into Eq. (4), gives rise to the so called spin AM (SAM)

$$\mathbf{L}_C = \mathbf{e}_z \mathcal{L}_C = \frac{1}{c^2}\int\left[\mathbf{r}\times\mathbf{S}_C\right](dr). \tag{13}$$

The second part contains third and fourth summands of (11), where contributions of different polarizations are separated, and can be represented as

$$\mathbf{S}_O = \mathbf{e}_Z S_{\|} + \mathbf{S}_{\perp X} + \mathbf{S}_{\perp Y}. \tag{14}$$

It produces two energy flow constituents: first, the longitudinal flow (properly, the beam intensity) with absolute value

$$S_{\|} = \frac{c}{8\pi}\left(A_X^2 + A_Y^2\right), \tag{15}$$

and second, the transverse flow formed by the sum of transverse flows of both polarization components in the common form

$$\mathbf{S}_{\perp j} = \frac{ic}{16\pi k}\left(u_j \nabla u_j^* - u_j^* \nabla u_j\right), \tag{16}$$

which, ultimately, produces the orbital AM (OAM) of a paraxial beam

$$\mathbf{L}_O = \mathbf{e}_z \mathcal{L}_O = \frac{1}{c^2}\int\left[\mathbf{r}\times\left(\mathbf{S}_{\perp X} + \mathbf{S}_{\perp Y}\right)\right](dr). \tag{17}$$

Note that, with allowance for (9), it follows from Eqs. (12) and (13) that the beam SAM is constant during the beam propagation in a homogeneous medium; analogous statement about the OAM (and, consequently, the total AM of the beam) results from Eqs. (17) and (16). This conclusion expresses the mechanical system AM conservation in application to a light beam [39].

In more detail, the meaning of the orbital and spin parts of the beam AM will become clear from further consideration of their properties. Now we merely remark that, strictly speaking, only the whole AM of the beam is a consistent physical quantity and its division presented above is rather conventional. What is more, such a division is forbidden by requirements of gauge invariance [40] but in the paraxial approximation, which is the exclusive subject of the present review, considering the SAM and OAM separately is quite possible [37, 38] and provides some substantial methodic advantages.

To finalize this section, it is important to mention that a light wave transporting the electromagnetic energy and momentum, can also transport the AM. This process is characterized by the AM flow [38, 41, 42]. From the theoretical point of view, using the AM flow (instead of the AM itself) for the description of "rotational" properties of light fields is more natural and consistent; in particular, in terms of the AM flow, the irreproachable separation of spin and orbital parts is possible [42]. However, in most practical cases application of the AM flow gives no advantages over the simpler approaches grounding on the beam AM or its linear density.

### 3. SPIN ANGULAR MOMENTUM

As follows from the definitions (12), (13), the SAM linear density can be determined by expression

$$\mathcal{L}_C = -\frac{i}{16\pi\omega}\int \mathbf{r}\cdot\nabla\left(u_X u_Y^* - u_X^* u_Y\right)\left(dr\right). \tag{18}$$

Hence, it is seen that the SAM depends on the correlation between the orthogonal polarization components (6), i.e. its existence essentially results from the vectorial nature of the optical field. For example, in a fully polarized beam where $u_Y = \beta u_X$,

$$\mathcal{L}_C = -\frac{\operatorname{Im}\beta}{8\pi\omega}\int \mathbf{r}\cdot\nabla\left(A_X^2\right)\left(dr\right). \tag{19}$$

Particularly, in case of linear polarization, when $\beta$ is real, $\mathcal{L}_C$ vanishes. Therefore, the SAM appears only if the beam contains certain "admixtures" of elliptic or circular polarization – in other words, due to the non-zero spin of photons. This explains the origin of term "spin angular momentum".

#### 3.1. Paradoxes associated with the spin angular momentum

Let us consider a transversely homogeneous CP beam (for example, a plane wave). For it, the integrand of (18) reduces to zero, which seems to witness that a plane wave carries no SAM – in complete contradiction to the everyday practice. Usually this contradiction is explained by reasoning that every real beam is transversely limited and, consequently, cannot be absolutely homogeneous: somewhere at the "periphery", for large enough $|\mathbf{r}|$, its intensity falls down to zero, and the regions of variable intensity are responsible for the whole SAM of the beam [18, 36–38]. However, in this case, if the CP beam impinges a small absorbing object that is situated "entirely" within the region of the beam homogeneity, one would have to recognize that the object obtains no rotational influence, since the absorbed part of the beam contains no AM. At the first glance, this objection is removed when taking into account the beam diffraction on the object due to which the "necessary" inhomogeneity emerges "by itself" [43]. However, one can propose additional arguments to its support, considering the beam interaction with a "combined" object composed of an internal circle and a tightly adjacent concentric external annulus (Fig. 3), which can independently rotate around axis $z$ [44]. In this situation, the conditions can be easily realized when the slit diffraction by the clearance gap between the circle 2 and the annulus 3 is absent, and then the circle seems to get no AM: at any relationship between sizes of the circle and the annulus, only the latter can "feel" the rotary action! In order to avoid this absurd conclusion, authors of Ref. [45] resort to imaginary decomposition of the beam 1 into separate partial beams 4 and 5, falling onto the central circle and external annulus, respectively (Fig. 3). Each partial beam contains regions of inhomogeneity (properly speaking, of the rapid intensity fall-off), and it is presumably these regions that cause the corresponding rotatory action. This explanation does not seem convincing as, in reality, the beam is not decomposed, so the inhomogeneities that could carry the necessary AM exist only in imagination.

In our opinion, the correct interpretation must be based on allowance for special features of the CP beam spatial structure schematically depicted in Fig. 4. The situation looks as if the transverse energy circulation takes place within microscopic "cells"; herewith, contributions of the adjacent cells compensate each other and the macroscopic energy flow is absent. This corresponds to zero value of the SAM density expressed by the integrand in Eq. (18). The compensation is not complete if the adjacent cells differ, and this explains the non-zero SAM density of a transversely inhomogeneous beam. At last, the compensation vanishes at all if the cell series breaks, i.e. at the beam boundary. This must not obligatory be a real physical boundary; no matter how a certain part of the beam cross section is "isolated", its "near-boundary" cells will be "uncompensated" and the resulting energy circulation will appear along this boundary, which gives rise to the SAM. Obviously, just such "isolation" takes place when the wave meets an object: its projection "cuts" a part of the beam and deprives some cells of their compensatory neighbors.

From the considerations presented, one can understand why, in spite of the absence of macroscopic energy circulation, a homogeneous CP wave carries the SAM, and every part of its cross section carries the corresponding part of the total SAM. It is also seen that there is no need to employ inhomogeneity or diffraction for explanation of the rotatory action of CP beams. However, we should also formalize this conclusion, since the direct application of expressions (18) or (19) gives incorrect results.

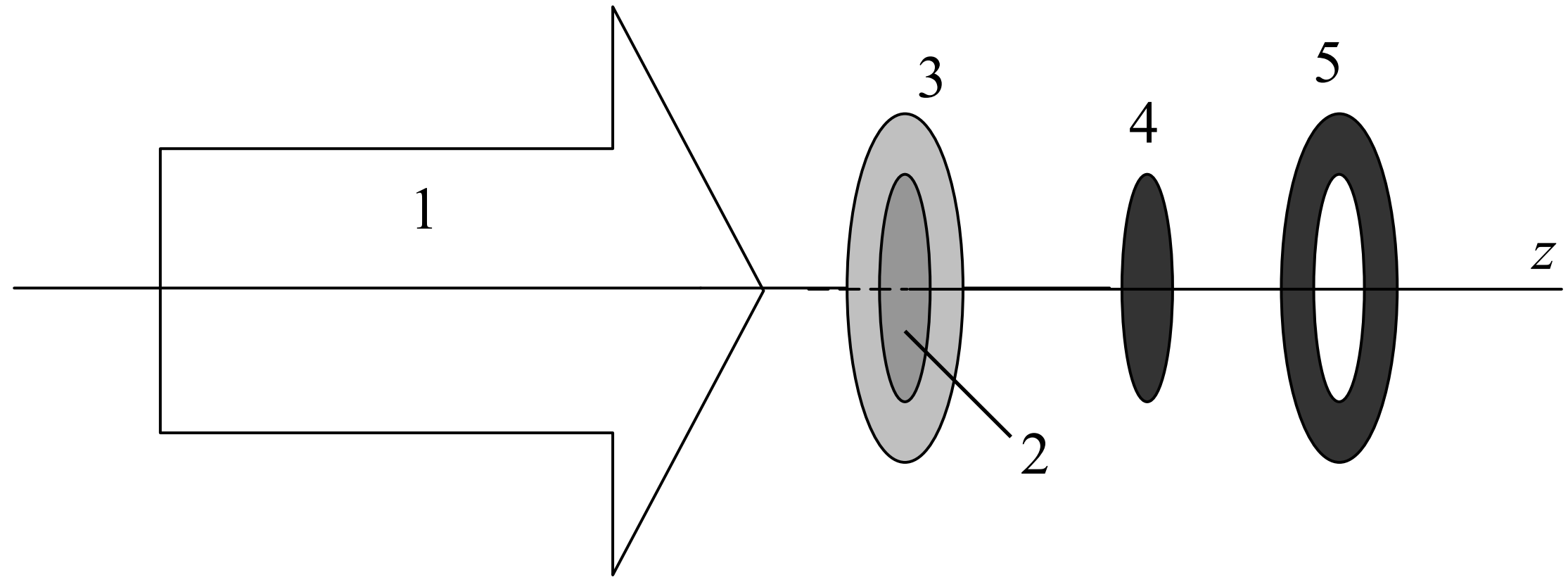

Fig. 3. Interaction of a light beam 1 with an object consisting of the circle 2 and the coaxial annulus 3; 4 – a part of the beam interacting with the circle, part 5 interacts with the annulus.

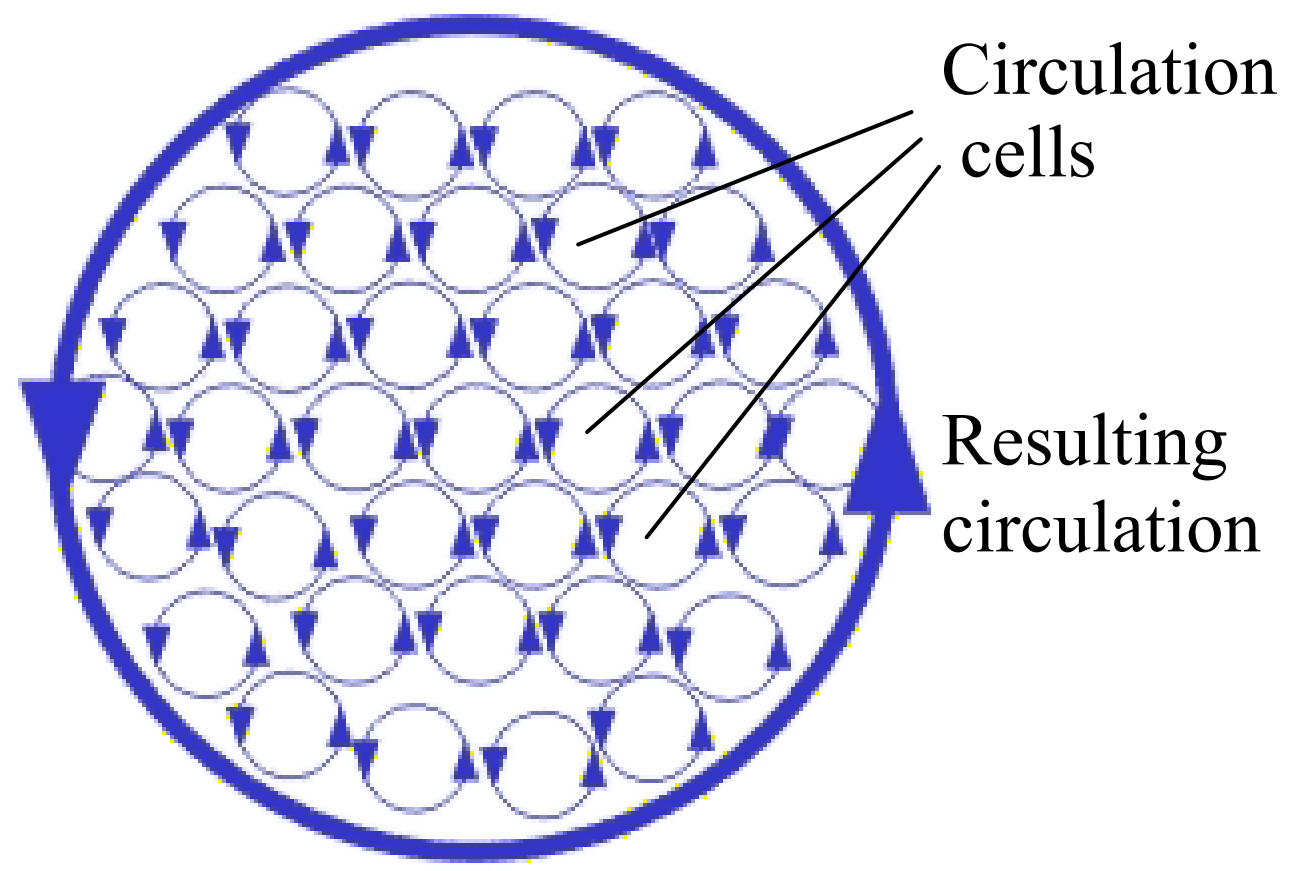

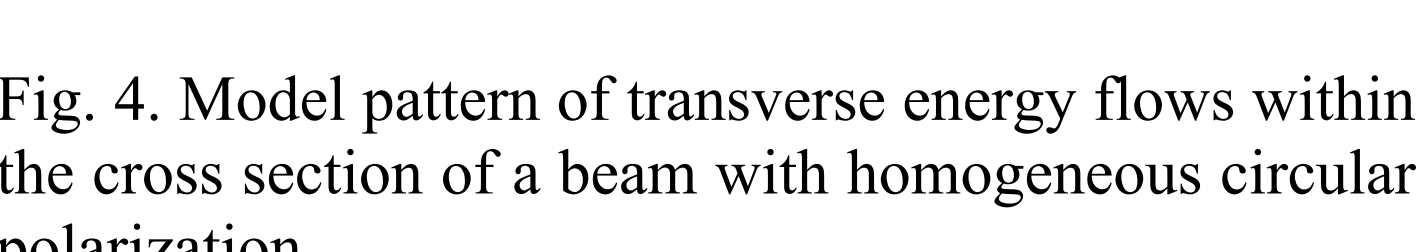

Fig. 4. Model pattern of transverse energy flows within the cross section of a beam with homogeneous circular polarization.

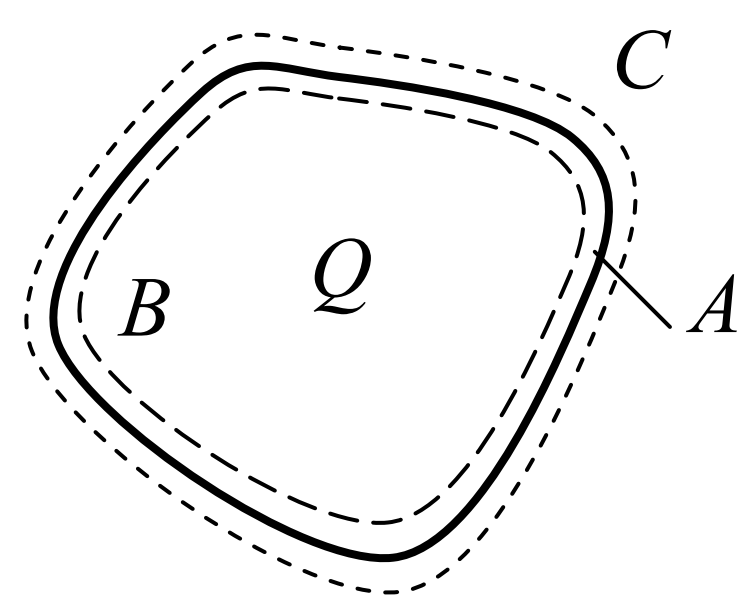

Fig. 5. Fragment of the beam cross section (region $Q$ with boundary $A$). Contours $B$ and $C$ approach $A$ from the inside and outside.

The model of Fig. 4 makes it clear that the SAM of an "isolated fragment" of the beam cross section $Q$ (see Fig. 5) contains not only the "bulk" contribution given by Eq. (18) but also the contribution of this fragment boundary $A$ [46]. To determine the latter, let us compare the SAM of two regions: the first one bounded by contour $B$, belonging to $Q$, and the second one bounded by contour $C$ that slightly oversteps the limits of $Q$; evidently, if $B \to A$ "from inside" and $C \to A$

"from outside", the difference between both values of SAM will tend to the sought boundary contribution. In the first case, applying the Green theorem [47] to the integral in (18), one obtains

$$\mathcal{L}_C(Q\,|\,B) \to \frac{i}{8\pi\omega}\int\limits_Q \left(u_X u_Y^* - u_X^* u_Y\right)(dr) - \frac{i}{16\pi\omega}\oint\limits_A \left(u_X u_Y^* - u_X^* u_Y\right)(xdy - ydx);$$

in the second case, since there is no optical field at the contour $C$, analogous transformations give almost the same result but with vanishing second summand,

$$\mathcal{L}_C(Q\,|\,C) \to \frac{i}{8\pi\omega}\int\limits_Q \left(u_X u_Y^* - u_X^* u_Y\right)(dr).$$

Hence, the boundary SAM contribution amounts to

$$\mathcal{L}_C(A) = \mathcal{L}_C(Q\,|\,C) - \mathcal{L}_C(Q\,|\,B) = \frac{i}{16\pi\omega}\oint\limits_A \left(u_X u_Y^* - u_X^* u_Y\right)(xdy - ydx),$$

and, adding this contribution to the "bulk" one (18), we find the universal SAM expression correct for the whole beam as well as for its arbitrary transverse fragment

$$\mathcal{L}_C(Q) = \frac{i}{16\pi\omega}\left[-\int\limits_Q \mathbf{r}\cdot\nabla\left(u_X u_Y^* - u_X^* u_Y\right)(dr) + \oint\limits_A \left(u_X u_Y^* - u_X^* u_Y\right)(xdy - ydx)\right] = \int\limits_Q \mathcal{L}_C'(dr), \quad (20)$$

where (see Eq. (10))

$$\mathcal{L}_C' = \frac{i}{8\pi\omega}\left(u_X u_Y^* - u_X^* u_Y\right) = \frac{A_X A_Y}{4\pi\omega}\sin\left[k\left(\varphi_Y - \varphi_X\right)\right] \quad (21)$$

is the SAM volume density. For usual transversely limited beams with smooth intensity fall-off this conclusion fully coincides with the known result [18, 36, 37].

### 3.2. Normalized angular momentum of a light beam

Considering the SAM, it is convenient to introduce some general concepts concerning the AM of paraxial beams. First of all let us compare the SAM density (21) with the volume density of electromagnetic energy of the beam that can be presented in the form [29, 36, 37, 48]

$$w' = \frac{1}{16\pi}\left(|\mathbf{E}|^2 + |\mathbf{H}|^2\right) = \frac{1}{8\pi}\left(|u_X|^2 + |u_Y|^2\right). \quad (22)$$

It is seen that the ratio

$$\Lambda_C = \frac{\mathcal{L}_C'}{w'}\omega = \frac{i\left(u_X u_Y^* - u_X^* u_Y\right)}{|u_X|^2 + |u_Y|^2} = \frac{s_3}{s_0} \quad (23)$$

is a dimensionless quantity that reduces to ratio of the Stokes parameters $s_3$ and $s_0$ which determines the degree of circular polarization [1, 28, 48]. Amount $\Lambda_C$ is suitable because it is "free" from influences of the inhomogeneous intensity distribution and immediately characterizes the beam structure responsible for its SAM. This concept can be generalized for characterizing any beam with AM regardless of the latter's nature. Then, since the AM linear density $\mathcal{L}$ corresponds to the linear density of the beam energy

$$w = \int w' dx\, dy, \tag{24}$$

which, in turn, is expressed through the total energy flow (the beam power)

$$\Phi = \int S_{\|}(x, y)\, dxdy = wc, \tag{25}$$

an analogue of Eq. (23) defines the quantity

$$\Lambda = \frac{\mathcal{L}}{w}\omega = \frac{\mathcal{L}}{\Phi}\omega c \tag{26}$$

which can be called "normalized", or "specific" AM [37] (accordingly, quantity (23) is the specific SAM). Its physical meaning is very demonstrative: every photon of the beam with the specific AM $\Lambda$ carries, in average, the AM of $\hbar\Lambda$, that is, $\Lambda$ quantitatively coincides with the AM measure in units "$\hbar$ per photon" [18].

In particular, absolute value of the normalized SAM (23) cannot exceed 1 and is a measure of presence and relative "strength" of the CP. In case of pure circular polarization (in (19) $\beta = \pm i$)

$$\mathcal{L}_C = \Phi/\omega c, \quad \Lambda_C = \pm 1, \tag{27}$$

i.e., the SAM amounts to $\pm\, \hbar$ per photon, which is just expected from the quantum mechanical reasoning.

## 4. ORBITAL ANGULAR MOMENTUM

Due to Eqs. (10), (16) and (17), the OAM linear density can be written as

$$\mathcal{L}_O = \frac{1}{8\pi c}\Big[\int A_X^2 \mathbf{e}_z \cdot \left[\mathbf{r} \times \nabla\varphi_X\right](d\mathbf{r}) + \int A_Y^2 \mathbf{e}_z \cdot \left[\mathbf{r} \times \nabla\varphi_Y\right](d\mathbf{r})\Big]. \tag{28}$$

The OAM is determined by the sum of contributions of both orthogonal polarizations and, in contrast to the SAM, can exist in a linearly polarized or non-polarized beam; its value is completely conditioned by the complex amplitude spatial inhomogeneity and can serve as its characteristic. Both contributions of (28) can be considered, to a great extent, independently (at least in a homogeneous and isotropic medium), and general properties of the OAM can be studied with the help of any one of summands of formula (28). This means that the OAM consideration can be based on the scalar model of light waves that adequately characterizes the evolution of beams with arbitrary but everywhere the same polarization. Therefore, in further OAM analysis we will rely on the main formulas (7), (8), (10), (16), (28) and so on, using only one of polarization components and omitting indices $X$, $Y$.

Unlike the SAM associated with microscopic circulatory flows existing "in every point" of the beam cross section (Fig. 4), the OAM is caused by the macroscopic energy circulation near the reference axis $z$. Hence, it follows that the beam AM depends on mutual disposition of the beam and the reference axis and if the latter experiences small (in paraxial limits [34]) transverse shift $\mathbf{r}_S$ i and deflection $\mathbf{p}_S$ (see Fig. 6) the OAM transforms as [36, 50]

$$\mathcal{L}_O \rightarrow \mathcal{L}_O + \frac{\Phi}{c^2} \mathbf{e}_z \cdot \left[ \mathbf{r}_S \times \mathbf{p}_S \right]. \qquad (29)$$

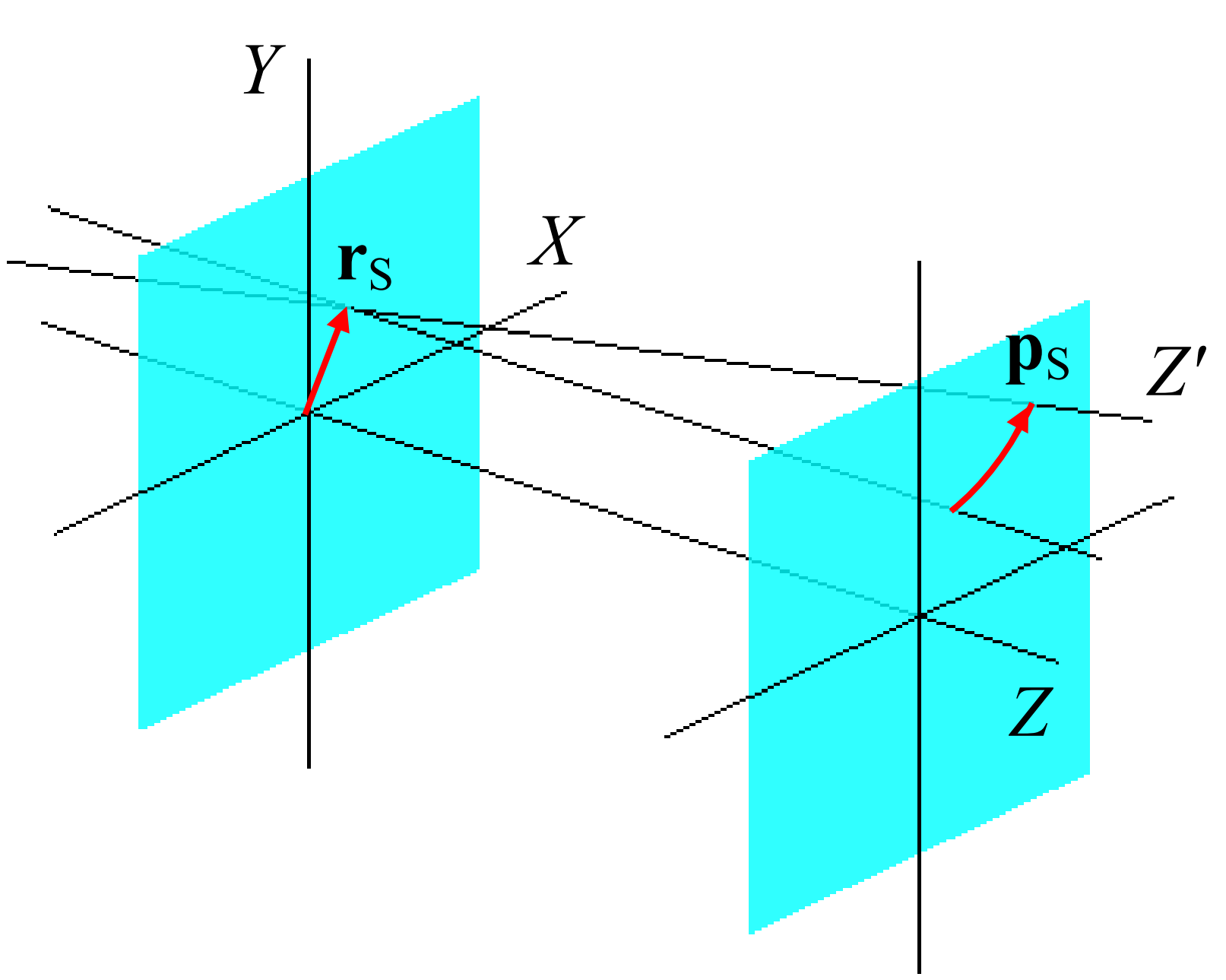

Fig. 6. Transverse shift $\mathbf{r}_S$ and deflection $\mathbf{p}_S$ of the reference axis

Such transformation is quite understandable since, upon shifting and deflecting the axis, the whole AM of the beam with respect to this axis changes; in accord with the AM decomposition into the spin and orbital parts (13), (17), this tells only on the OAM. This means that the OAM characterizes not only "proper" peculiarities of spatial structure of the beam but, sometimes, inessential details of its spatial position in a conditional coordinate frame. In order to avoid the indeterminacy, it is expedient to adhere certain preferential reference frame. One of the physically grounded variants demands that the moment axis coincide with the local trajectory of the beam "center of gravity" [51], whose local position and slope in an arbitrary frame are described by equations

$$\mathbf{r}_0(z) = \int A^2(\mathbf{r},z)\mathbf{r}(d\mathbf{r}), \quad \mathbf{p}_0(z) = \frac{d\mathbf{r}_0(z)}{dz}, \tag{30}$$

that is, axis $z$ should be chosen so that quantities (30) vanish. Such an OAM is called "intrinsic" [52], and then the amount described by the second summand of Eq. (29) expresses the "extrinsic" OAM.

### 4.1. Helical beams

So far, we discussed the light beam AM in its abstract aspects; now let us address more specific examples. Perhaps, first objects with OAM, realized and investigated in the laboratory [17–19, 53, 54], were beams whose transverse complex amplitude distribution is given by the special case of formula (10):

$$u(r,\phi) = \psi(r)\exp(il\phi), \tag{31}$$

where $r$ and $\phi$ are polar coordinates in the beam cross section, $\psi(r)$ is, generally, complex function that falls rapidly enough with $r \to \infty$, and $l$ is an integer azimuthal index. In view of the importance of such beams in the whole OAM doctrine, it would be expedient to assign them a common collective name. It is seen from Eq. (31) that the transverse intensity distribution of such beams possesses a circular symmetry, and the second multiplier of (31) determines the "spiral" growth of phase upon a round trip near the axis; by these two attributes, beams of this form will further be referred to as "circularly-spiral" (CS) beams.

Note that if in Eq. (31) $l \neq 0$, the continuity of the beam electric field requires $\psi(0) = 0$, i.e., at least in a very small vicinity of the axis, function $\psi(r)$ has to fall down to zero. For CS beams, the polar coordinate representation is especially convenient. In this sotuation, formula for OAM that follows from Eqs. (16), (17) and (28) takes on the suitable form

$$\mathcal{L}_O = \frac{1}{16\pi kc}\int\left[u\left(-i\frac{\partial}{\partial\phi}u\right)^* + u^*\left(-i\frac{\partial}{\partial\phi}u\right)\right] r\,dr\,d\phi\,, \tag{32}$$

from which a very important result for CS beams (31) follows [18, 37]

$$\mathcal{L}_O = \frac{l}{8\pi kc}\int |u|^2\, r\,dr\,d\phi = l\frac{\Phi}{c\omega} = l\frac{w}{\omega}. \tag{33}$$

Therefore, OAM of a CS beam does not depend on the special form of the radial amplitude distribution and is determined by the second (azimuthal) factor alone. Moreover, Eq. (33) means that the normalized OAM (26) for such a beam exactly equals to the azimuthal index

$$\Lambda_O = l, \tag{34}$$

which is usually treated as the OAM "quantization".

After all, the arguments presented suggest many "quantum allusions" [38]. Besides what was already noted, one can mention the similarity of differential operators in (32) to the angular momentum operator [13–15] so that Eq. (34) can be considered as its "eigenvalue", employment of the "state operators" and Pauli matrices [36], manipulations with spin and orbital "quantum numbers", etc. These notions are useful in calculations and intuitive analyses [36–39, 55, 56] but, within the classical scope accepted throughout this review, have no absolute physical meaning. In particular, result (34) really means that each photon of beam (31) carries the OAM $l\hbar$, but this does not deny other variants. For example, integer value of index $l$ in Eq. (31) is necessary for the electromagnetic field continuity, but one can imagine (and even realize in practice [57, 58]) CS beams with non-integer $l$, for which Eqs. (33) and (34) lead to a fractional value of $\Lambda_O$. Beams (31) with integer $l$ are really distinguished by the fact that they are eigenfunctions of the operator $\exp(\Delta\theta\, \partial/\partial\phi)$ of rotation by angle $\Delta\theta$, and form hereupon a basis of irreducible representations for the group of rotations about axis $z$ [59]. Therefore, they form a system of angular (azimuthal) harmonics, just like plane waves form a system of spatial (translational) harmonics. A physical distinction of such beams is their ability to preserve the form (31) during propagation in a homogeneous space: although the radial function $\psi(r)$ can modify essentially, it preserves the circular symmetry, and azimuthal dependence of the complex amplitude is always the same.

These features make the considered CS beams similar to the CP waves and stimulate to look for the common physical sources of their properties. Let us begin with the important remark relating the spatial-temporal behavior of the CP beams described in the previous section. For example, for $\beta = -i$ (see (19)), the electric field (7) of a homogeneous wave obeys the rule

$$\boldsymbol{E}_X \sim \mathbf{e}_x \cos(kz - \omega t), \quad \boldsymbol{E}_Y \sim \mathbf{e}_y \sin(kz - \omega t), \tag{35}$$

i.e., in a given cross section the electric vector rotates with angular velocity $\omega$ (in this case, clockwise when seeing against the beam propagation[1]), and the instantaneous three-dimensional

[1] Following the convention used in optics [28, 49] we will call this case the "right" polarization,

configuration of the electric field forms a right screw (helicoid) whose pitch equals to the wavelength $\lambda = 2\pi/k$ (Fig. 7a). Even without detailed analysis of the situation, one can affirm that it is this helicoidal structure that is responsible for the "vortex" properties of the beam, particularly, for its SAM.

It is natural to expect something alike in case of beams (31). In order to examine the instant spatial pattern of the electromagnetic field of such a beam, we substitute expression (31) into Eqs. (7), (8); then for the scalar value of the transverse component of vector $\boldsymbol{E}$ we obtain the expression

$$E(r, \phi, z, t) \sim \psi(r) \cos(l\phi + kz - \omega t). \tag{36}$$

In a fixed cross section ($z$ = const) this relation describes the two-dimensional field distribution $\psi(r)\cos(l\phi)$ that rotates with velocity $\omega/l$ while its three-dimensional shape is a helicoidal structure with pitch $|l|\lambda$ (Fig. 7c illustrates the situation when $l = -1$).

Behavior of the averaged (over the light oscillation period) characteristics agrees with the noticed facts. Beams of type (31) possess circularly symmetric intensity distribution that is formed due to rapid rotation of the Fig. 7c pattern; the WF, determined by condition $l\phi + kz$ = const, represents a system of $|l|$ helicoids with the same pitch $|l|\lambda$. For example, the WF shape of a CS beam with $l = -1$ (Fig. 7b) has no principal distinctions from the helicoid of Fig. 7a circumscribed by the electric vector of a "right" CP beam. These specific features stipulate another term for the CS beams – "optical vortices" (OV) [19, 26]; the index $l$ is sometimes called "order of OV". It looks rather convenient that all the introduced definitions can be generalized also for the "non-vortex" (zero-order vortex) case when $l = 0$ in Eq. (31), and the zero-pitch helicoid naturally degenerates into a smooth surface.

Therefore, the CP beams as well as scalar CS beams possess a clearly expressed helical structure that is a source of their vortex properties, in particular, of the mechanical AM with respect to the propagation axis. This structure constitutes a sort of the "configuration skeleton" of the beam and, besides already mentioned ones, stipulates a series of other beam peculiarities that will be discussed below. Now we make an important remark that evolution of such structures upon propagation can be described in two equivalent ways: as the rotation around axis $z$ or as the translation along it. The physical indistinguishability of these motions is an attribute of any helical objects and exists due to their helical symmetry [60].

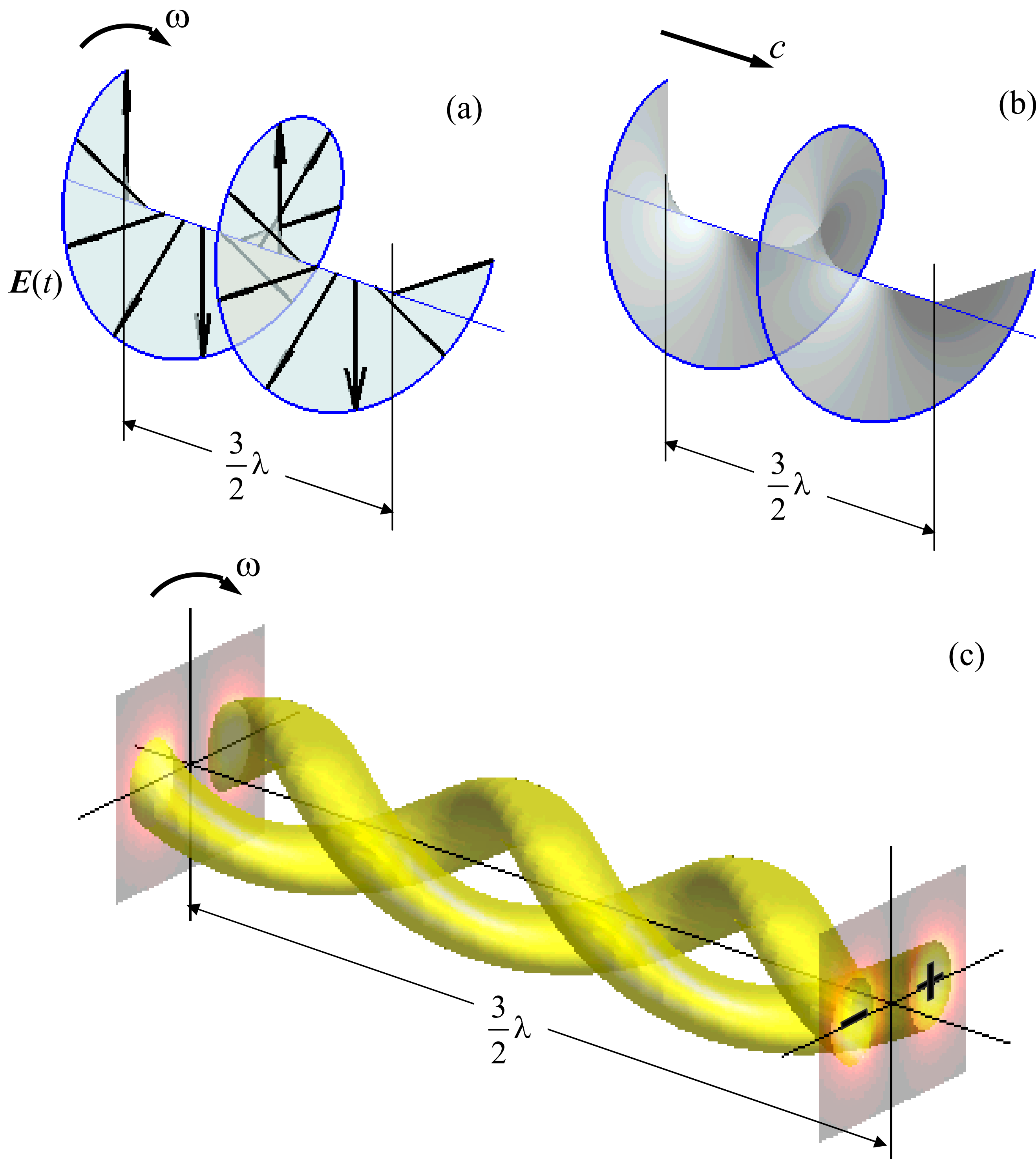

Fig. 7. Spatial patterns of helical beams: (a) instantaneous configuration of the electric field vectors in a CP beam with "right" polarization (Eq. (35)); (b) wavefront of a CS beam (31) with $l = -1$; (c) instantaneous pattern of the electric field spatial distribution of this circular beam (signs "+" and "–" denote regions where the instantaneous electric field possesses different signs). Structures (a) and (c) rotate with the oscillation frequency ω, wavefront (b) translates with the light velocity $c$.

Similarity of some fundamental properties of CP waves and CS beams allows to classify them together as "helical beams" and to consider them, in many cases, following the single scheme [36]. For example, result (27) can be presented as the analog of (34)

$$\Lambda_C = \sigma,$$

where $\sigma$ is an integer "spin" index ("spin quantum number " [39]), which equals to $\pm 1$ for the two CP signs and 0 for the plane polarization. This is especially suitable for beams where the CP structure is combined with the CS spatial configuration, i.e. for the CP beams with the spatial shape described by Eq. (31). In such rather widespread cases it would be expedient to use the "quantum number of full AM" [36, 39, 48]

$$J = \sigma + l \tag{37}$$

that possesses the physical meaning of full normalized AM of a helical beam.

However, despite the common symmetry, the physical discrepancies between the CP waves and CS beams (31) should not be neglected. The vortex qualities of CS beams are related with specialities of their "macroscopic" spatial construction, and this gives rise to much more reach morphology. First, for them the helicoid pitch is not fixed by the wavelength and can theoretically amount any multiple value (OVs with $l \le 200$ were observed [61]); correspondingly, absolute value of their normalized AM can be arbitrary integer (at least, in theory).

Second, their transverse profile has a singular point where the phase is not defined (in this case, at the beam axis $x = y = 0$). Upon a round trip near this point the field phase does not reproduce its initial value but changes by $2\pi l$; this corresponds to the longitudinal shift of the WF by $-l\lambda$. Such a structure is called "screw WF dislocation" [23, 26]. As other singularities [25, 26], wave dislocation can be characterized by the so-called topological charge $m$ defined as integer number $m = (1/2\pi)\oint df$ where $f(x, y)$ is a certain function of transverse coordinates describing the wave field properties and integration is performed along a contour enclosing the singular point. In case of an OV, usual choice is $f = k\varphi$ (see Eq. (10)) and then for a CS beam (31) the topological charge of singularity is $m = l$. However, regarding the WF dislocation, function $f(x, y)$ may be defined so that it should characterize the WF shape (the surface of constant phase), that is as a solution of equation $k\varphi + f = \text{const}$. Then, topological charge of the dislocation (or of the OV) equals to the Burgers vector length expressed in units of $\lambda$ (see Fig. 8), which entails $m = -l$. In this definition, positive topological charge $m$ corresponds to the right WF helicoid, positive index $l$ – to right "screw" of the energy circulation. The topological charge is rather a geometrical than physical characteristic of an OV; it is useful in studies of beams with multiple OVs providing a means for description of transformation, creation and annihilation of singularities (topological reactions) [24–26].

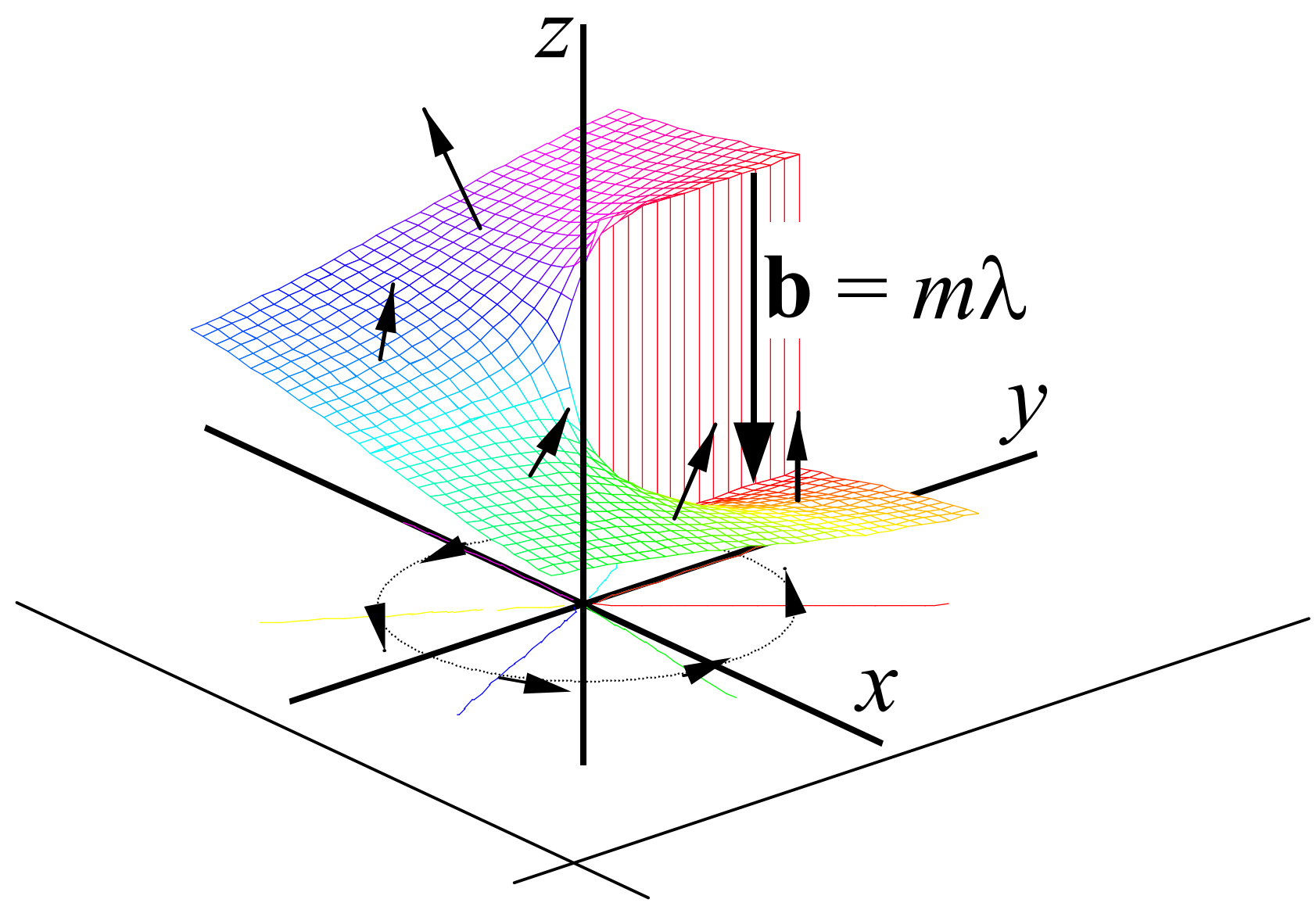

Fig. 8. Screw wavefront dislocation of a CS beam (31) with positive $l$. The whole WF consists of $|l|$ identical helicoids distanced by λ along axis $z$; the picture shows only one of them

Due to connection with the WF singular points, the whole theory of OAM appears to be in close relation with the general topology of wave fields [24, 25] and constitutes an important instrument of the so called singular optics [26] that deals with various phase singularities (or catastrophes [27]) of light waves.

Contrary to expectations, rotation of the instantaneous structures shown in Fig. 7 has no dynamical meaning and cannot be treated as the immediate reason for the beam AM: this follows, for example, from the fact that, for OV of higher orders, the velocity of rotation decreases while the OAM increases (see Eqs. (33), (34)). The true nature of the non-zero OAM is revealed regarding the WF shape near a screw dislocation (Fig. 8). In every point, the local energy flow is directed along the normal to WF [35] (in Fig. 8 the normals are shown as arrows "growing" from the WF surface). Due to the helical WF shape, the normals possess transverse azimuthal components that form a closed contour in plane $(x, y)$ and provide the transverse energy circulation.

### 4.2. Examples and methods for generation of circularly-spiral beams

For further references, we now present some important examples of CS beams. The most known models of such class are Laguerre-Gaussian (LG) beams [31, 62] that can be formed in lasers with stable resonators, being their eigen modes of oscillation. For them, in Eq. (31) one should accept

$$\psi_{pl}^{\mathrm{LG}} = \sqrt{\frac{8\Phi_{pl}}{c}} \sqrt{\frac{p!}{\left(p+|l|\right)!}} \frac{1}{b} \left(\frac{r}{b}\right)^{|l|} L_p^{|l|}\left(\frac{r^2}{b^2}\right) \exp\left(-\frac{r^2}{2b^2}\right) \exp\left[ik\frac{r^2}{2R} - i\left(2p+|l|+1\right)\chi\right]. \tag{38}$$

For every value of $l$ there can exist different beams of this family, which differ by the radial index $p > 0$, $\Phi_{pl}$ is the power of the $\mathrm{LG}_{pl}$ mode with indices $p$, $l$. Upon the beam propagation, its transverse radius $b$, radius of the WF curvature $R$ and the additional phase shift due to finite transverse beam size (Gouy phase [26, 63]) $\chi$ change; if the beam waist is situated in the initial cross section ($z = 0$), they obey the equations [31, 53]

$$R\left(z\right) = \frac{z_R^2 + z^2}{z}, \quad b^2\left(z\right) = \frac{z_R^2 + z^2}{kz_R}, \quad \chi\left(z\right) = \arctan\left(\frac{z}{z_R}\right), \tag{39}$$

where

$$z_R = kb_0^2 \tag{40}$$

is the confocal parameter (Rayleigh length) of the beam [64], $b_0$ – its radius in the waist cross section[2].

Formally, the family (38) includes also "non-vortex" beams with $l = 0$, e.g., Gaussian ones with $p = l = 0$. Beams (38) are suitable due to the fact that they constitute a full orthogonal set and any paraxial beam can be represented as a superposition of LG modes. In concrete problems, there are often used other families, in particular, Bessel-Gaussian beams [65–67], which are determined by general expression

$$\psi_l^{\mathrm{BG}} = \sqrt{\frac{8\Phi}{c}} \frac{1}{b} \exp\left(-\frac{q^2}{4}\right) \left[I_l\left(\frac{q^2}{2}\right)\right]^{-1/2} I_l\left(\frac{q}{1+iz/z_R}\frac{r}{b}\right) \exp\left(-\frac{r^2}{2b^2}\right) \exp\left(ik\frac{r^2}{2R} - i\chi\right). \tag{41}$$

Parameters $b$, $R$, $\chi$ are determined by Eq. (39), parameter $q$ is free. Another important model is "non-diffracting" Bessel beams which form a family with parameter $k_r$ [68, 69]

---

[2] Coincidence with the notation of the transverse scale of a paraxial beam spatial variations (Sec. 2) is not occasional, $b_0$ can also characterize the transverse scale of the beam (38) inhomogeneity.

$$\psi^{\mathrm{B}} = AJ_l\left(k_r r\right)\exp\left(-i\frac{k_r^2}{2k}z\right). \qquad (42)$$

Theoretically, Bessel beams possess infinite energy and one cannot attribute them a certain power; their practical realization is only possible approximately and in a limited spatial range [69, 70].

The simplest method for practical generation of the CS beams is based on the spontaneous occurrence of type (38) modes in laser resonators with absent or compensated sources of astigmatism. At the same time, this method is the least reliable because it is very sensitive to the smallest misalignments and the topological charge of the obtained OV is determined occasionally [71]. The stable and reproducible generation of modes with rectangular symmetry and smooth WF (Hermite-Gaussian [62]) is reached much easier, that is why many approaches were proposed for creation of the screw WF dislocations outside a laser resonator.

Due to "affinity" of the rectangular Hermite-Gaussian and LG modes [72], they can be mutually transformed in the so called mode converters constructed with astigmatic lenses [18, 53, 54, 73] or with astigmatic light-transmitting fibers [74]. Such converters are able to perform efficient transformations of a wide class of the rectangular modes into practically ideal LG beams (38). More specialized but simpler in construction and application are the interference converters [75], in which an LG mode is formed due to the superposition of an input HG beam with a certain auxiliary beam formed inside the device. Interestingly, schemes of mutual transformation between the rectangular and circular beams turn out to be quite similar to the linear/circular polarization transformations [37, 53, 76, 77].

Another method consists in the immediate "introduction" of the screw WF deformation during the transmission of a beam with regular WF through a "spiral" phase plate with special relief to provide the necessary azimuthal dependence of the output beam phase [78, 79]. Of course, such dependence can be realized only approximately because of the phase singularity at the axis, however, beams formed by this way show quite "decent" vortex properties.

A simple and efficient procedure of obtaining regular beams with WF dislocations is based on combining several (at least two) Gaussian beams spatially skewed with respect to each other and to a certain reference axis $z$ [80–82]. Every such beam possesses the extrinsic OAM relative to the reference axis (see Sec. 4, Eq. (29)); when interfering, they form a combined beam propagating along axis $z$ and carrying the OAM which is the sum of OAMs of all the component beams, but now it is the intrinsic OAM of the combined beam. Importantly, the necessary set of "skewed" beams can be simply obtained during the Gaussian beam diffraction on a dielectric wedge [80, 81]. In this process, the initial beam generates a series of secondary beams whose axes are mutually skewed due to difference in the propagation conditions (in a free space or through the wedge, with additional possibilities of several internal reflections). Depending on the constituent beams' parameters, the combined beam can carry a single OV or an array of OVs; at certain conditions it represents a relatively good CS beam with nearly ideal circular structure [82].

However, the most flexible and universal methods of producing not only the helical WFs but, in general, WFs of any complex shape are based on the holographic principles. As is well known [83], if some recording medium carries the pattern of interference between arbitrary "object" and a standard "reference" beams, then after illuminating this structure with the reference beam, the WF of the object beam can be restored in the diffracted field. More often, the plane or spherical waves (properly, Gaussian beams or their fragments with plane or spherical WF) are employed as the reference fields. Their interference with helical CS beams of Eq. (31) is explained by Figs. 9, 10.

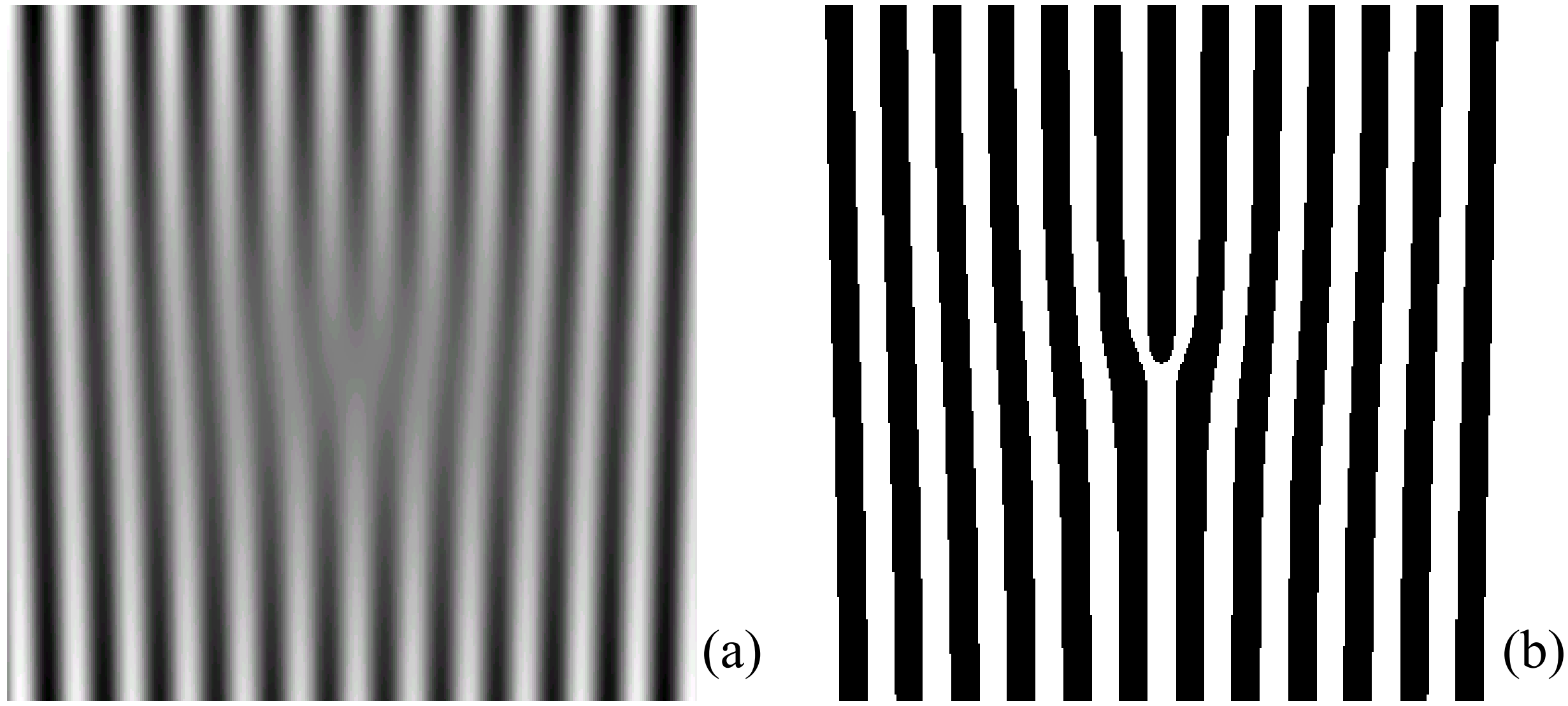

Fig. 9. (a) Pattern of interference between a plane wave and a CS beam (31) with $l = 1$ and (b) corresponding binary grating.

Fig. 9a presents the central fragment of the interference pattern formed by superposition of a plane wave and a beam of the sort (31) ($l = -m = 1$) with a small angular disalighnment of the propagation directions in the horizontal plane; in the center, the pattern contrast falls down because $\psi(r) \to 0$. The characteristic attribute of this pattern is the famous "fork" [22] – a bifurcated interference fringe, or an "odd" fringe to one side from the center, that serves the most widespread "indicator" of the helical WF. The same pattern can be considered as a transparency whose transmittance near the center is described by function

$$T(\mathbf{r}) = \frac{1}{2}\left[1 + \frac{r}{R_0}\sin(qx + m\phi)\right] = \frac{1}{2} - \frac{ir}{4R_0}\exp\left[i(qx + m\phi)\right] + \frac{ir}{4R_0}\exp\left[-i(qx + m\phi)\right], \qquad (43)$$

where $R_0$ is a certain radius of an "active zone" to provide positiveness of the function $T$, $m$ is the topological charge of the vortex "imbedded" within the transparency. Hence it is seen that if a plane wave illuminates the transparency (43), in the diffracted field there appear, besides the invariable "zero order" beam associated with the first summand, the two beams deflected by angles $\pm q/k$ from the axis $z$ and carrying OVs with indices $\pm m$. Of course, such transparencies should not obligatory be prepared by real hologram recording process, it is much more suitable to synthesize them by means of calculations (so called "computer generated holograms" [22, 84]). It is often expedient to employ the binary transparencies whose transmission is obtained by rounding the function (43) values to closest integer, i.e., can only equal to 0 or 1 [78] (Fig. 9b). Besides the obvious simplicity and possibility of precision manufacture, this variant provides the advantage that the transmission function contains not two, as in Eq. (43), but an infinite set of harmonics of the common form $\exp[iN(qx + m\phi)]$ for arbitrary integer $N$, which allows to have many OVs of higher orders simultaneously. Of course, this scheme does not generate "pure" LG modes or some other modes from the known families but every order contains a superposition of modes with the same $l = Nm$. However, in most cases this is not very important [84].

Superposition of the CS beam (31) and a spherical wave with different WF curvatures produces typical spiral patterns one of which (for case $|l| = 1$) is presented in Fig. 10a. The transmission function of the corresponding transparency has the form

$$T(\mathbf{r}) = \frac{1}{2}\left[1 + \frac{r}{R_0}\cos\left(\frac{kr^2}{2f} + m\phi\right)\right] = \frac{1}{2} + \frac{r}{4R_0}\exp\left[i\left(\frac{kr^2}{2f} + m\phi\right)\right] + \frac{r}{4R_0}\exp\left[-i\left(\frac{kr^2}{2f} + m\phi\right)\right]. \qquad (44)$$

The binary version of this transparency which can be termed "spiral zone plate" (SZP) [85], presented in Fig. 10b, is described by the sum of harmonics in the form $\exp\left[iN\left(kr^2/2f + m\phi\right)\right]$, i.e. for an input plane wave all diffracted beams have the same directions (and this somewhat complicates their utilization), however, curvatures of their WFs differ. In positive orders ($N > 0$), diverging waves are formed, in negative orders – converging ones for which the transparency act as a collecting lens with the focal distance $f/|N|$. This can be used for separating necessary orders with the help of a pin-hole filter, as is shown in Fig. 10c [86].

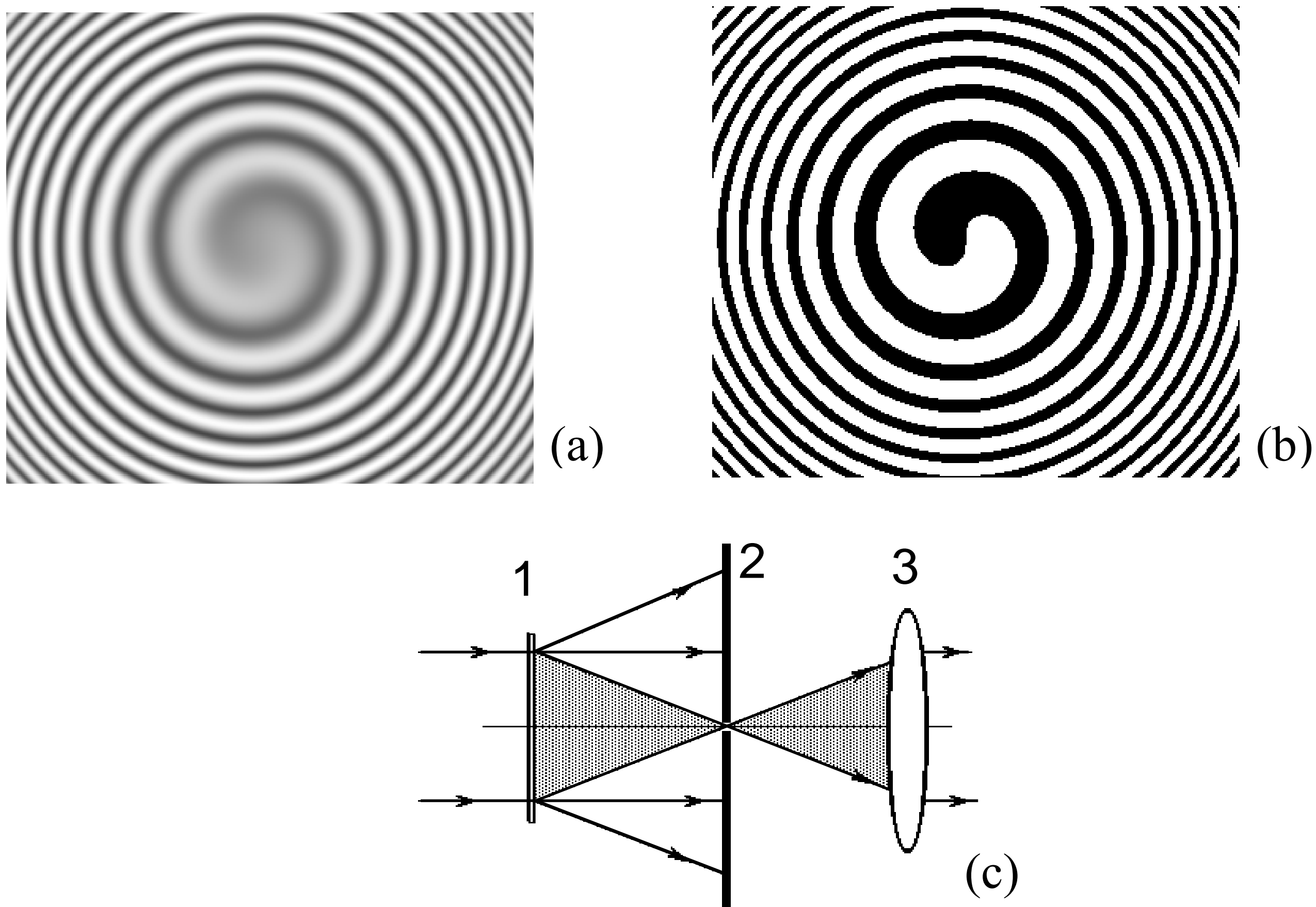

Fig. 10. (a) Interference pattern of a spherical wave and a CS beam (31) with $l$ = 1, (b) corresponding binary grating and (c) separation of the necessary diffraction order scheme: 1 – hologram of the form presented in Fig. 10b (spiral zone plate), 2 – pin-hole screen, 3 – collimating lens.

In principle, possibilities of holographic formation of OV beams are rather comprehensive. Interaction of different OV beams between each other and with model beams of various special forms is characterized by extremely reach variety of configurations [26, 37], and almost all of them can be used for practical obtaining light fields with vortex structure.

All the above-presented methods deal, in fact, with scalar beams. The wave polarization is inessential and conserves during the whole process so that the vortex beam creation can be considered as exchange of OAM between light and matter. Recently, an interesting approach is proposed that employs the beam polarization transformations in spatially inhomogeneous birefringent elements [206–209]. Such elements can be built on the base of discrete or continuous space-variant subwavelength dielectric gratings [206–208] (this technique is now applicable only to the midinfrared spectral domain, e.g., at a wavelength of 10.6 μm) or with using the liquid crystal technologies [209]. By continuously controlling the local birefringent properties of the element, any desired phase transformation can be performed.

In particular, the vortex wavefront is generated in an inhomogeneous half-wave plate whose optical axis orientation α depends on the transverse coordinates in accord with relation

$$\alpha(r,\phi) = q\phi + \alpha_0$$

where $q$ and $\alpha_0$ are constant parameters. Then, an input CP beam with Jones vector $\begin{pmatrix} 1 \\ \pm i \end{pmatrix}$ [49], after passing the plate, is transformed to the form $\exp\left(\pm 2iq\phi\right)\,\exp\left(\pm 2i\alpha_0\right)\begin{pmatrix} 1 \\ \mp i \end{pmatrix}$, i.e., the output beam is oppositely polarized and obtains the vortex phase factor of Eq. (31) with $l = \pm 2q$. Unlike other considered methods where the topological charge of the output beam is fully predetermined by the transforming optical system, here it is governed by the input CP handedness and thus can be controlled dynamically. In case of $q = 1$, the total AM of the beam does not change, which enables to treat this transformation as spin-to-orbital AM conversion [209].

### 4.3. Orbital angular momentum and the intensity moments

Before advancing further, briefly consider a relatively new efficient approach to characterization of paraxial beams, whose application has recently allowed to substantially extend understanding the nature of the OAM for beams with complicated spatial configuration. The matter concerns the optical Wigner distribution function [87] that for coherent paraxial beams is determined by expression [88, 89]

$$I\left(\mathbf{r},\mathbf{p}\right) = \frac{k^2}{4\pi^2}\int u\left(\mathbf{r}+\frac{\mathbf{r}'}{2}\right)u^*\left(\mathbf{r}-\frac{\mathbf{r}'}{2}\right)\exp\left[-ik\left(\mathbf{p}\cdot\mathbf{r}'\right)\right]\left(dr'\right). \qquad (45)$$

It generalizes the geometric-optical notion of ray intensity [90] and characterizes, from the consistent wave point of view, the beam energy distribution over the spatial **r** as well as angular **p** coordinates (directions of propagation). A concise description of the beam spatial structure is supplied by means of the Wigner function moments (intensity moments) [91–94]. The first moments form a vector

$$\begin{pmatrix} \mathbf{r}_0 \\ \mathbf{p}_0 \end{pmatrix} = \int \begin{pmatrix} \mathbf{r} \\ \mathbf{p} \end{pmatrix} I(\mathbf{r},\mathbf{p})(dr)(dp) \quad (46)$$

that combines the components of the center of gravity trajectory (30), the second moments form a positive definite symmetric 4×4 matrix

$$\mathsf{M} = \begin{pmatrix} \mathsf{M}_{11} & \mathsf{M}_{12} \\ \tilde{\mathsf{M}}_{12} & \mathsf{M}_{22} \end{pmatrix} \propto \int \begin{pmatrix} \mathbf{r} \\ \mathbf{p} \end{pmatrix} (\tilde{\mathbf{r}},\tilde{\mathbf{p}})\, I(\mathbf{r},\mathbf{p})(dp)(dr) \quad (47)$$

(symbol ~ denotes the matrix transposition).

Separate 2×2 blocks of the moment matrix possess the immediate physical meaning. The block

$$\mathsf{M}_{11} = \begin{pmatrix} (\mathsf{M}_{11})_{xx} & (\mathsf{M}_{11})_{xy} \\ (\mathsf{M}_{11})_{xy} & (\mathsf{M}_{11})_{yy} \end{pmatrix}$$

characterizes general features of the intensity distribution within the current cross section of the beam. As every positive definite symmetric matrix, it has the associated ellipse, so called "intensity ellipse" (Fig. 11), defined by equation

$$\left(\mathbf{r}\cdot\mathsf{M}_{11}^{-1}\mathbf{r}\right) = \frac{(\mathsf{M}_{11})_{yy}\, x^2 + (\mathsf{M}_{11})_{xx}\, y^2 - 2(\mathsf{M}_{11})_{xy}\, xy}{\det \mathsf{M}_{11}} = 1\,. \quad (48)$$

The matrix $\mathsf{M}_{11}$ eigenvalues equal to the squared semiaxes of the ellipse $b_1$ and $b_2$, and its orientation is determined by angle θ (see Fig. 11)

$$\theta = \frac{1}{2}\arctan\left( \frac{(\mathsf{M}_{11})_{xx} - (\mathsf{M}_{11})_{yy}}{2(\mathsf{M}_{11})_{xy}} \right). \quad (49)$$

Block $\mathsf{M}_{22}$ similarly characterizes the beam far field while the off-diagonal block $\mathsf{M}_{12}$ appears to be linked with the transverse energy flow (that in the considered scalar case can be expressed by Eq. (16) with dropped index $j$):

$$\mathsf{M}_{12} = \frac{1}{\Phi}\int \mathbf{r}\tilde{\mathbf{S}}_{\perp}(\mathbf{r})(dr)$$

and thus contains some information on the transverse energy circulation. In particular, it immediately determines the linear density of the beam OAM [50, 95] so that the normalized OAM (see Sec. 3.2, Eq. (26)) gets the representation

$$\Lambda_O = -k\,\mathrm{Sp}(\mathsf{M}_{12}\mathsf{J}) = \frac{1}{2}k\,\mathrm{Sp}\left[\left(\tilde{\mathsf{M}}_{12} - \mathsf{M}_{12}\right)\mathsf{J}\right], \quad (50)$$

where $\mathsf{J}=\begin{pmatrix} 0 & 1 \\ -1 & 0 \end{pmatrix}$ the simplest antisymmetric 2×2 matrix, $\mathrm{Sp}$ is a symbol of the matrix trace [47]. The OAM representation obtained above is quite general and includes also the extrinsic contribution stipulated by disposition of the beam as a whole (see the beginning of Sec. 4). To have just the intrinsic OAM, one should use Eq. (50) with elements of the matrix of central moments, which implies that the reference axis is chosen so that the first moments (46) vanish [50, 93].

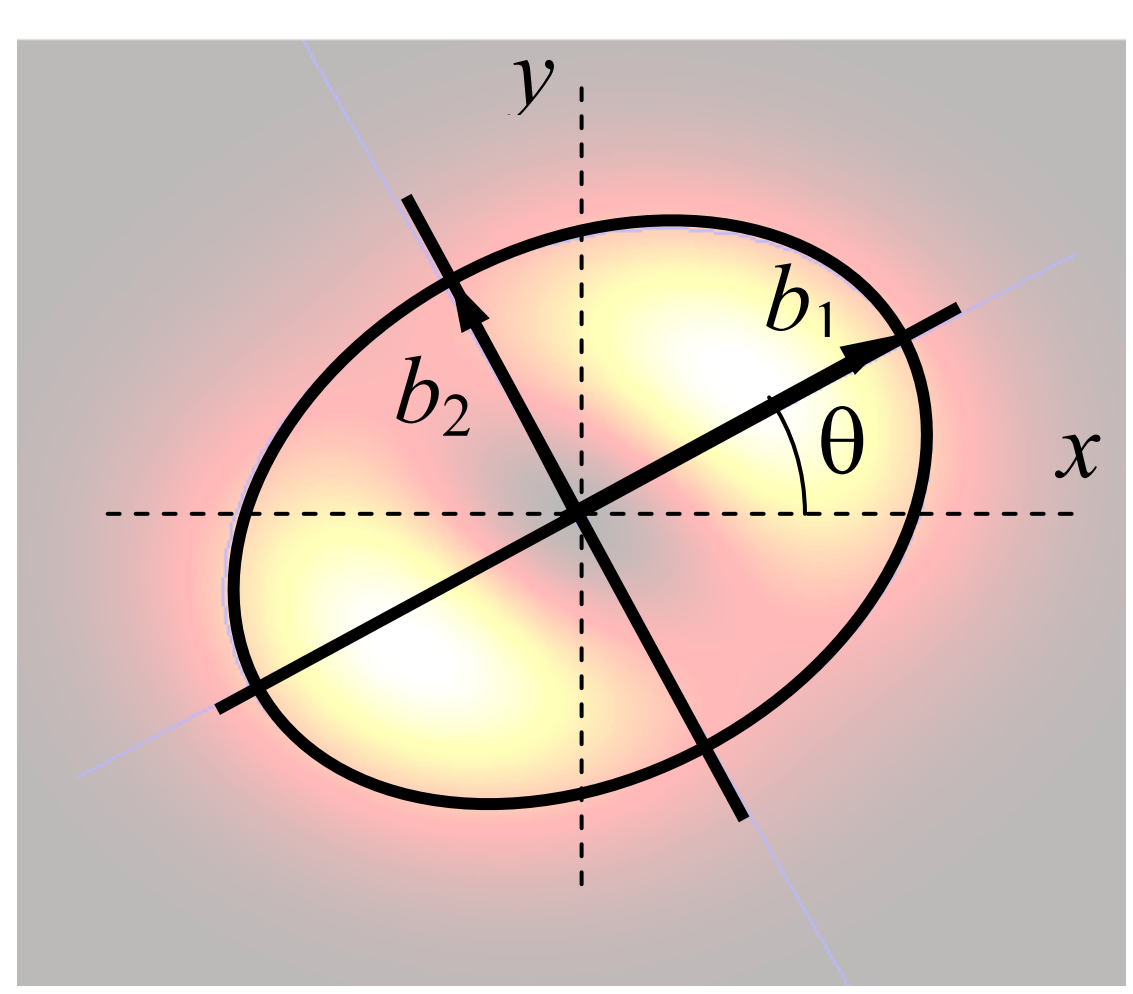

Fig. 11. Intensity ellipse on the background of the transverse intensity distribution (the beam "spot"): $b_1$, $b_2$ – semiaxes, θ – angle of the azimuthal orientation.

Eq. (50) presents a rather simple result that however turns out to be fairly important. Due to Eq. (50), the beam OAM finds a place within the well developed scheme of the beam characterization by means of the intensity moments [89, 93, 94], which allows to study the OAM and its properties from a more general point of view, employing all the associated advantages: methodological and metrological definiteness, availability of standard means for the beam investigation and corresponding software [94, 96–98]. This opens up additional facilities, some of which will be described below. Now we consider application of the intensity moments for the study of vortex beams whose form is more complicated than one specified by Eq. (31).

### 4.4. Symmetry breakdown and decomposition of the orbital angular momentum

It was noticed rather long ago that there can exist two qualitatively different forms of the transverse energy circulation in light beams [99, 100]. The "vortex" form is inherent in vortex beams (31) that were analyzed in detail above. Their peculiar feature is that the energy circulation caused by the helical WF is "hidden"; during the free propagation of the beam, no rotational redistribution of the light energy is visible (see, e.g., the left column in Fig. 12). The second form of the circulation explained by Fig. 13 occurs in asymmetric beams with mismatching main axes of the amplitude and phase distributions, for example, in an astigmatic Gaussian beam with the complex amplitude distribution

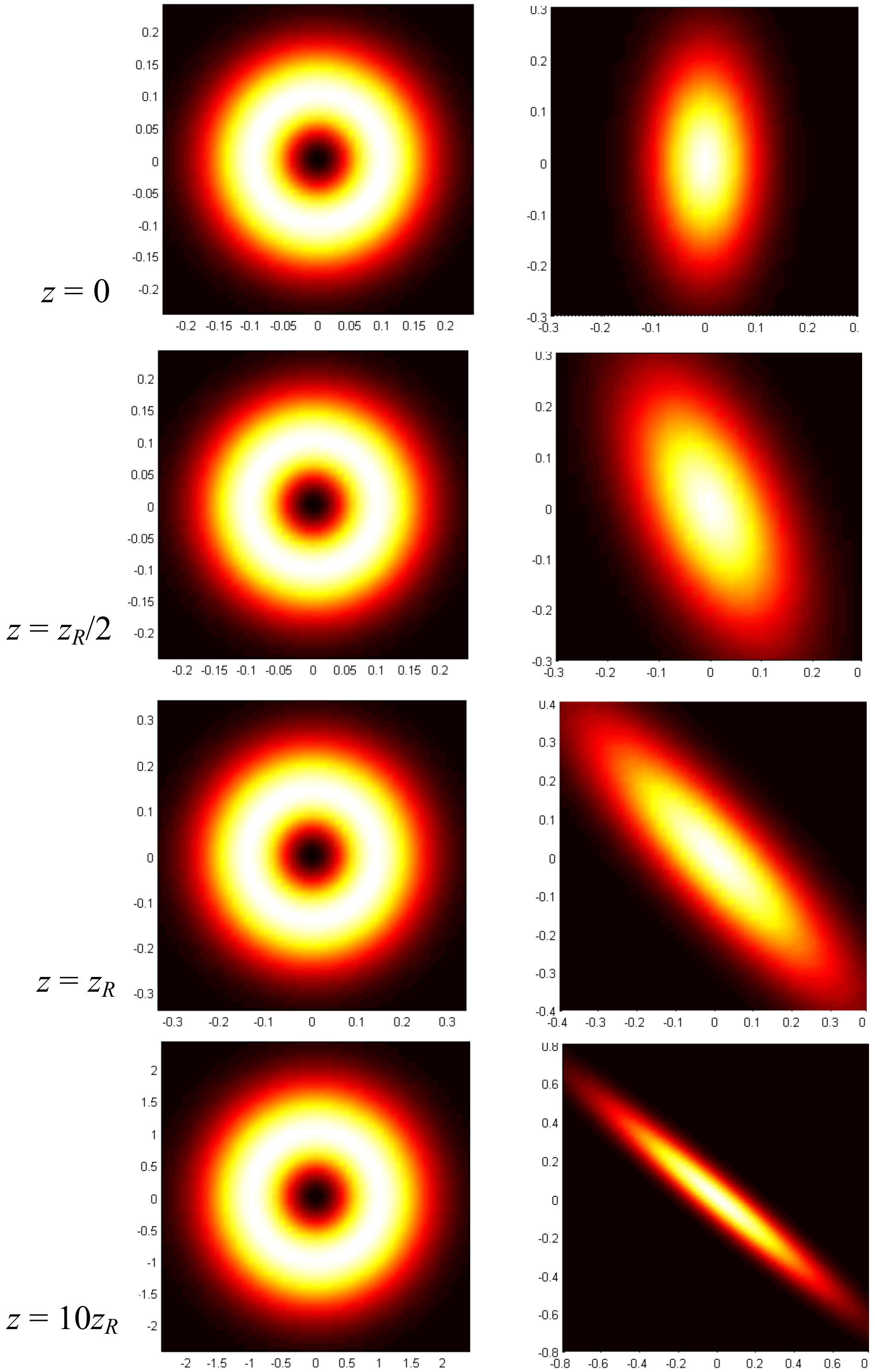

Fig. 12. The transverse profile evolution of a light beam generated by the He-Ne laser ($k = 10^5$ cm$^{-1}$): (left column) $LG_{01}$ beam (38) with the initial (waist) radius $b_0 = 0.1$ cm, Rayleigh length $z_R \approx 10^3$ cm; (right column) astigmatic Gaussian beam (51) with the initial transverse sizes $b_x = 0.1$ cm, $b_y = 0.2$ cm passed an astigmatic lens with focal distances $f_x$ = 80 cm, $f_x$ = 200 cm oriented at angle 0.46 rad = 27° with respect to the initial beam axes. Distances from the initial cross section are indicated.

$$u(\mathbf{r}) = \sqrt{\frac{8k\Phi}{c}}\left(\det \mathsf{D}_i\right)^{1/4} \exp\left[\frac{ik}{2}\left(\mathbf{r}\cdot \mathsf{D}\mathbf{r}\right)\right], \qquad (51)$$

where $\mathsf{D} = \mathsf{D}_r + i\mathsf{D}_i$ is a symmetric complex matrix [50]. As the right column of Fig. 12 witnesses, the visible rotation of the transverse structure takes place upon propagation of such beams. On the other hand, Eqs. (7), (8) allow to find the "true" instantaneous field of the beam with complex amplitude (51)

$$\boldsymbol{E}(\mathbf{r}, z, t) \sim u(\mathbf{r}) \cos[i\,(kz - \omega t)],$$

that is, contrary to the "vortex" case (Sec. 4.1, Fig. 7c), here is no rotation of the instant beam pattern and the circulation only appears in the large (relatively to the wavelength and the oscillation period) spatial-temporal scale. This form of transverse energy circulation can be called "non-vortex" or "asymmetry" circulation.

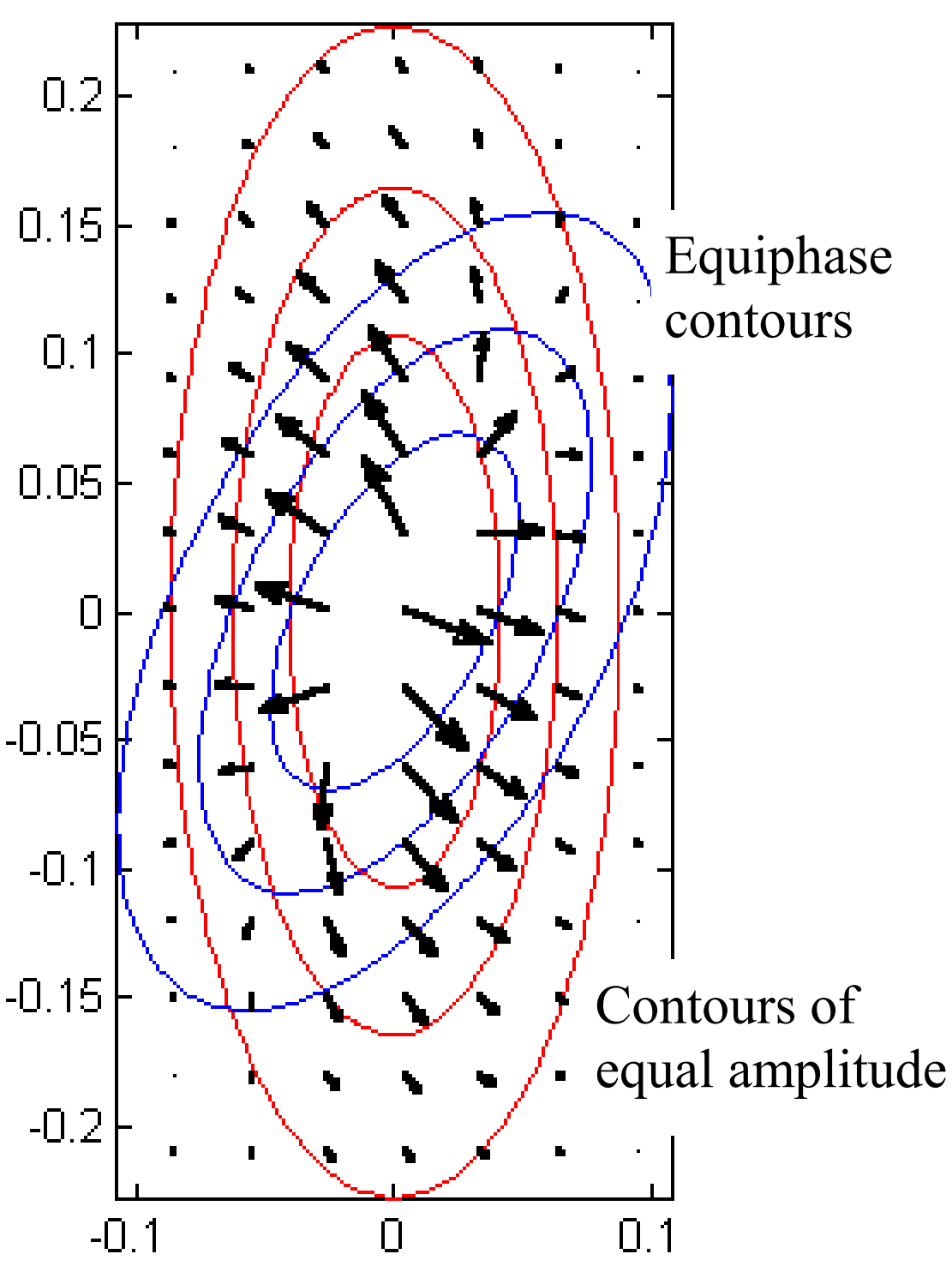

Fig. 13. Origin of the "asymmetry" form of energy circulation. Transverse components of the Poynting vector (black arrows) are directed orthogonal to the equiphase contours, their magnitudes are affected by the local amplitude. In the course of the beam propagation, the lower part of the beam cross section deviates a bit to the right, the upper one – to the left, and the beam as a whole rotates.

These are, so to speak, "pure" cases but the most frequent in practice are "mixed" situations when the beam simultaneously contains the WF singularities and asymmetry (for example, anisotropic screw WF dislocations [101, 102], "non-canonical" OVs [103, 104] etc.), which is often connected with the existence of several OVs arbitrarily situated within the inhomogeneous carrier beam [105–109]. In this case, a problem arises of separating the contributions of different nature, which can be made on the base of the following observations [50, 95].

The mentioned forms of circulation behave differently upon the beam transformations in thin phase correctors [31] (spherical or astigmatic lenses or mirrors). One can easily ascertain that in this process the vortex circulation of the LG mode (38) never changes while the asymmetry circulation of the Gaussian beam (51), after passing the corresponding corrector, can theoretically acquire any value, including the complete vanishing [50, 99]. This fact has been used in Refs. [50, 95] for introduction of the "vortex"

$$\Lambda_v = -2k \frac{\mathrm{Sp}\left(\mathsf{M}_{12}\mathsf{M}_{11}\mathsf{J}\right)}{\mathrm{Sp}\mathsf{M}_{11}} \tag{52}$$

and the "asymmetry"

$$\Lambda_a = \Lambda_O - \Lambda_v = k\left[\frac{2\,\mathrm{Sp}\left(\mathsf{M}_{12}\mathsf{M}_{11}\mathsf{J}\right)}{\mathrm{Sp}\mathsf{M}_{11}} - \mathrm{Sp}\left(\mathsf{M}_{12}\mathsf{J}\right)\right] \tag{53}$$

parts of the total OAM of the beam as quantitative characteristics of corresponding forms of the transverse energy circulation. Amounts (52) and (53) reflect the main properties of these forms: for example, $\Lambda_v$ does not change upon passing the quadratic phase correctors, and for a CS beam coincides with the whole OAM (50), and $\Lambda_a$ characterizes the asymmetry and the visible rotation of the beam in terms of the eccentricity $\varepsilon = \sqrt{\left(b_1^2 - b_2^2\right)/b_1^2}$ and orientation angle $\theta$ of the intensity ellipse (see (48), (49) and Fig. 11):

$$\Lambda_a = k\frac{\left(b_1^2 - b_2^2\right)^2}{b_1^2 + b_2^2}\frac{d\theta}{dz} = kb_1^2\frac{\varepsilon^4}{2-\varepsilon^2}\frac{d\theta}{dz}. \tag{54}$$

The introduced classification of the transverse energy circulation forms and the corresponding OAM constituents are useful instruments for studying the vortex beams with lack of circular symmetry and their transformations. The simplest example is the beam with anisotropic WF dislocation formed by superposition of two LG modes (38) with opposite signs of $l$ and different powers whose complex amplitude in the case $p = 0$, $l = \pm 1$ obtains the form

$$u(\mathbf{r}) = \frac{4}{b^2}\sqrt{\frac{\Phi}{c}}\left(x\cos\gamma_a + iy\sin\gamma_a\right)\exp\left(-\frac{x^2+y^2}{2b^2}\right), \tag{55}$$

where $\gamma_a$ is the anisotropy parameter determining the amplitude ratio of the superposition members. The transverse intensity profile of this beam possesses broken circular symmetry and, on a round trip near the axis, the phase vary non-uniformly, so the maximum rate of the phase change takes place in regions with the minimum intensity [110] (Fig. 14). For this beam, Eqs. (45), (47), (50), (52) and (53) give for the OAM and its constituents independent on $z$ values:

$$\Lambda(z) = \Lambda_v(z) = \sin 2\gamma_a, \quad \Lambda_a(z) = 0. \tag{56}$$

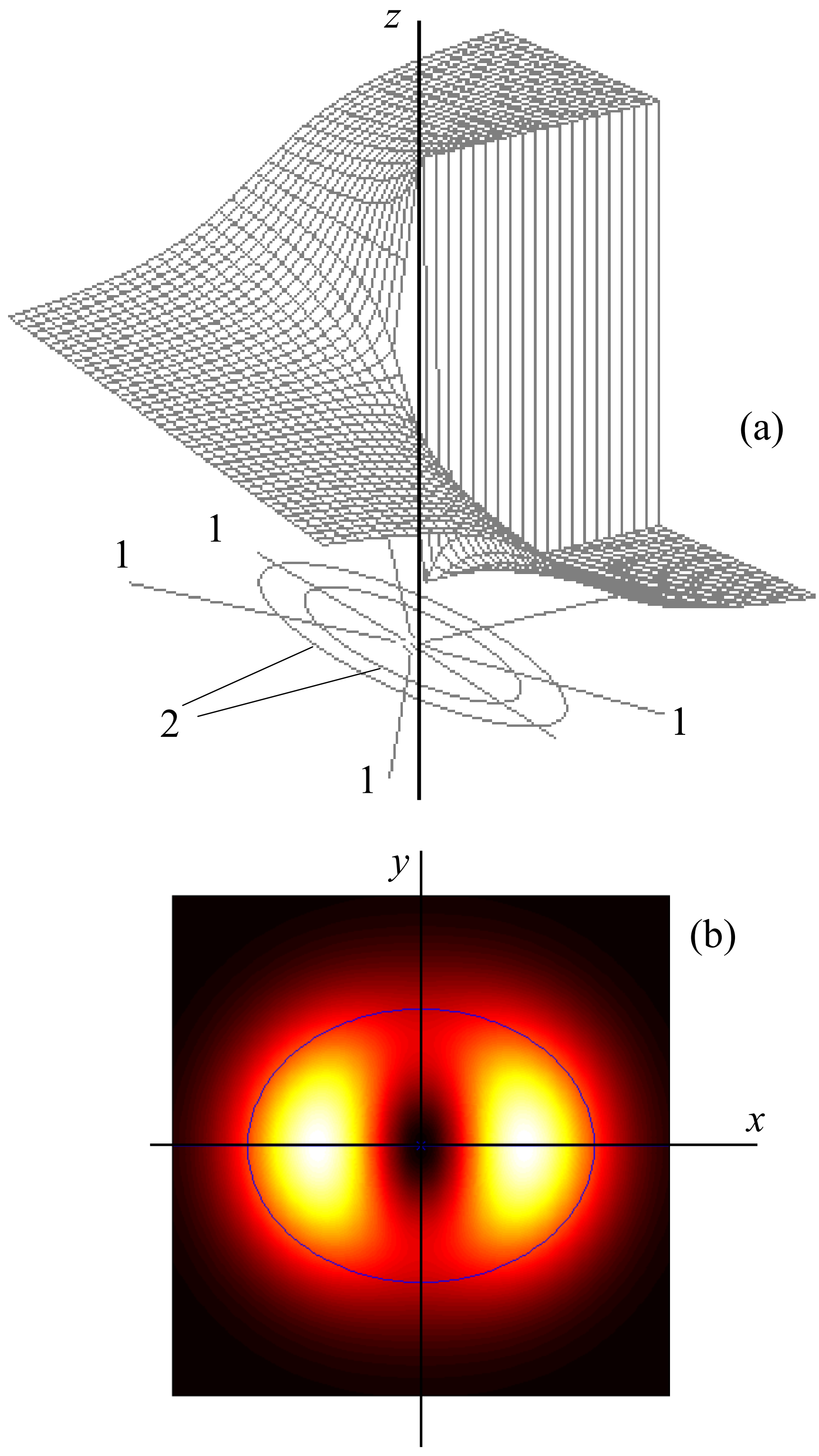

Fig. 14. Geometric structure of a beam with anisotropic screw WF dislocation, $|l| = 1$. The WF shape (a) is helical with pitch $\lambda$, but the phase changes with azimuth non-uniformly (higher rate of the phase change corresponds to higher closeness of projections of the equiphase lines 1 on the horizontal plane). Near the beam axis, lines of equal amplitude 2 are ellipses stretched along $y$ direction, if the whole intensity distribution (b) is stretched along $x$. In Fig. (b), contour and axes of the intensity ellipse are shown.

Therefore, despite the lack of symmetry, the whole OAM possesses completely vortex character while the asymmetry component is absent. This agrees with the fact that the beam "spot" (transverse intensity distribution) preserves its shape during the propagation [50], that is, the energy circulation is hidden.

This example also introduce us into the group of questions concerning the OV morphology, i.e. the description of optical field in the vicinity of the WF dislocation in more detail. Not being absorbed in the subject, exhaustively considered elsewhere [105–108], we only remark that the amount equivalent to $\sin 2\gamma$ plays an important role in the corresponding theory and is called "anisotropy parameter" of the OV [105]. Therefore, the first equation (56) determines the additional meaning of the vortex OAM which appears to be a natural measure of the OV anisotropy.

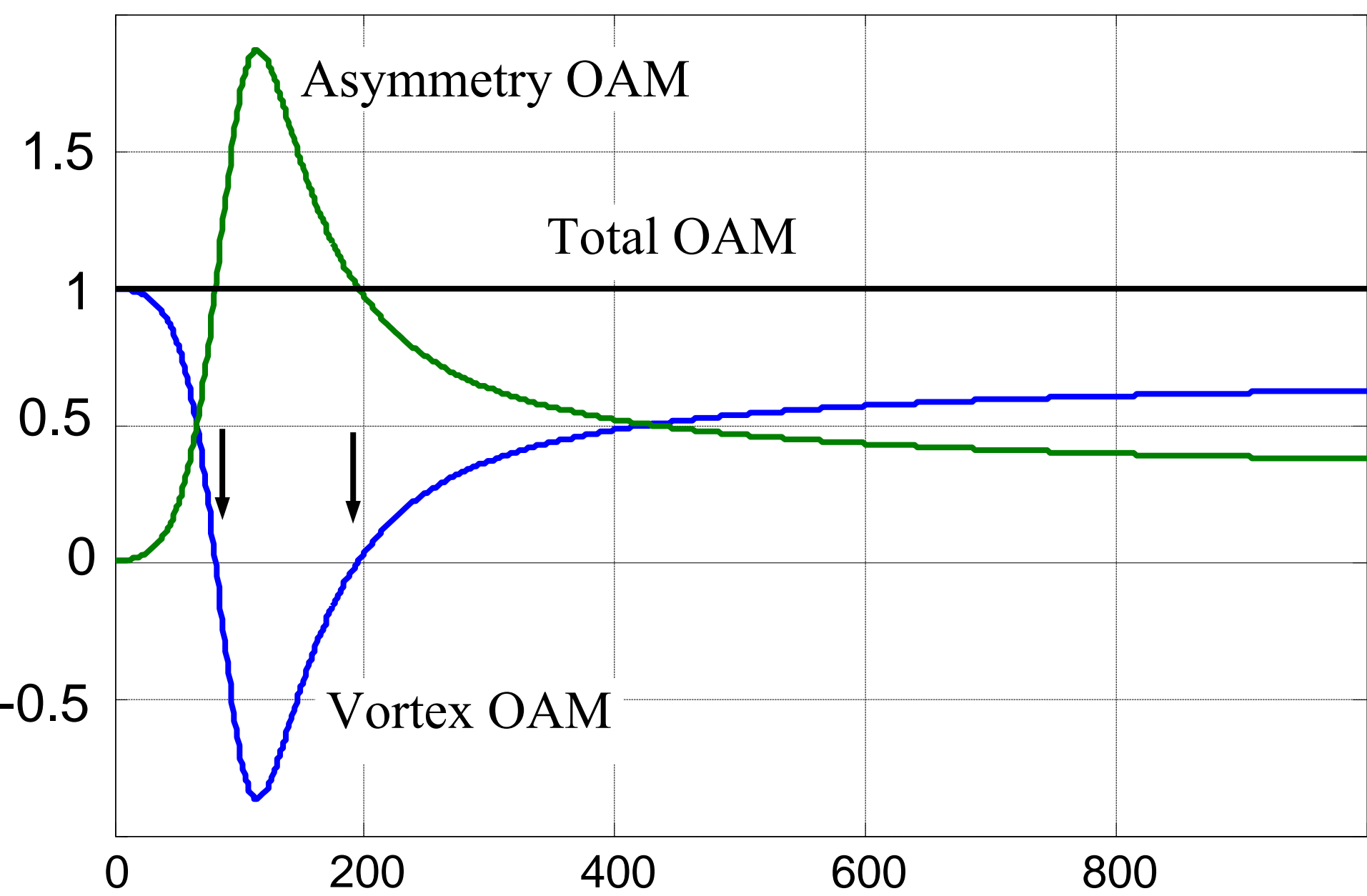

Fig. 15. Evolution of the beam OAM and its constituents upon propagation of the $LG_{01}$ beam of a He-Ne laser ($k = 10^5$ cm$^{-1}$) with the initial (waist) radius $b_0 = 0.1$ cm behind the astigmatic lens with focal distances $f_x = 80$ cm, $f_x = 200$ cm. Arrows indicate the inversion planes.

Another instructive example relates to transformations of a circular LG beam after passing a thin astigmatic lens [50, 109]; in Fig. 15, 16 the results, obtained for mode (38) with $p = 0$, $l = 1$, are presented. Just behind the lens the spot preserves its circular symmetry, the initial OAM possesses completely vortex character, the WF near the axis has the form of a left helicoid somewhat deformed due to focusing. But afterwards, the vortex OAM gradually decreases; instead, the asymmetry component appears (Fig. 15). Accordingly, the beam transverse profile becomes asymmetric and starts to visibly rotate (Fig. 16). At certain distances behind the lens – in the so called "inversion planes" ($z$ = 80 см i $z$ = 200 см in Fig. 16) – the vortex OAM vanishes totally;

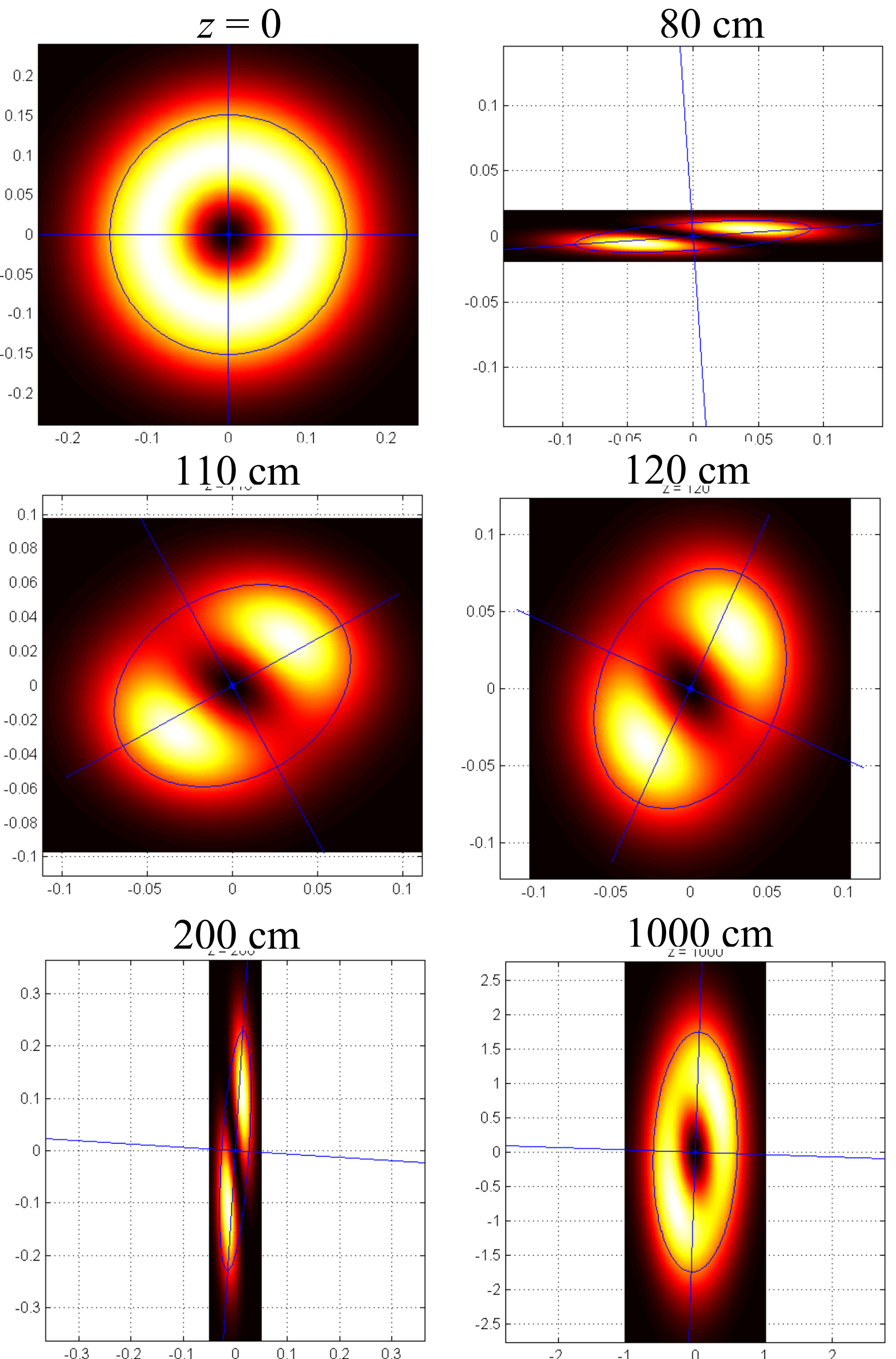

Fig. 16. Transverse profile transformation of a $LG_{01}$ beam after passing the astigmatic lens (the initial beam and the lens parameters are the same as in Fig. 15). Every image is marked by the corresponding distance from the lens and contains current contour and axes of the intensity ellipse. Scale labels at the transverse axes are given in centimeters.

simultaneously, the screw WF dislocation disappears (degenerates into an edge dislocation [26, 110]). Within the interval between these planes, the OAM sign changes and the WF shape also inverses to the right helicoid. However, the full OAM remains constant (which complies with the AM conservation, see notes at the end of Sec. 2) and all these changes are "tracked" by the asymmetry component. In the inversion region, the asymmetry OAM reaches its maximum absolute values, which, in accord with Eq. (54), is accompanied by the maximum rate of the visible beam rotation.

These and other examples [50, 109] clearly witness that in the beam transformations, accompanied by the symmetry breakdown, the vortex and asymmetry forms of the energy circulation can mutually convert and the corresponding OAM constituents are able to adequately characterize these processes. In particular, due to such transformations (besides the considered case of astigmatic focusing, there are possible astigmatic telescopic transformations, the slit diffraction, edge diffraction etc. [111]), the hidden vortex circulation converts into the visible asymmetry form, which enables to propose suitable methods of visualizing and determining the sign of an OV [50, 112].

## 5. MECHANICAL MODELS OF THE VORTEX LIGHT BEAMS

### 5.1. Spiral light beams

Numerous mechanical analogies and evident mechanical properties make the beams with OVs a demonstrative example of mechanical aspect of the electromagnetic field, supplying stimuli as well as opportunities for the deeper insight into the nature of such phenomena. The following questions appear: (i) is it possible to assimilate a vortex light beam to a mechanical rotating body; (ii) to what degree the mechanical ideas are applicable to such beams and, at last, (iii) what physical consequences they entail? To get the answers, one should carefully trace the "rotatory" evolution of the beam upon its propagation. Unfortunately, this task is complicated by two circumstances: in the propagation of CS beams of Eq. (31), the mentioned evolution cannot be seen because of the perfect circular symmetry of the intensity distribution, while the propagating asymmetric beams experience rather complicated deformation that can hardly be interpreted unambiguously (see, e.g., right column of Fig. 12). However, there exist paraxial beams that propagate "self-similarly", that is, preserving the shape of transverse intensity distribution and, at the same time, possessing a distinct transverse structure allowing to watch changes in the azimuthal orientation [113–121]; following to Refs. [113, 114], we will call them "spiral beams". Like any other paraxial beams, they can be represented as superpositions of LG modes (38)

$$u\left(r,\phi,z\right)=\sum_{p=0}^{\infty}\sum_{l=-\infty}^{\infty}C_{pl}\psi_{pl}^{\mathrm{LG}}\left(r,\phi,z\right) \tag{57}$$

($C_{pl}$ are weight coefficients). This equation represents a spiral beam if any pair of the superposition members with sets of indices ($p$, $l$) and ($p'$, $l'$) satisfy the relation [113, 114, 116–119]

$$B=\frac{2\left(p-p'\right)+\left|l\right|-\left|l'\right|}{l-l'}=\text{const}\,; \tag{58}$$

then the beam intensity pattern $S_{\|}(r,\phi,z) \propto |u(r,\phi,z)|^2$ rotates with "angular velocity"

$$\Omega_z=B\frac{d\psi\left(z\right)}{dz}=B\frac{z_R}{z^2+z_R^2}\,. \tag{59}$$

However, to have the complete notion on the self-similar evolution one should also take into account the beam divergence due to which any selected point of the beam cross section with the initial (at the waist $z = 0$) polar coordinates $r = r_0$, $\phi = \phi_0$ moves away from the axis so its current radial position $r(z)$ obeys the relation $r\left(z\right)=\left(r_0/b_0\right)b\left(z\right)$. Together with Eqs. (59) and (39), this gives the point "trajectory"

$$\phi\left(r\right)=\phi_0+B\arccos\left(\frac{r_0}{r}\right). \tag{60}$$

This equation describes spiral-like curves (which explains the name of spiral beams). If $|B| = 1$, the spirals degenerate into straight lines so that the "spot" evolution looks like centrifugal expansion of the beam "body" [115, 116, 122]. This picture seems quite demonstrative; however, it cannot be provided with a direct dynamical meaning (as it is possible, for example, in some non-linear cases of the rotational beam evolution [123–125]). Really, one can search the dynamical source of this rotation in the transverse energy circulation brought in superposition (57) by the vortex components with $l \neq 0$. However, Eq. (60) witnesses that the rotation can occur (theoretically) with arbitrary rate and direction, actually, regardless of the real weight of vortex components. More clearly this can be seen in confrontation with the beam OAM (3), equalling to the sum of OAMs of all the superposition members in (57) [121]

$$\mathcal{L}=\frac{1}{\omega c}\sum_{p,l}l\left|C_{pl}\right|^2\Phi_{pl}=\sum_{p,l}\left|C_{pl}\right|^2\mathcal{L}_{pl} \tag{61}$$

(here $\mathcal{L}_{pl}$ is the OAM of the individual LG mode $\psi_{pl}^{\mathrm{LG}}$ (38), see Eq. (33)). This expression shows that among the superpositions satisfying equations (57), (58), there can exist spiral beams with zero OAM (for example, two-term superpositions (1) with $p \neq p'$ and $l = -l'$) and even positively rotating beams ($B > 0$) can possess the negative OAM $\mathcal{L}$ and vice versa. Consequently, the OAM, known to characterize the overall energy circulation, appears to be inapplicable for description of the visible rotation of spiral beams.

Nevertheless, the mechanical interpretation can still make sense but in this situation it requires the deeper study of energy flows within the beam cross section, which, generally, are described by the transverse Poynting vector component (16). To show this [126], let us consider two examples of spiral beams (Fig. 17), formed by simple two-term superpositions of the class (57), (58), i.e. with all coefficients in (57) equaling to zero, except: $C_{12} = C_{00}\sqrt{\Phi_{00}/\Phi_{12}}$ (upper row) and $C_{02} = C_{20}\sqrt{\Phi_{20}/\Phi_{02}}$ (lower row). Rotation of the beams is seen immediately; their transverse expansion can be noticed due to the scale labels. Both beams carry the same total OAM independent on $z$ and equaling to $+\hbar$ per photon. This positive value corresponds to positive direction of the "gross" energy circulation (i.e., counter-clockwise in Fig. 17). Visual rotation of the $LG_{12} + LG_{00}$ beam ($B = 2$) agrees with this circulation while the $LG_{02} + LG_{20}$ beam ($B = -1$) rotates oppositely.

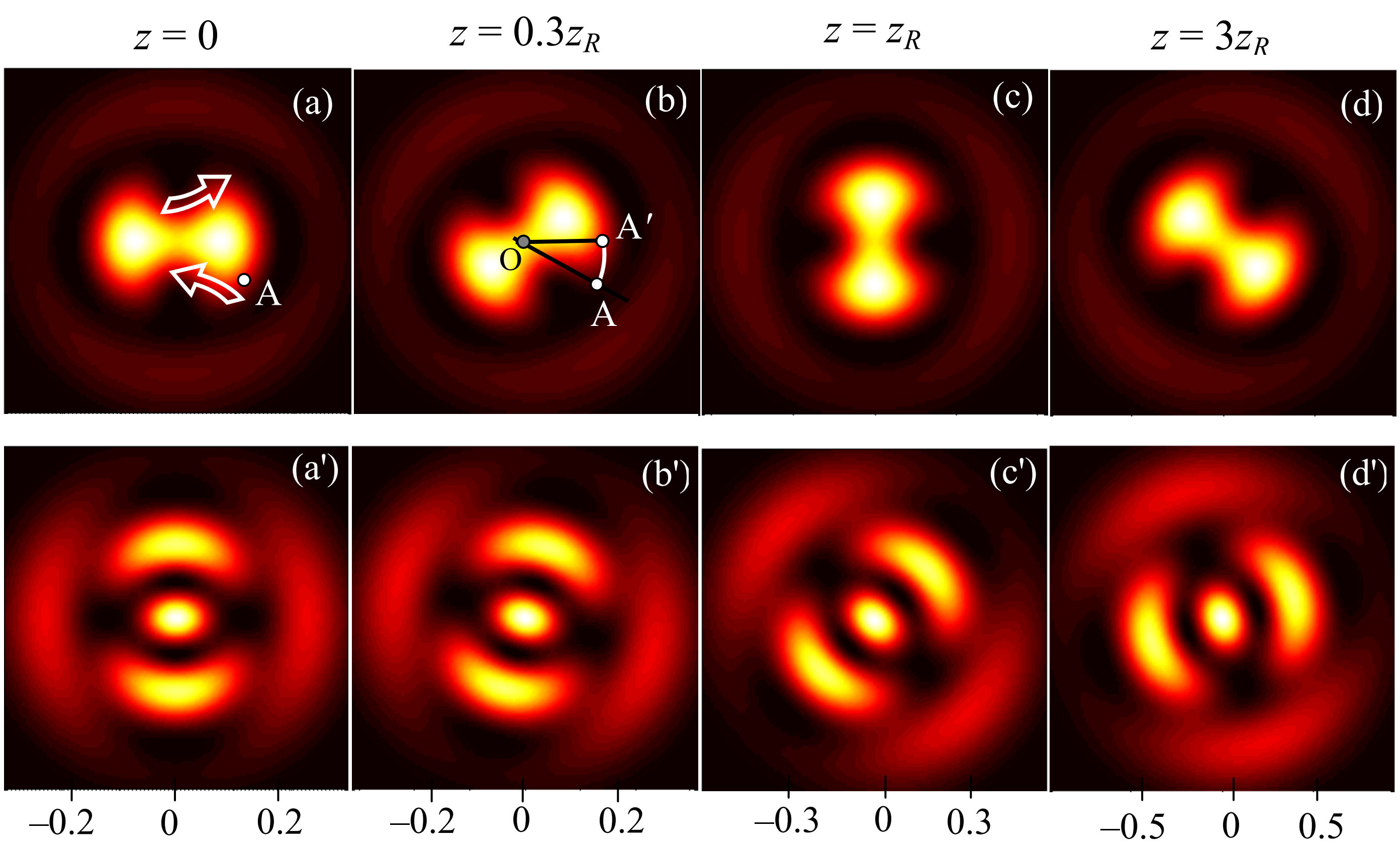

Fig. 17. Evolution of the transverse intensity pattern of model spiral beams viewed against the beam propagation: (a)–(d) superposition $\mathrm{LG}_{12} + \mathrm{LG}_{00}$ ($B = 2$); (a')–(d') superposition $\mathrm{LG}_{02} + \mathrm{LG}_{20}$ ($B = -1$). Distances from the initial (waist) plane are indicated; the beam broadening can be traced by the scale labels in the lower panels (in centimeters). Details in panels (a), (b) are explained in the text; the beam parameters accepted in calculations are: $b_0 = 1$ mm, $k = 10^5$ cm$^{-1}$ (He-Ne laser).

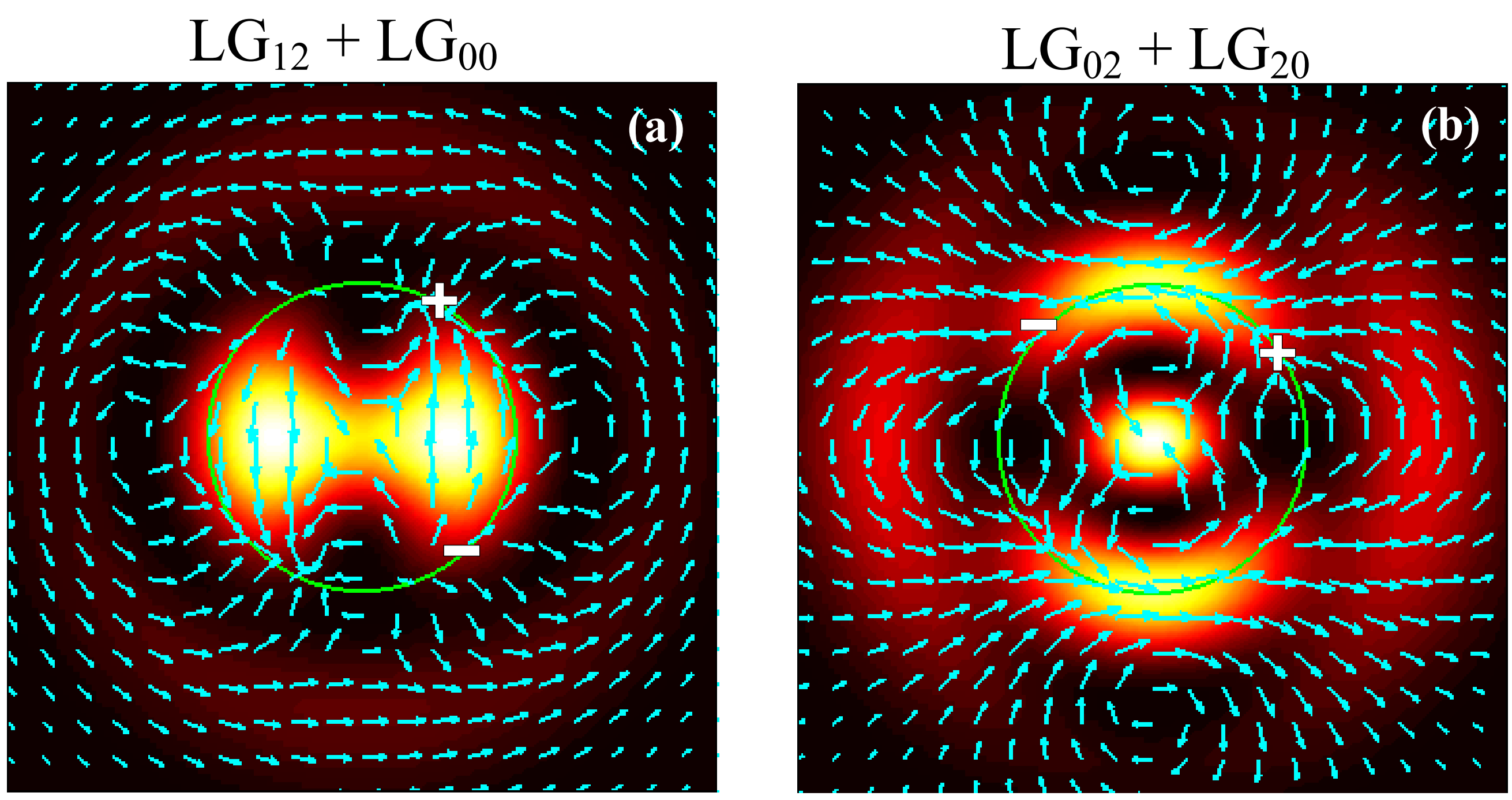

Fig. 18. Spot patterns of spiral beams with energy flow vectors, signs "+" ("–") denote regions where flow lines converge (diverge). Light circumferences are examples of closed contours surrounding the beam axis (see Ref. [126]).

The physical reasons for such behavior can be understood from Fig. 18 displaying maps of the transverse energy flows in these beams. Really, in both cases the energy circulation flow is directed counter-clockwise, which seems to testify that the whole pattern must have been "transported" also counter-clockwise. But in Fig. 18b, the energy current lines "converge" when approaching the intensity maximum and "diverge" on coming out of it. As a result, the energy concentrates in the "rear" and dissipates near the "front" of the bright spot. This process is superimposed on the "normal" energy transport, and the summary effect determined by the competition is the "backward" rotation. In Fig. 18a, the energy flow convergence and divergence act in agreement with the "normal" energy transport and increase the positive rotation velocity.

The described picture of transverse energy flows enables to explain apparent discrepancies between the visual beam rotation and its OAM. Let an observer watch, for example, the beam transformation from pattern of Fig. 17a to that of Fig. 17b. "From the outside", it looks as if the light energy moves from a certain initial point, say, A to its current position at point A' along the arc AA'. Such way of reasoning implicitly supposes that the beam "rotates" like a rigid body, which, in view of the above remarks, is generally incorrect and leads to false conclusions. However, we can still use a mechanical model for the beam rotational evolution but then a beam should be assimilated to a fluent medium with flexible internal motion characterized by the energy flow pattern. In particular, the visible transformation of A into A' can be realized if some "portion" of the beam energy moves from A towards the beam axis O, and the equivalent portion goes from the beam central area to A', approximately as is shown by white arrows in Fig. 17a. Comparison of Figs. 17a, 17b and Fig. 18a shows that real picture of the energy transfer is close to this schematic which, of course, is coupled with much smaller OAM with respect to the beam axis than the "direct" motion along AA'.

Quantitatively, transformation of the intensity pattern of a propagating beam is described by relation

$$\frac{\partial S_{\parallel}}{\partial z} = -\operatorname{div}\mathbf{S}_{\perp}, \tag{62}$$

that follows directly from the equation of paraxial propagation (9). By taking into account Eqs. (15) and (22), this result can be presented in terms of the energy density (22)

$$c\frac{\partial w'}{\partial z} = -\operatorname{div}\left(w'\mathbf{v}_{\perp}\right) \tag{63}$$

where the "transverse energy flow velocity" $\mathbf{v}_{\perp} = \dfrac{\mathbf{S}_{\perp}}{w'}$ is introduced. Since the spot transformation is accompanied by the longitudinal translation with the light velocity *c*, any longitudinal transformation of the beam can be provided with the temporal sense by relation

$$c\frac{\partial}{\partial z} = \frac{\partial}{\partial t}, \tag{64}$$

and then the meaning of result (63) as the continuity equation for the energy density becomes obvious:

$$\frac{\partial w'}{\partial t} = -\operatorname{div}\left(w'\mathbf{v}_\perp\right). \tag{65}$$

Equations (62), (63) form a ground for the "hydrodynamic" approach to the beam evolution that appears especially helpful in complex situations of stochastic wave propagation [127–129]. Here we see that in case of regular beam evolution it can also be fruitful. Notice that by introducing the electromagnetic mass [12, 29] with the density

$$m_e' = \frac{w'}{c^2} \tag{66}$$

we can formulate the law (62) of the beam transformation exactly in terms of the fluid mechanics; however, this analogy is not full because, except the continuity equation (62)–(65), the electromagnetic "fluid" obeys also the Maxwell equations that differ from the Navier–Stokes equations of the fluid motion [130].

### 5.2. Mechanical structure of a helical beam

The presented qualitative analysis of the spiral beam evolution in terms of mechanical motion shows the consistency and heuristic value the mechanical argumentation. The picture becomes especially clear and admits the exhaustive analytical description in application to pure LG modes of the form (38) [131]. To show this, we at first consider the simplest spiral beam of the family (57), (58) formed by superposition of a Gaussian beam and an $LG_{01}$ mode (the so called "off-axis OV"), whose evolution is shown in Fig. 19. Its rotation is seen due to the eccentric dark spot; the rotation rate $\Omega_z$ is determined by Eq. (59) where $B = 1$. The "temporal" velocity, due to Eq. (64), accepts the form

$$\Omega\left(z\right) = \frac{d\theta}{dt} = c\frac{d\theta}{dz} = c\Omega_z = \frac{c}{kb^2\left(z\right)}. \tag{67}$$

Note that, because of independence of the components' amplitudes, this relation is valid also in the limit case when the amplitude of the LG component vanishes. This means that one can consider the velocity (67) as the rotation velocity of a "pure" $LG_{01}$ mode.

Therefore, in addition to the AM with the linear density (33), the beam with OV obtains the new mechanical attribute – the rotation angular velocity (67). Besides, it possesses the electromagnetic mass (66) with the linear density that follows from Eqs. (24) and (25),

$$m_e = \frac{w}{c^2} = \frac{\Phi}{c^3}.$$

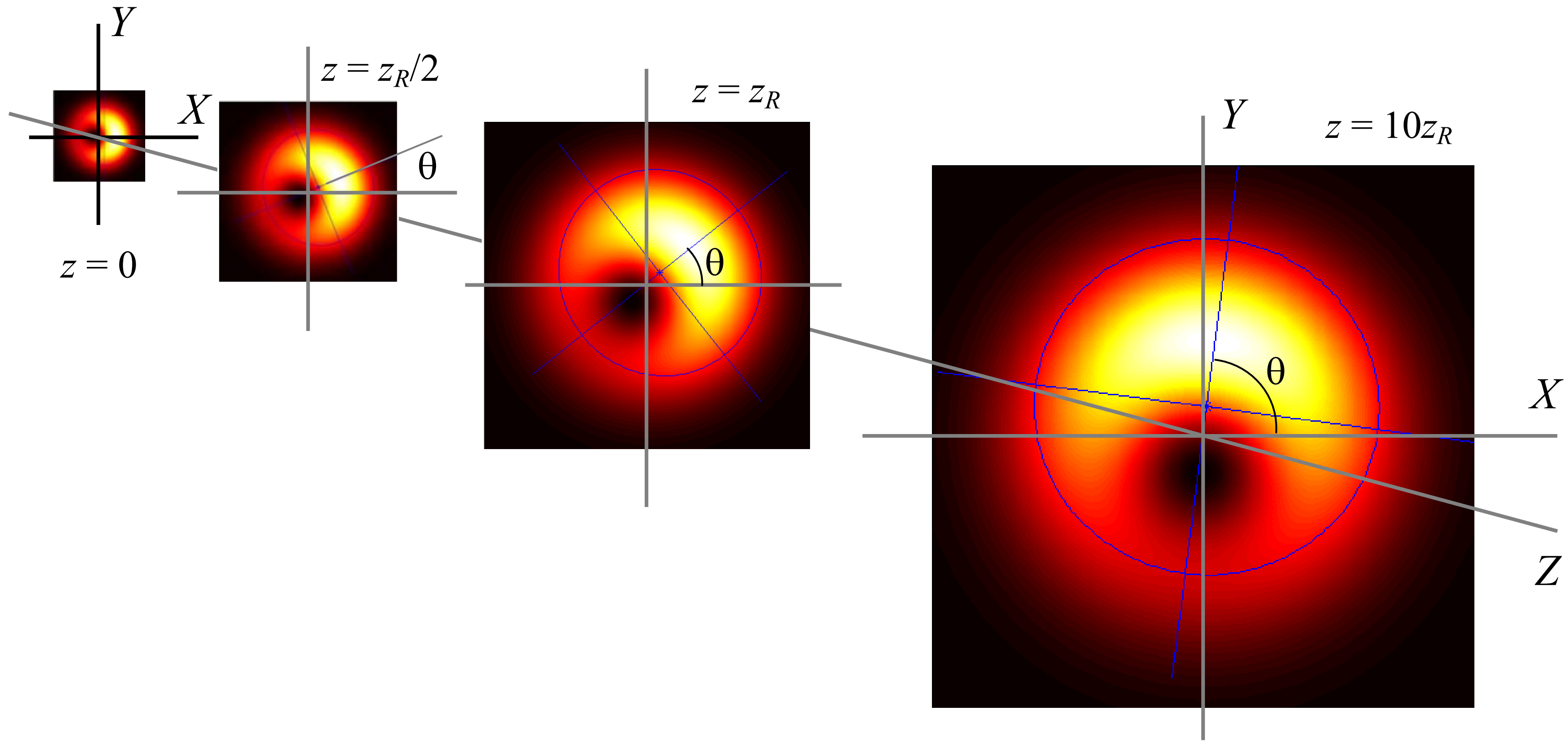

Fig. 19. Transverse profile transformation of the coherent mixture of Gaussian and $LG_{01}$ beams with relative part of the Gaussian component $\Phi_G/\Phi$ = 0.1 (off-axial vortex). Each image is labeled by corresponding distance from the beam waist. The beam broadening is shown schematically, without maintaining the scale.

With allowance for this equation, OAM (33) of the $LG_{01}$ beam can be presented in the "body-like" form

$$\mathcal{L} = J(z)\,\Omega(z), \qquad (68)$$

where amount $J(z) = m_e b^2(z)$ has a meaning of the moment of inertia of the beam unit length. Its expression follows from the confrontation of Eqs. (57, (68) and (33) and appears to be very similar to the known expression $J(z) = 2m_e b^2(z)$ for the moment of inertia of a mechanical body [132] with the same mass density distribution; the only difference is the additional coefficient 2. This difference is caused by the fact that in the "body", modeling a vortex beam, the OAM density is proportional to the intensity, that is, to the electromagnetic mass distribution while in a rigid body, points remote from the axis produce greater contributions since their velocities are proportional to the off-axis distances. We can explain the proportionality between the OAM and electromagnetic mass distributions only by supposing that in the mechanical model of an LG beam the velocities of points decrease with moving away from the axis (see Fig. 20).

This conclusion agrees with calculations of the helical energy transfer in such beams [133], which show increase of the energy flow azimuthal velocity with approaching the beam axis (see Fig. 21). This confirms that the "energetic body" of an LG beam is also not rigid; its radial layers "slip" with respect to each other. The vortex beam evolution can be treated as the vortex motion of a fluid "body" whose density is determined by the mass equivalent of the electromagnetic beam energy. It is worth noting that the velocity field of this motion is quite similar to the velocity field in the vicinity of a straight-line vortex filament in fluid or to the magnetic field distribution near a straight-line electric current [130]. This observation points out the deep analogy between vortex motions of different nature and serves an additional evidence for the universal character of physical laws.

In conclusion, we stress again (first it has been made in Sec. 4.1) on the connection of the considered mechanical models with field characteristics averaged over the oscillation period. A new argument against direct mechanical interpreting the rotation of instantaneous patterns of the sort presented in Fig. 7c consists in the fact that any point of this instantaneous "energetic body", remote from the axis by distance $\geq k^{-1}$, must have superluminal linear velocity, which is quite inappropriate for a mechanical model. Nevertheless, combination of mechanical analogies with the symmetry considerations appears to be useful [131]. As is shown in Sec. 4.1, helical beams possess the special symmetry due to which translation along axis $z$ with speed $c$ is physically identical to the rotation around the same axis with angular velocity $\omega/l$; in other words, these motions are not different but rather the same motion can be treated in two equivalent ways. So, its kinetic energy can be ascribed either to the rotation or to the translational motion:

$$\frac{1}{2}J\left(\frac{\omega}{l}\right)^2 = \frac{1}{2}m_e c^2 ,$$

whence the expression for the beam moment of inertia follows

$$J = m_e c^2 \left(\frac{l}{\omega}\right)^2 = w\left(\frac{l}{\omega}\right)^2 .$$

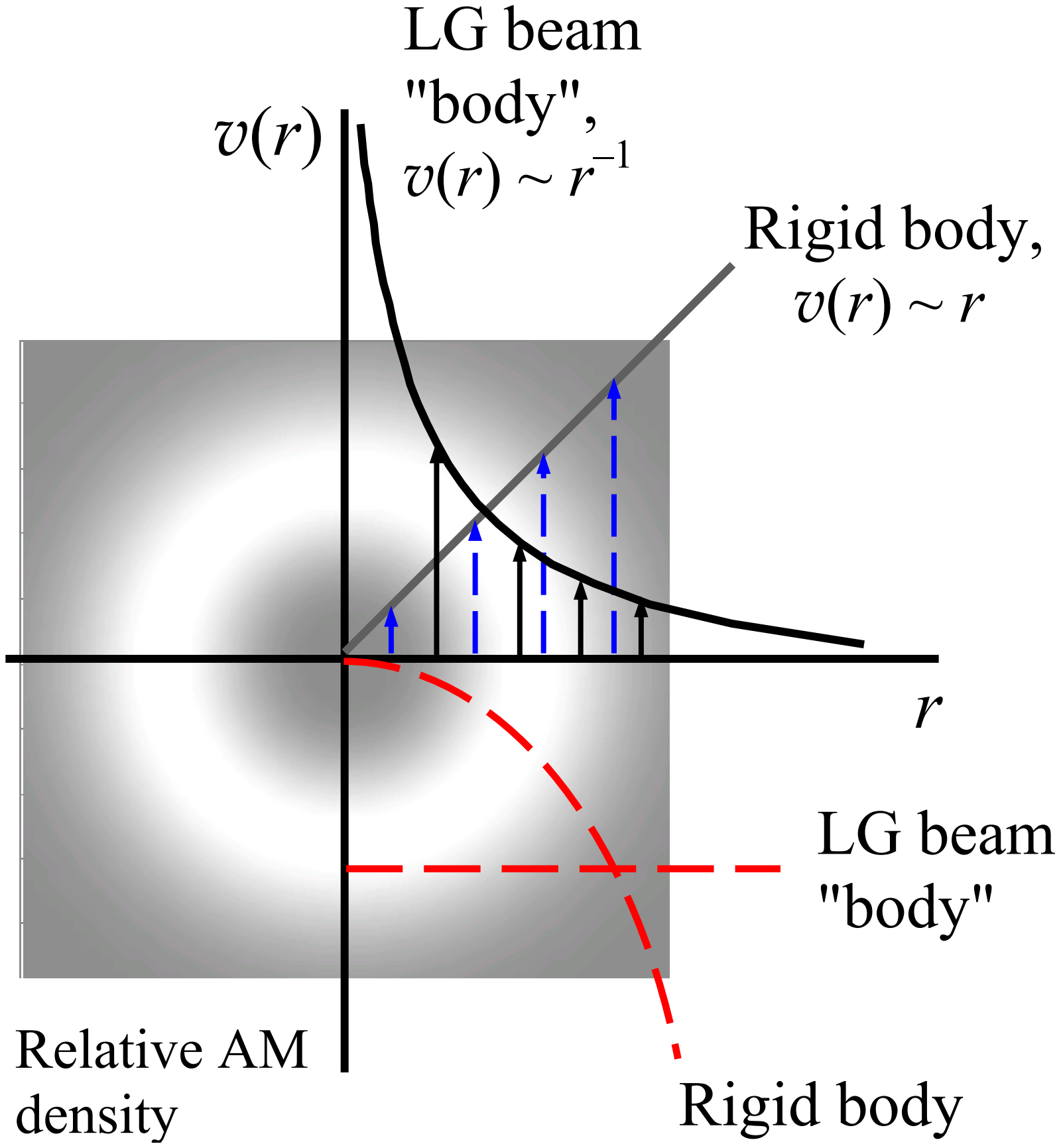

Fig. 20. Comparative transverse distributions of rotational characteristics of the $LG_{01}$ beam and of the rotating rigid body: solid lines – azimuthal velocity of points, dashed lines – relative AM density (ratio of the volume AM density to the mass density)..

This final result does not depend on the mass $m_e$ definition and can be considered as universal law for objects with this sort of symmetry. By the way, it determines the known relationship (33) between the energy and the angular momentum of the helical beam $\mathcal{L} = J\left(\omega/l\right) = l\, w/\omega$, which therefore obtains substantiation in the helical symmetry alone and can thus be considered a universal property of objects with such symmetry type.

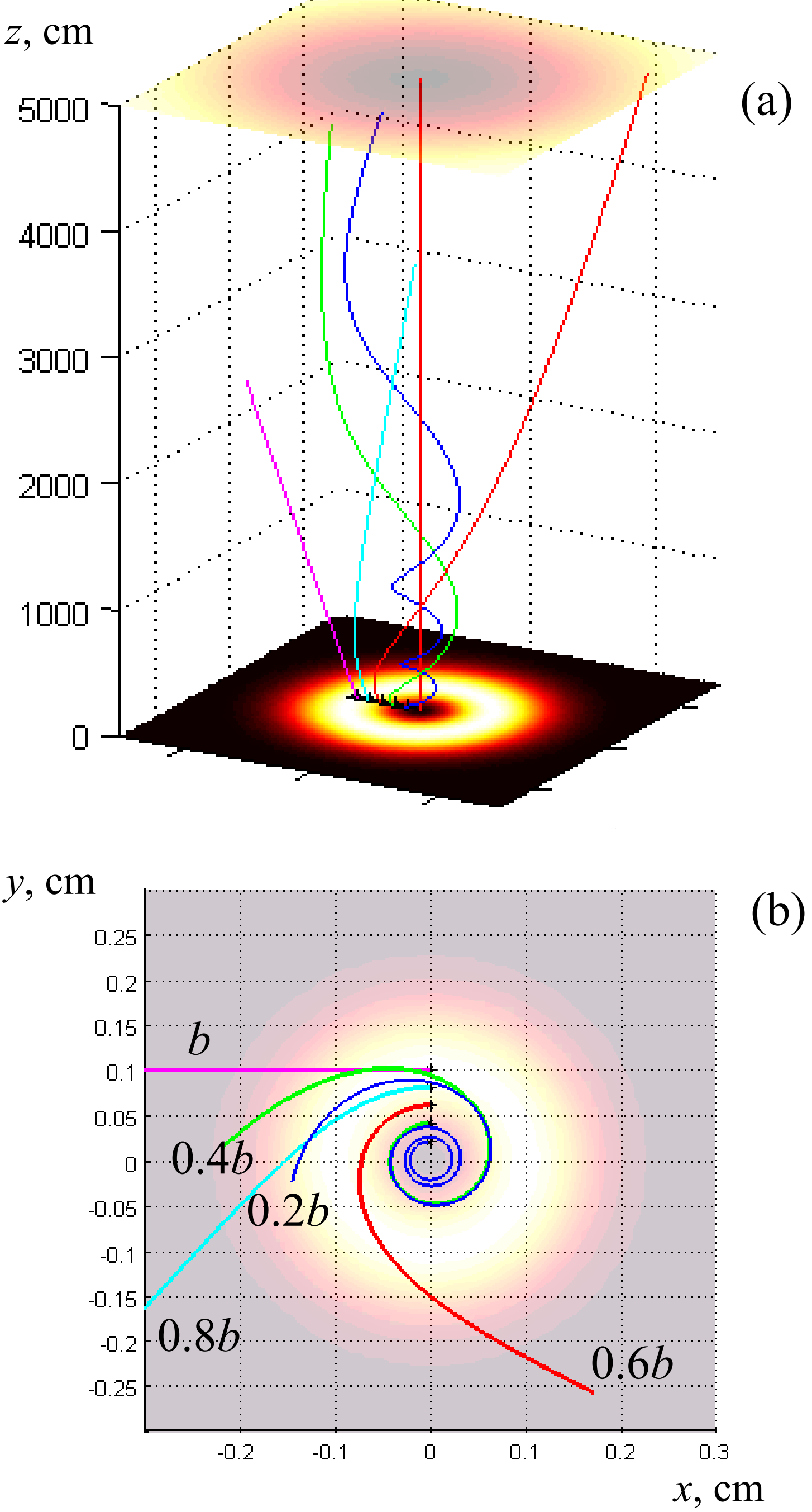

Fig. 21. Energy flow lines in the $LG_{01}$ beam with parameters indicated in caption to Fig. 15: (a) in the axonometric projection, (b) view from the positive side of axis *z*. Lines are labeled by corresponding distances from the beam axis in units of the beam radius *b*.

## 6. MECHANICAL ACTION OF THE BEAM ANGULAR MOMENTUM

Another feature that links light beams with OAM and mechanical rotating bodies is the capability to cause the rotational motion of other objects. Starting from general ideas of the dynamics, this effect can be described in a rather simple manner [134]. Let the beam possess the AM with linear density $\mathbf{L}_1$ before interaction with a certain optical system (at the input), and $\mathbf{L}_2$ after the interaction (at the output); the vector character of AM is important because the beam direction can change. This means that in each second the system receives the AM of amount $c\mathbf{L}_1$ and delivers the AM $c\mathbf{L}_1$; respectively, the torque

$$\mathbf{K} = c(\mathbf{L}_1 - \mathbf{L}_2), \qquad (69)$$

is applied to the system, and this is a manifestation of the mechanical rotatory action of the beam. From the general point of view, this effect is analogous to the light pressure [135], which also takes place in this process and stipulates the mechanical force acting on the system

$$\mathbf{F} = c(\mathbf{P}_1 - \mathbf{P}_2), \qquad (70)$$

where $\mathbf{P} = \int \mathbf{\Pi}(dr)$ is the beam momentum (see Eq. (2)). This force 'per se' is now not of our interest, but we can note that, in the order of magnitude, the torque $\mathbf{K}$ relates to the light pressure force $\mathbf{F}$ as the longitudinal energy flow relates to the transverse one, i.e. is of the order $\gamma|\mathbf{F}| \sim 10^{-3}|\mathbf{F}|$ (see Sec. 2). Extremely weak for the usual laboratory conditions, for a long time manifestations of the mechanical action of light were predominantly observed either in very large (astrophysics) or in very small (atomic) scales; only the invention of lasers changed the situation [136].

### 6.1. Beth's experiment (measurement of the spin angular momentum)

The first experimental confirmation of the vortex mechanical properties of a light beam was obtained as early as in 1936 but despite the long past time, the famous work by Beth [9] constitutes not only the historical interest. The ideas and experimental tricks used in it are still actual. They find applications, in particular, in the frame of the rotational Doppler effect that will be considered below, so we describe this work in more detail.

The experimental equipment included a half-wave plate, suspended with a quartz filament, and a fixed quarter-wave plate with reflecting layer on its back side (Fig. 22). The suspended plate formed a torsion pendulum with the oscillation period about 10 s. Let the input light beam possess the left CP ($\beta = i$ in Eq. (19)) and, if its power is $\Phi$, carries the SAM with linear density $\mathcal{L} = \Phi/\omega c$ (see Eq. (27)). After passing the plate, the beam polarization becomes opposite and, in accord with Eq. (69), the plate experiences torque

$$K_1 = c(2\Phi/\omega c) = 2\Phi/\omega.$$

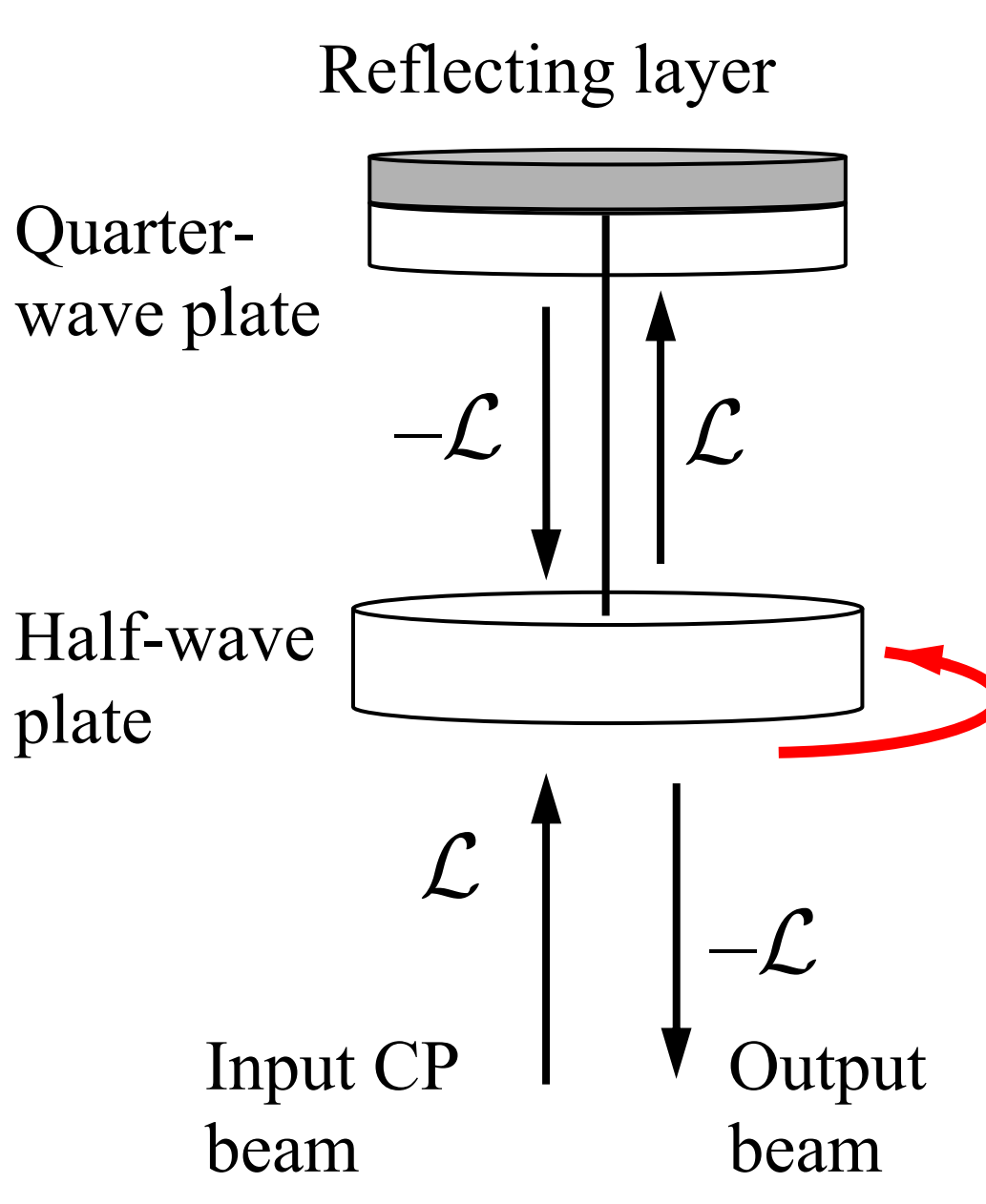

Fig. 22. Experiment for the CP beam measurement [9].

After reflection and second passing the quarter-wave plate, the beam polarization is restored, and thus, interacting with the suspended plate again, the additional torque of the same amount is exerted. So, the total torque felt by the half-wave plate amounts to $2K_1 = 4\Phi/\omega$. This value was still too small to cause a noticeable plate rotation, and the periodic toggling the CP sign of the input beam was employed, which gave possibility to agitate the torsion pendulum. The resulting plate oscillations were observed with the help of a telescope.

In such a way, this experiment had not only demonstrated the existence of the rotational action of CP beams but also had proved relation (27) between the AM and the beam energy.

### 6.2. Angular momentum transfer to small particles

Novel developments of the experimental technique provide substantial facilitations in observing the light beam mechanical action, in particular, the rotatory one. In most cases, the well-developed technique of keeping separate particles in optical levitation traps is employed [136–138], due to which it is possible to watch the motion of separate microparticles within the field of optical beam of necessary configuration [139]. Various experiments differ by certain details, sometimes of technical importance, but their principles are essentially the same, so we present the generalized scheme (Fig. 23).

Absorptive particles suspended in a liquid (ceramic particles in kerosene, copper oxide in water [140, 141], glass or teflon particles in various liquids [142]) are optically and/or mechanically trapped within a laser beam strongly focused by the microscope objective. Absorbing a certain part of the beam energy, the particles simultaneously obtain the associated part of its AM. This causes the rotational motion of particles that can be seen by the microscope or registered by means of a TV camera. Usually, possibility of changing parameters of the investigated "governing" laser beam is provided (for example, to supply an input beam in forms of $LG_{0l}$ modes with different $l$ and in various polarization states).

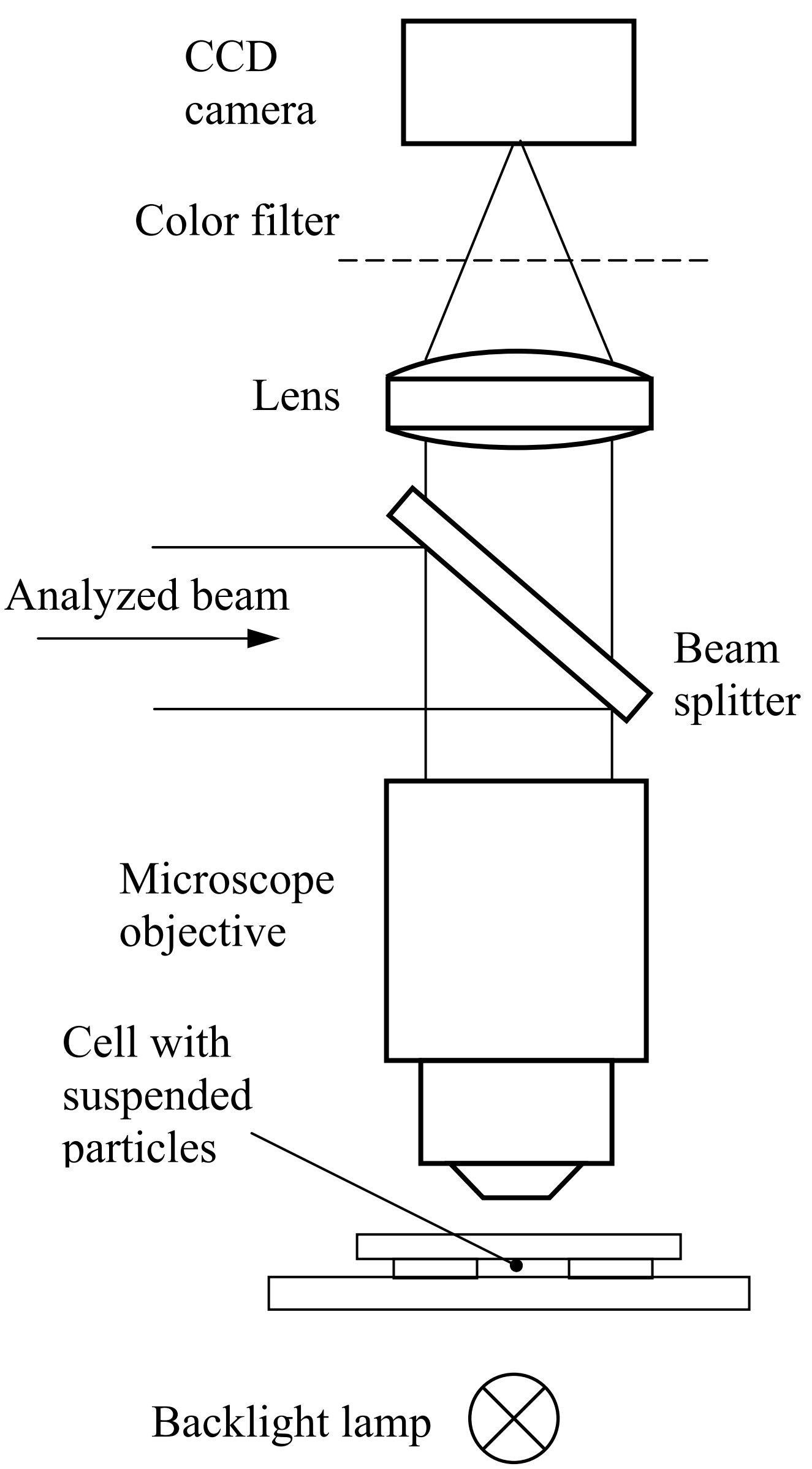

Fig. 23. Observation of rotatory action of light beams with AM

Results of experiments show that when a linearly polarized CS beam with $l = 0$ falls onto a particle, no rotation is observed, which is quite understandable since such a beam contains no AM. The situation becomes more interesting when vortex beams are employed. Regardless of the spin or orbital character of the beam AM, the particles begin to rotate in agreement with the AM sign; if the beam with non-zero $l$ is also circularly polarized, the rotation of a particle accelerates or slows down to the full stop, depending on whether signs of the beam SAM and OAM are the same or opposite [142].

These experiments convincingly not only demonstrate the existence of mechanical action of beams with AM but also prove the mechanical equivalence of the spin and orbital parts of the beam AM (although the latter statement can hardly be called in question in view of the nature of the beam AM decomposition as presented in Sec. 2). By analogy with the above-described Beth's experiment, one could expect such measurements to be also a means for determining the OAM value, but because of a number of uncontrolled factors affecting the motion of particles this appears to be impossible. For this purpose, experiments with macroscopic systems changing the beam OAM seem to be more promising (for example, those with astigmatic lens converters [53, 54] that appear as "orbital" analogs of half-wave plates [76]). However, to our knowledge, such experiments have not been realized so far, perhaps, because the lens systems are rather massive and mechanically complicated.

Nevertheless, the "suspended particle" technique allows detecting the presence of the beam AM [52, 143]. Experimental skills and experience, obtained in the course of investigations of the particle motions in the field of beams with AM, has been successfully used for development of so-called "micromachines", designed for manipulation, sorting and spatial concentrating of particles [144–146]. There are also propositions to utilize the beams with AM for governing the position of a particle within an optical trap with the aim of exhaustive study of its properties [147].

Problems of the OAM measurement are not the topic of this paper, but as long as it is concerned here, we mention works [148, 149, 150] where, instead of the beam mechanical action, direct determination of the beam spatial parameters responsible for the OAM is proposed. It turns out that upon the beam propagation through a refracting boundary or at diffraction on a grating into a non-zero order, the beam transverse profile experiences certain modifications [151] (in particular, the specific shift and deflection of the "center of gravity" trajectory, see Sec. 4) which are directly connected with its OAM. Despite the very small expected value, the effect is quite measurable and its first experimental witnesses have been reported [152].

## 7. ROTATIONAL DOPPLER EFFECT

### 7.1. Geometrical (kinematical) aspect

One of the most known wave phenomena, the Doppler effect consists in dependence of the observable oscillation frequency on the relative motion of a wave source and an observer [29, 132]. For example, in a plane wave, the spatial distribution of perturbations is determined by expression

$$E \sim \cos(\mathbf{kR} - \omega t), \tag{71}$$

where $\mathbf{k}$ is the wave vector, $\mathbf{R}$ is the 3D radius-vector of the point of observation (see Eq. (3)). In the usual situation, when the mentioned motion is translation with velocity $\mathbf{v}$, an observer "sees" the point coordinates to change in accord with the frame transformation law which in the non-relativistic case possesses the form [132]

$$\mathbf{R} \rightarrow \mathbf{R}' = \mathbf{R} + \mathbf{v}\mathrm{t}.$$

Then expression (71) takes on the form

$$E \sim \cos(\mathbf{kR}' - \omega' t),$$

where

$$\omega' = \omega + \mathbf{kv} \tag{72}$$

is the "shifted" oscillation frequency "felt" by the observer.

Rotational Doppler effect (RDE), instead of plane waves (71), deals with the helical beams (see Sec. 4.1). The specific symmetry of such beams due to which the translation along axis $z$ is identical to the certain rotation around it, requires that analogous frequency transformation should appear during the rotational motions around the axis connecting the source and observer (Fig. 24). Via transforming Eq. (36) to the frame rotating with angular velocity Ω by the substitution[3]

$$\phi = \phi' - \Omega t, \tag{73}$$

one will obtain

$$E(r, \phi, z, t) = \psi(r) \cos (l\phi' + kz - \omega' t),$$

where the frequency observed in the rotating frame is expressed as $\omega' = \omega + \Delta\omega$ with

$$\Delta\omega = l\Omega. \tag{74}$$

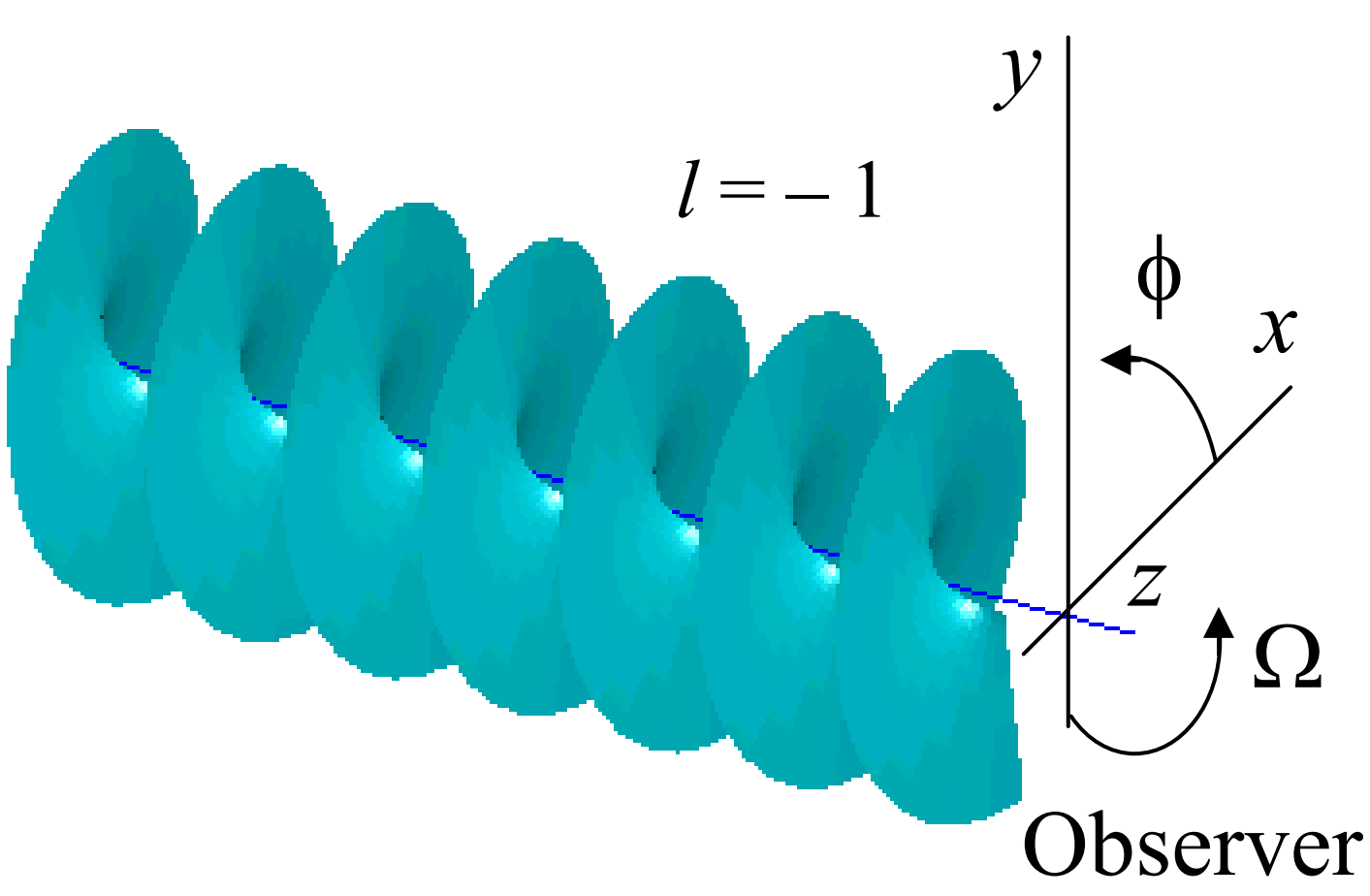

Fig. 24. Explanation of the rotational Doppler shift.

The frequency shift due to RDE can be pictorially explained with the help of geometric images presented in Figs. 7 and 24: upon the rotation shown in Fig. 24 (i.e., in accord with the sense of the beam helical structure and, therefore, oppositely to the energy circulation), the "frequency" of the CS beam "crests", that "fly" by the observer, decrease, and if the rotation is opposite – increase. Obviously, the same rule is also correct for the helical CP beams. From the quantum point of view, the RDE is a direct consequence of the special transformation properties of spiral photons for which the rotational displacement around the axis of propagation is equivalent to phase shift of the wave function [14, 15].

---

[3] Of course, this substitution is a very simplified form of the transformation to a rotating frame that disregards relativistic effects, which inevitably manifest themselves in considered optical processes; however, at small, as compared to ω, rotation rates it leads to correct quantitative and, to a great degree, qualitative conclusions.

Being of the common nature, the usual (translational) Doppler effect and RDE possess common as well as distinctive features. First, both versions of the Doppler effect are immediately applicable only to special model types of wave fields (the fact that plane waves are more usual and widespread objects than helical beams is not principal). As a special property of helical beams whose characteristic feature is existence of the preferential longitudinal direction, RDE is "fastened" to this direction (and in case of a CS beam (31), even to axis $z$). This fact restricts the multitude of possible mutual motions of the beam and the observer (registering system): while the translational Doppler effect (72) takes place in any translatory motions, for observation of the RDE only rotation near the beam axis is important. This can be formulated as a requirement that an observer always be situated in a certain plane orthogonal to the beam axis; additionally, there must exist a certain reference frame (for example, the frame $(x, y)$ in Fig. 24) that enables to describe the observer orientation in this plane. At last, the rotational frequency shift (74) differs from (72) by that its value does not depend on the initial frequency and thus on the beam wavelength.

### 7.2. Interaction (energy) aspect

The Doppler frequency shift can be also explained in terms of energy exchange between the light beam field and optical elements, which is possible due to mechanical properties of light [153–155]. Practically, under every beam transformation, elements of an optical system experience the ponderomotive action from the light wave, so the motion or deformation of the elements is accompanied by producing the mechanical work. Ads a result, the beam energy changes, and this leads to the frequency change [134, 156, 157].

Consider, for example, a scheme of the CS beam rotation with the help of a Dove prism A or an equivalent mirror system B (Fig. 25). Let an input beam with the structure of left helicoid possess OAM with linear density $\mathbf{L}_1 = -(l\Phi/\omega c)\mathbf{e}_z = -(\mathcal{N}l\hbar)\mathbf{e}_z$ where $\mathcal{N}$ is the number of photons contained within a unit length of the beam. At the system output, the beam OAM is $\mathbf{L}_2 = -\mathbf{L}_1$, that is, the system undergoes the action of torque (69) while the opposite recoil torque is applied to the beam

$$\mathbf{K} = c(\mathbf{L}_2 - \mathbf{L}_1) = 2\,c\mathbf{L}_2 = 2(l\Phi/\omega)\mathbf{e}_z = 2(\mathcal{N}l\hbar c)\mathbf{e}_z. \qquad (75)$$

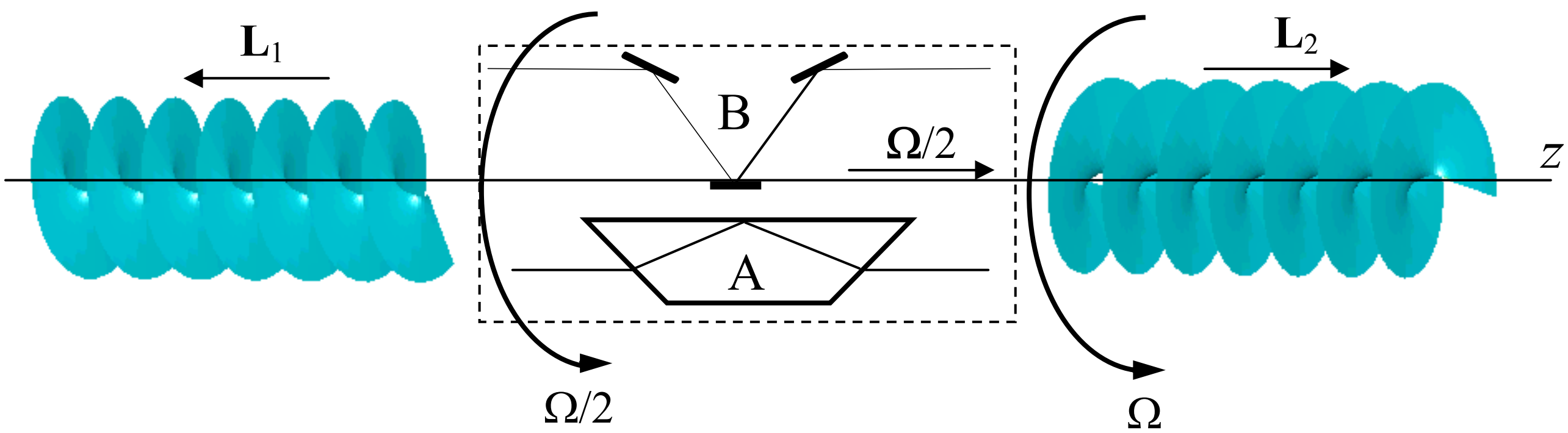

Fig. 25. Beam transformation in a rotational optical system (explanations in text).

If the system rotates with angular velocity Ω/2, the output beam rotates with the double velocity Ω and the torque performs the mechanical work changing the beam energy. In particular, for the beam unit length,

$$\Delta\mathcal{E} = (\mathbf{K}\cdot\mathbf{\Omega}/2)/c = (\mathbf{L}_2\cdot\mathbf{\Omega}) = |\mathbf{K}|\,(\Omega/2c) = \mathcal{N}l\hbar\Omega \qquad (76)$$

(in the first two equalities, the notation **Ω** for angular velocity vector is used), that is, every photon of the beam gets additional energy $\Delta(\hbar\omega) = l\hbar\Omega$. The same arguments can be applied for the CP beam rotation. Therefore, if a helical beam with AM **L** rotates with the vector rate **Ω**, this is equivalent to the shift of its frequency by $\Delta\omega = (\mathbf{L}\cdot\mathbf{\Omega})/\hbar$; in particular, for a helical beam with normalized AM $J = \sigma + l$ (see Eq. (37))

$$\Delta\omega = (\sigma + l)\,\Omega, \qquad (77)$$

which obviously generalizes Eq. (74).

Result (77) is quite general, despite that it was derived for a very special mechanism of energy exchange. It can be reinforced, e.g., by means of a direct analysis of modulation of the helical beam spatial pattern under the interaction with a rotating lens system [54]. Identical conclusions follow from the most "physical", to our opinion, approach that is based on studying the modifications of atomic and molecular spectra under conditions of rotational motion of the radiation sources. Such discourses, even in the elementary form [154], and especially when grounded on the rigorous quantum-mechanical calculations [158], essentially enrich the notions on the RDE nature.

One should remark that, just like its translational "prototype", the RDE belongs to the most "many-sided" physical phenomena and can be treated from different points of view. Among other things, it demonstrates intelligible dynamical manifestations [159, 160] of the Berry topological phase [160–162].

### 7.3. Experimental observations

A fundamental feature of any experiments with RDE is the necessity to determine very small relative shifts of frequency in systems with moving optical elements where unavoidable instabilities and fluctuations can strongly mask the sought effects. That is why the whole history of its observations is conjugated with the "fight" against the harmful influences of such instabilities. Historically, first RDE observations dealt with spatially homogeneous CP beams. As far as we can judge, the corresponding frequency shift was first registered in the microwave range [163] and only well later an all-optical experiment had been realized [164]; this is quite understandable because requirements to the measurement accuracy rapidly fall down with the wavelength growth.

First experiments with CS beams also dealt with millimeter waves [165, 166]; in this range, elements made of the polyethylene are used whose refraction index is approximately 1.52 [79]. A helical beam was formed by transmitting a plane polarized beam, produced by the Gunn generator, through a spiral phase plate whose thickness linearly depends on the azimuthal angle (see Sec. 4.2). In Ref. [165], the frequency shift was registered with the help of a Mach–Zehnder interferometer (Fig. 26), in one arm of which the beam was forcedly rotated by means of a Dove prism (as is shown in Fig. 25) while the other arm contained the similar immovable prism in order to compensate the accompanying mirror-like beam transformation. At the interferometer output, there appears a superposition of two spatially identical waves with different frequencies; the detector signal represents beats with the difference frequency and unambiguously demonstrates the presence and quantitative value of the effect. In Ref. [166], the CP of the input beam was additionally introduced and a rotatable half-wave plate was placed in series with the Dove prism, which enabled to observe the "spin" RDE component as well. The frequency shift was detected by means of immediate measurement of the output beam frequency. In this experiment, additivity of the "spin" and "orbital" RDE components expressed by the two summands of Eq. (77) was convincingly proven.

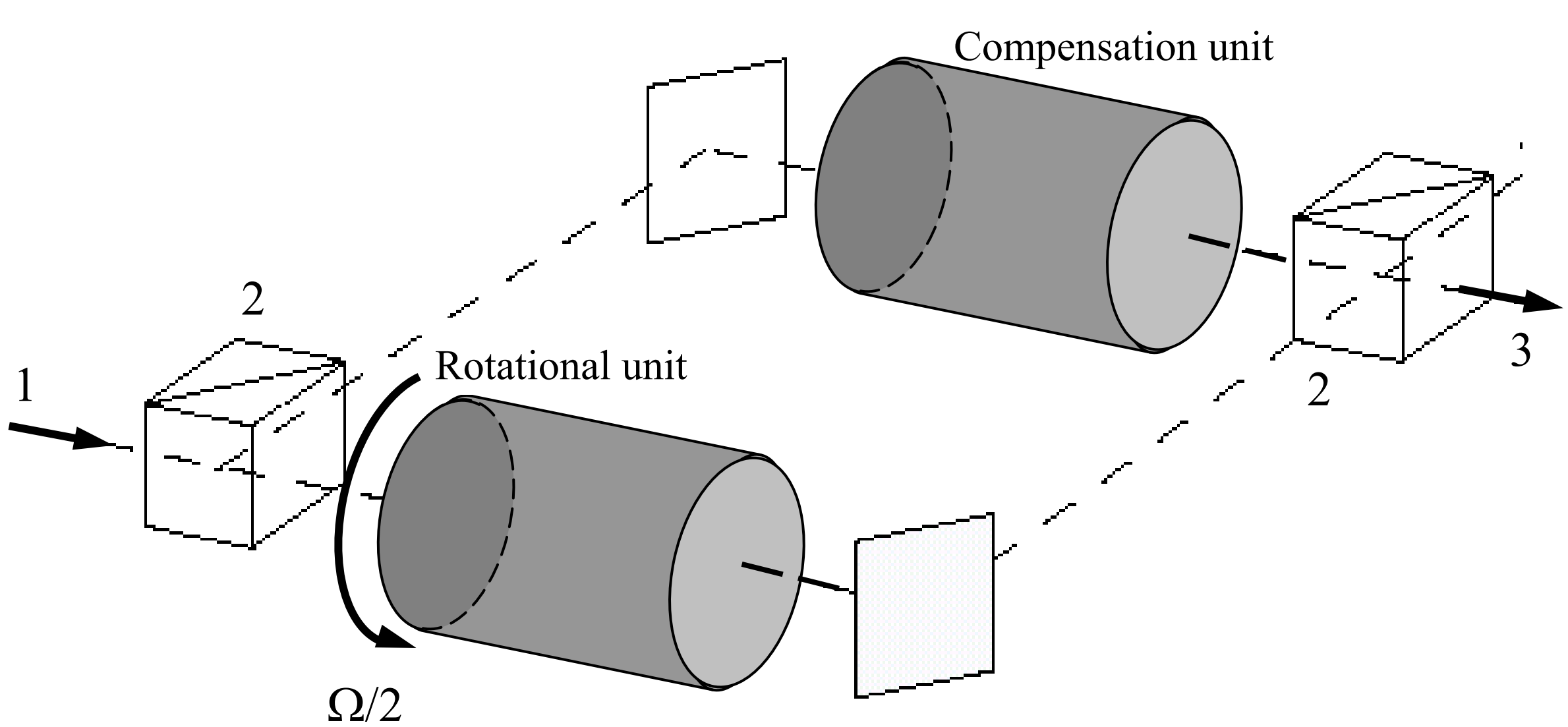

Fig. 26. Interferometric scheme of the RDE observation: 1 – input beam, 2 – light splitters, 3 – output beam entering the detector. The compensation unit in the reference arm produces the same beam transformations as the rotational unit but does not move.

Attractive simplicity and efficiency of the approach used in Refs. [165, 166] is lost in optical wavelength range because of the necessity to align and stabilize massive rotatable elements with the sub-wavelength accuracy. This problem is, to a certain extent, eliminated in approach based on the spiral zone plates (SZP) [85] that can be considered as "generators" of helical beams (see Sec. 4.2, Eq. (44)). If such a plate rotates about the system axis with constant rate $\Omega$, the phase factor of $N$-th diffracted beam is transformed as

$$\exp\left[iNm\left(\phi-\Omega t\right)-i\omega t\right]=\exp\left[iNm\phi-i\left(\omega+Nm\Omega\right)t\right], \tag{78}$$

which signifies the change of the observable frequency by $\Delta\omega = Nm\Omega$.

In Ref. [86], the binary SZP with "embedded" singularity of the order $|m| = 1$ (Fig. 10b) was used together with the spatial filter (Fig. 10c) adjusted for separation of the wave with $N = -1$ (see Eq. (44)); as a result, an input Gaussian beam of a He-Ne laser produced a CS beam with helical WF at the collimating lens output. This beam was directed to the measurement arm of a Mach-Zehnder interferometer in which the reference wave was created from the initial beam by separating a small part of its cross section with consecutive broadening and collimation. At the interferometer output, a system of interference fringes was formed that permitted to see the beam transformations upon the SZP rotation. The frequency shift induces regular fringe motion while possible parasitics (the rotation axis beatings, etc.) manifest themselves only by inessential changes of the interference pattern period and can thus be efficiently filtered out. With a photodetector of aperture size less than the fringe width, it was possible to determine their motion velocity which exactly equaled to the expected frequency shift $\Delta\omega = 3\ \mathrm{c}^{-1}$. The analogous experiment with a "vortexless" beam of zero diffraction order gave zero result; in case of a SZP with the "embedded vortex" $m = 2$, the double frequency beats of the detector signal were observed.

An interesting situation takes place when the input beam already possesses the OV with certain index $n$. Then the phase factor in (78) takes the form $\exp\left[i\left(Nm+n\right)\phi - i\left(\omega + Nm\Omega\right)t\right]$, which, in particular, means that at $n = -Nm$ the frequency shift will occur even in a vortexless output beam. This can be interpreted as an "inverted" RDE and explained by the specific picture of the energy exchange in a SZP. The incident bean brings to it photons with OAM $n\hbar$; behind the SZP, each photon in the $N$-th diffraction order carries OAM of $(n + Nm)\hbar$. Hence, each photon obtains the AM $Nm\hbar$ and the recoil torque (see Eq. (75)) is applied to the SZP. The mechanical work of this torque during the SZP rotation changes the photon energy by amount $Nm\hbar\Omega$, whence the frequency shift follows $\Delta\omega = Nm\Omega$ regardless of the diffracted beam OV index.

Another way of eliminating noise following from vibrations and irregular displacements of optical elements in the RDE investigation consists in such experimental conditions that both the signal and reference beam should pass the same rotatory elements [167, 168]. In this approach, as a reference beam, one must take a beam insensitive to RDE (e.g., one belonging to the class (31) with $l = 0$); then, all the incontrollable influences will be the same for both beams and their interference will present the effect of rotation "in a pure form". The formation of co-axial signal and reference beams with required properties and identical parameters $b_0$, $R$ and the common waist plane (see Eq. (38)) turns out to be rather simple. Such possibility is again supplied by the holographic technique of creating helical WFs (see Sec. 4.2), particularly, upon transmitting a Gaussian beam through the holographic grating presented in Fig. 9a, provided that its center slightly mismatch with the beam axis. If this shift takes place along axis $x$ and equals $x_0$, then, in the first diffraction order behind the grating a beam with complex amplitude distribution

$$u\left(r,\phi\right) = \frac{x_0 + r\exp\left(i\phi\right)}{R_0}\exp\left(-\frac{r^2}{2b_0^2}\right) \tag{79}$$

is produced (notations see Eqs. (38), (43)). One can easily see here the presence of a signal LG beam with $p = 0$, $l = 1$ and a reference Gaussian beam ($p = l = 0$); their "common" existence within a single beam gives possibility of treating this combined beam as a rigid "one-beam interferometer". Combination (79), as is known (see Sec. 5, Fig. 19), constitutes an "off-axis OV" whose azimuthal position depends on the phase difference between the LG and Gaussian components [120].

Now consider how Eq. (79) transforms if the beam in one way or another is set in rotation around its own axis. Then, according to Eq. (77), frequency of the LG component changes so that it obtains an additional phase factor $\exp(-i\Omega t)$; respectively, the whole expression modifies to

$$\frac{x_0 + r\exp\left(i\phi - i\Omega t\right)}{R_0}\exp\left(-\frac{r^2}{2b_0^2}\right),$$

that is, in consequence of the RDE, the transverse profile of beam (79) experiences transformation $u\left(r,\phi\right) \to u\left(r,\phi - \Omega t\right)$. This is nothing else than the transverse profile of a beam that rotates with angular velocity Ω. In other words, in the course of coercive rotation of the combined beam (79) its helical component gets the rotational frequency shift, and the external manifestation of this shift in the one-beam interferometer is such interference pattern that has no differences from the geometric pattern of the rotating beam! Therefore, the beam rotation itself can be considered as the manifestation of the RDE for its helical component. This conclusion is a consequence of the repeatedly mentioned helical symmetry of such beams that stipulates the deep intrinsic interrelation between their spatial and phase properties due to which the rotation with angular rate Ω and the frequency shift $\Delta\omega = l\Omega$ are completely equivalent.

### 7.4. Rotational Doppler effect and the image rotation

The latter observation of the previous subsection can be essentially generalized. It paves the way to interesting and far-reaching consequences which concern the immanent interdependences between the rotation and the frequency spectrum of arbitrary optical fields [167–172].

As was said in Sec. 4.1, the CS beams form the system of azimuthal harmonics; this means that any function of coordinates in the beam cross section can be presented in the form of a superposition

$$f\left(r,\phi\right) = \sum_l a_l\left(r\right)\exp\left(il\phi\right). \tag{80}$$

As a matter of fact, this is the Fourier expansion of a function of argument ϕ, defined on the finite interval (0, 2π), with $r$-dependent coefficients $a_l(r)$. Such a representation of function $f(r,\phi)$ is called sometimes the helical (spiral, azimuthal) harmonic expansion [169], or the expansion in OAM spectrum [67, 170]. Considering (80) as a transverse profile of a monochromatic light beam with frequency ω, one can write the full wave field in this cross section similarly to Eq. (7) in the form $\mathrm{Re}[f(r,\phi)\exp(-i\omega t)]$. Upon the beam rotation, the frequency of each azimuthal harmonic changes in agreement to (74), which entails

$$f\left(r,\phi\right)\exp\left(-i\omega t\right) = \sum_l a_l(r)\exp\left[i\left(l\phi - \omega t\right)\right] \to$$

$$\to \sum_l a_l(r)\exp\left\{i\left[l\phi-\left(\omega+l\Omega\right)t\right]\right\}=\sum_l a_l(r)\exp\left\{i\left[l\left(\phi-\Omega t\right)-\omega t\right]\right\}= \tag{81}$$

$$= f\left(r,\phi-\Omega t\right)\exp\left(-i\omega t\right). \tag{82}$$

Now let us look closely at the expressions obtained. Line (82) describes a beam with complex amplitude distribution (80) rotating around axis $z$ with angular velocity $\Omega$, and this rotation is caused by those frequency shifts that were made in line (81). Therefore, transition from (81) to (82) demonstrates the absolute reversibility of the RDE: not only the beam rotation changes the frequencies of its helical components but also the certain matched, "coordinated" combination of such shifts evokes the beam rotation. More precisely, interference of the helical harmonics with shifted frequencies produces exactly the rotating beam (by the way, such conditions are realized sometimes in deformed optical fibers, which results in the so called optical Magnus effect [172]). Examples of numerical analyses of situations, when a certain image "decomposes" in azimuthal harmonics that afterwards obtain the frequency shifts and this entails the image rotation, are presented in [46, 67].

Comparison of Eqs. (81) and (82) shows that, when applied to RDE, statements "the beam rotation is the reason of the frequency shift" or, to the contrary, "the frequency shifts give rise to the beam rotation" are one-sided and, in essence, meaningful – both process are inherently equivalent. In application to the combined beam (80) that can represent an arbitrary optical object this implies that any beam rotation around its axis is absolutely equivalent to certain coordinated frequency shifts of its helical components, and any rotation of an optical image within its own plane is accompanied by the RDE, though in a hidden form. And since the RDE shifts of frequency do not depend on the wavelength, this statement is valid also for polychromatic images.

From Eq. (81), it follows yet another important observation: rotation of a monochromatic optical beam makes it polychromatic [54], there appear "side frequencies" differing from the main frequency $\omega$ by $l\Omega$ where $l$ stands for indices of all azimuthal harmonics of the beam. Positions and intensities of the side frequencies depend on the OAM spectrum of the beam; they can be observed with the help of the usual temporal spectral analysis of the beam electromagnetic field. Of course, only beats of the side frequencies between each other and with external coherent signals can be a subject of the direct observation. For example, the mutual interference of the side bands is manifested in variations of the rotating beam intensity

$$I(r,\varphi,t) \sim \left| f(r,\phi - \Omega t) \right|^2 = \left| \sum_l a_l(r) \exp(il\phi) \exp(-il\Omega t) \right|^2, \qquad (83)$$

which can be easily watched by a spatially fixed photodetector with a finite aperture [169–171]. Therefore, a variable signal emerging when spatially inhomogeneous beam moves with respect to the detector aperture, seeming to be a purely kinematical effect, can be treated as the Doppler effect manifestation[4] [171].

### 7.5. Spectrum of helical harmonics and associated problems

The displayed regularities can be used for experimental investigation of the spectrum of helical harmonics (OAM spectrum) of arbitrary optical signal [169–171] (in other words, retrieval of coefficients of expansion (80)), which is proposed to utilize in systems of the information encoding and processing [169, 173]. Practical realization of this procedure, however, meets some indeterminacy. First, the Fourier-spectrum of intensity (83) contains only the difference frequencies $(l - l')\Omega$ where $l$ and $l'$ are arbitrary indices of the azimuthal harmonics "belonging" to sum (80). In principle, this is not fatal because absolute values of indices of the beam harmonics can be found from its interference with a known reference CS beam [170]. But this does not remove another difficulty, which follows from dependence of signal (83) harmonics on $r$. An example of such situation is presented in Fig. 27. If, in the relative motion of the image and a pinhole aperture, the latter circumscribes the contour indicated by the solid circumference in Fig. 27a, power spectrum of the detector signal possesses the form shown in Fig. 27b. In its structure, a special significance of eighth harmonic attracts the attention which seems to witness for the importance of helical harmonics with $l = \pm 4$. Nevertheless, if the aperture is slightly shifted in radial direction (dashed contour) the signal spectrum changes radically and other peaks become predominant (Fig. 27c).

Therefore, besides some simple cases in which sufficient a priori information about the analyzed beam composition exists (for example, mixtures of Gaussian and $LG_{0l}$ beams [171] or combinations of Gaussian, $LG_{01}$ and $LG_{0,-1}$ beams [170]), the problem of practical retrieval of the OAM spectra for real beams is not solved as yet. A possible way of its solution exploits application of a "sectorial" aperture that can supply the "integrated" weight of the helical harmonics (80)

$$A_l = \int_0^\infty a_l(r) r dr .$$

---

[4] This conclusion is not a special feature of the RDE. Applying the corresponding arguments to the translational Doppler effect, one can notice that any spatially inhomogeneous light beam is composed with a set of plane waves, and the time-dependent signal arising upon the mutual translation of the beam and observer is a result of beats between the plane-wave harmonics with shifted frequencies.

On the other hand, more exact specification of the expansion (80) coefficients is desirable, for example, their representation as a series in certain "radial harmonics". In such cases, it seems the most natural to use LG functions (38)

$$a_l(r) = \sum_p C_{pl} \psi_{pl}^{\mathrm{LG}}(r);$$

then the set of coefficients $C_{pl}$ constitutes the "LG spectrum" of the beam [171], which completely characterizes its spatial profile. One can expect that usual schemes of the RDE analysis including off-axis apertures (Fig. 27) will be valid for the LG spectrum determination, perhaps, with addition of gradual radial displacement of the aperture.

To reach the success in usage of the off-axis apertures, one should know the position of the investigated beam axis with high accuracy. In this context, the role of inevitable "disagreements" between the nominal and real beam axes becomes important. Corresponding transformations of the OAM spectra of CS beams were considered by the example of LG modes experiencing transverse shifts and deflections [67]. In cases of "pure" shifts (Fig. 28) or deflections (Fig. 29), the azimuthal harmonic spectrum broadens symmetrically. The growth of broadening with increase of "disagreement" is quite understandable; the only unexpected consequence is that the central component "looses" its domination if the $LG_{01}$ beam is shifted strongly enough (Fig. 28c).

It is interesting to track the relation between OAM of a whole beam and OAMs of its individual components. First consider the problem in its general formulation. Let expansion (80) describe the complex amplitude of a certain paraxial beam. Employing formula (32), one finds the beam OAM

$$\mathcal{L}_O = \frac{1}{16\pi kc} \sum_{l,l'} (l + l') \int_0^\infty a_{l'}^*(r) a_l(r) r\, dr \int_0^{2\pi} \exp\left[i(l - l')\phi\right] d\phi = \frac{1}{4kc} \sum_l l \int_0^\infty |a_l(r)|^2 r\, dr \qquad (84)$$

(integration over the azimuthal angle yields $2\pi$ if $l = l'$ and zero in other cases). In accord with Eqs. (15) and (25) the energy flow of $l$-th helical component amounts to

$$\Phi_l = \frac{c}{4} \int_0^\infty |a_l(r)|^2 r\, dr ,$$

and then result (84) is represented in form [121]

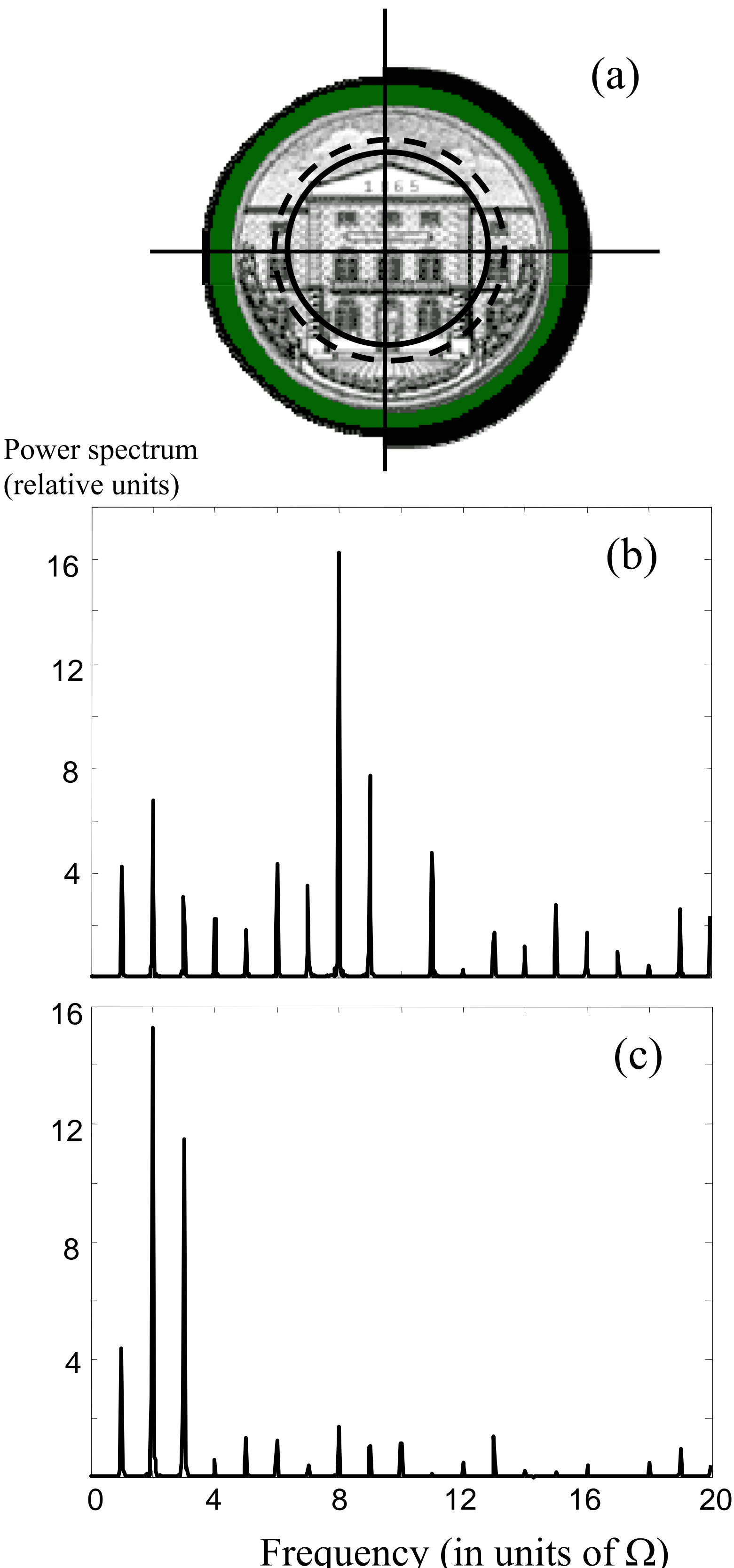

Fig. 27. A generic optical image (a) and power spectra of the beat signal appearing during its rotation if the detector aperture trace is presented by: (b) solid circumference in panel (a), (c) dashed circumference in panel (a).

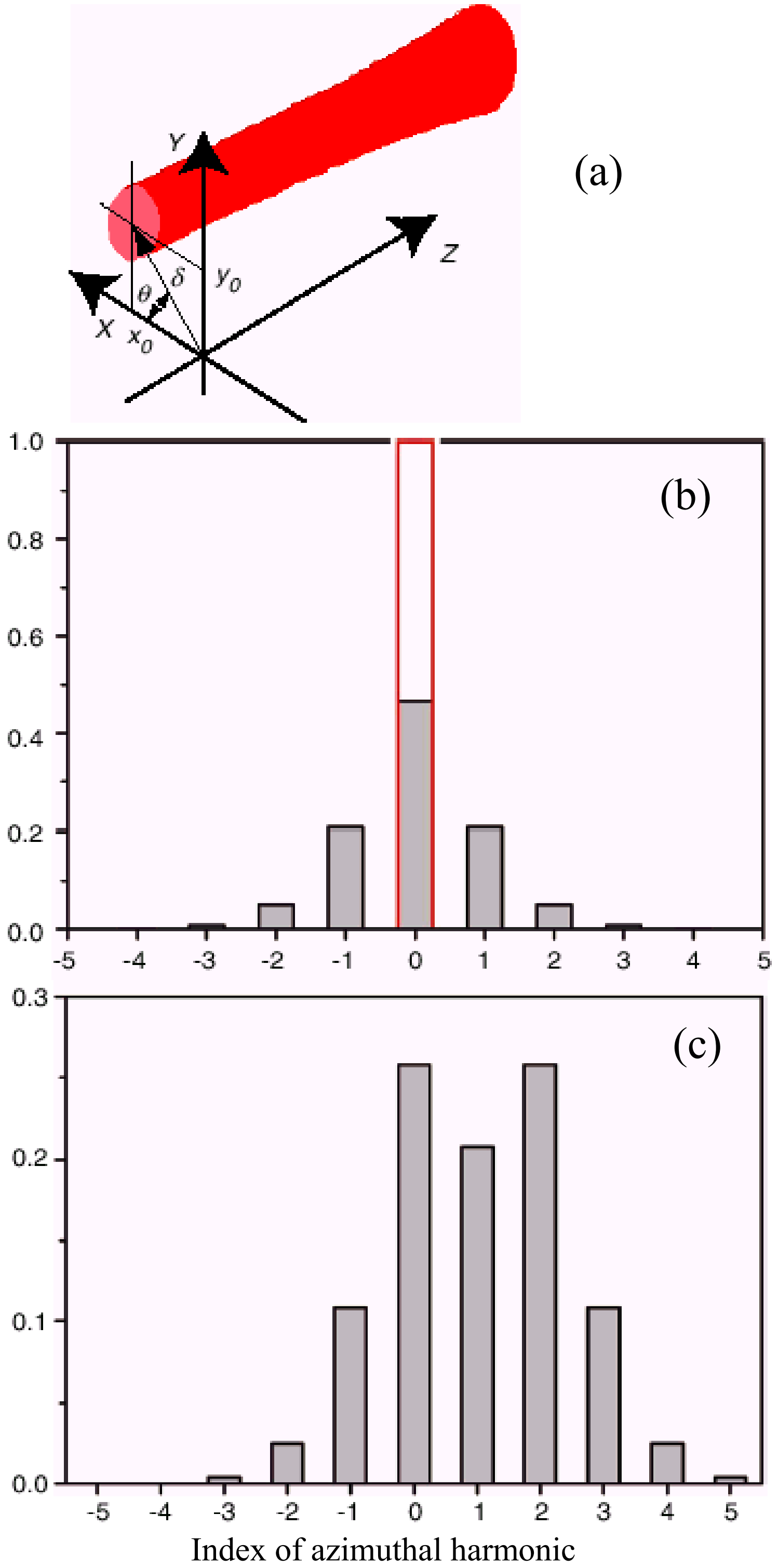

Fig. 28. (a) Schematic arrangement and (b, c) the OAM power spectra of transversely shifted (b) Gaussian and (c) $LG_{01}$ beams. Axis of propagation is parallel to axis *Z*, its shift is $\delta = 1.4b_0$, the peak height denotes a part of the total beam power belonging to corresponding helical component. "Empty" column in panel (a) presents the "initial" spectrum of the non-shifted beam.

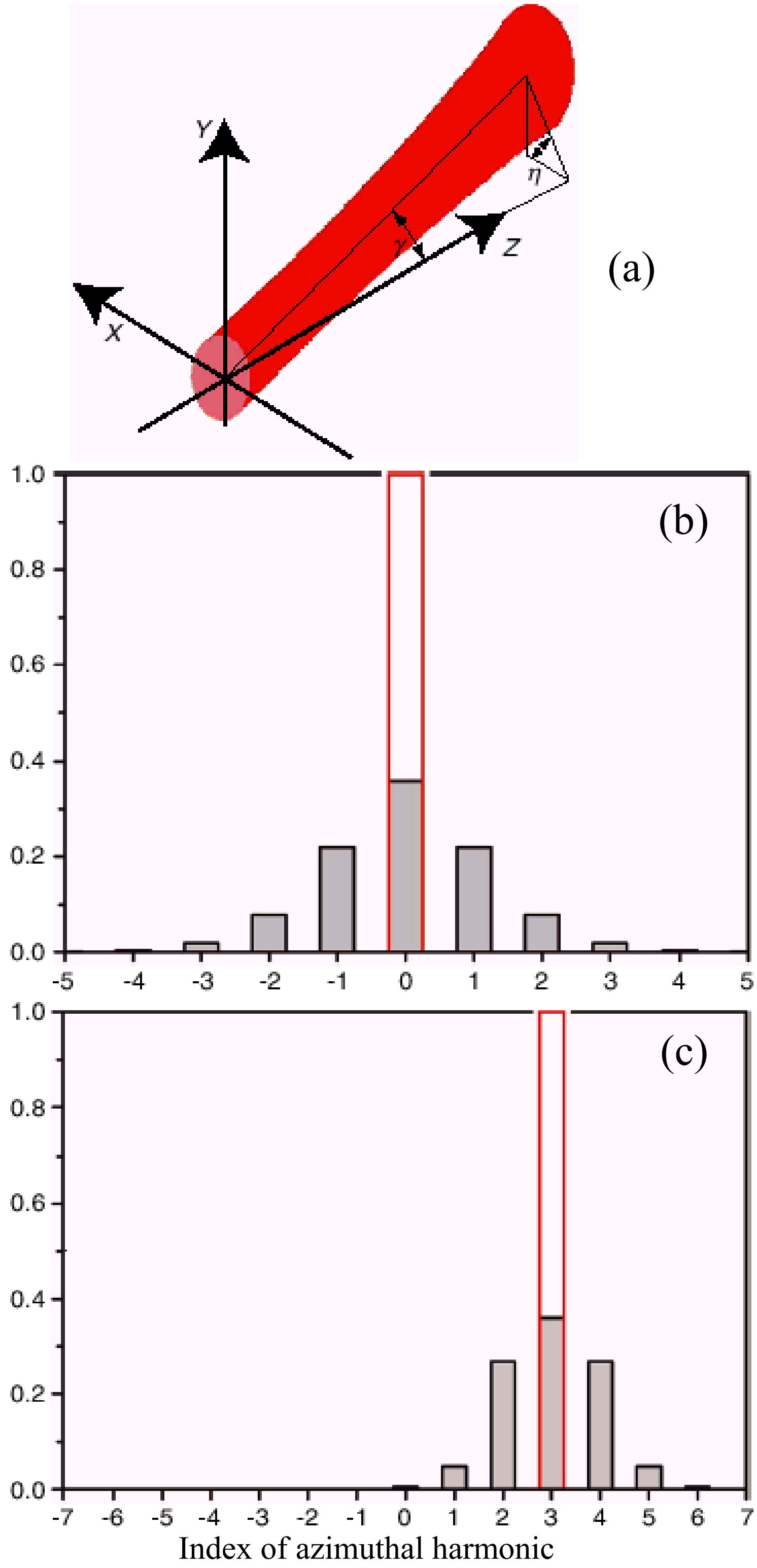

Fig. 29. (a) Schematic arrangement and (b, c) the OAM power spectra of deflected (b) Gaussian and (c) $LG_{03}$ beams. Angular displacement of the beam axis from axis *Z* amounts $\gamma = 2.5\times10^{-4}$, the peak height signifies relative weight of the corresponding component. "Empty" columns represent the spectrum of the non-deflected beam.

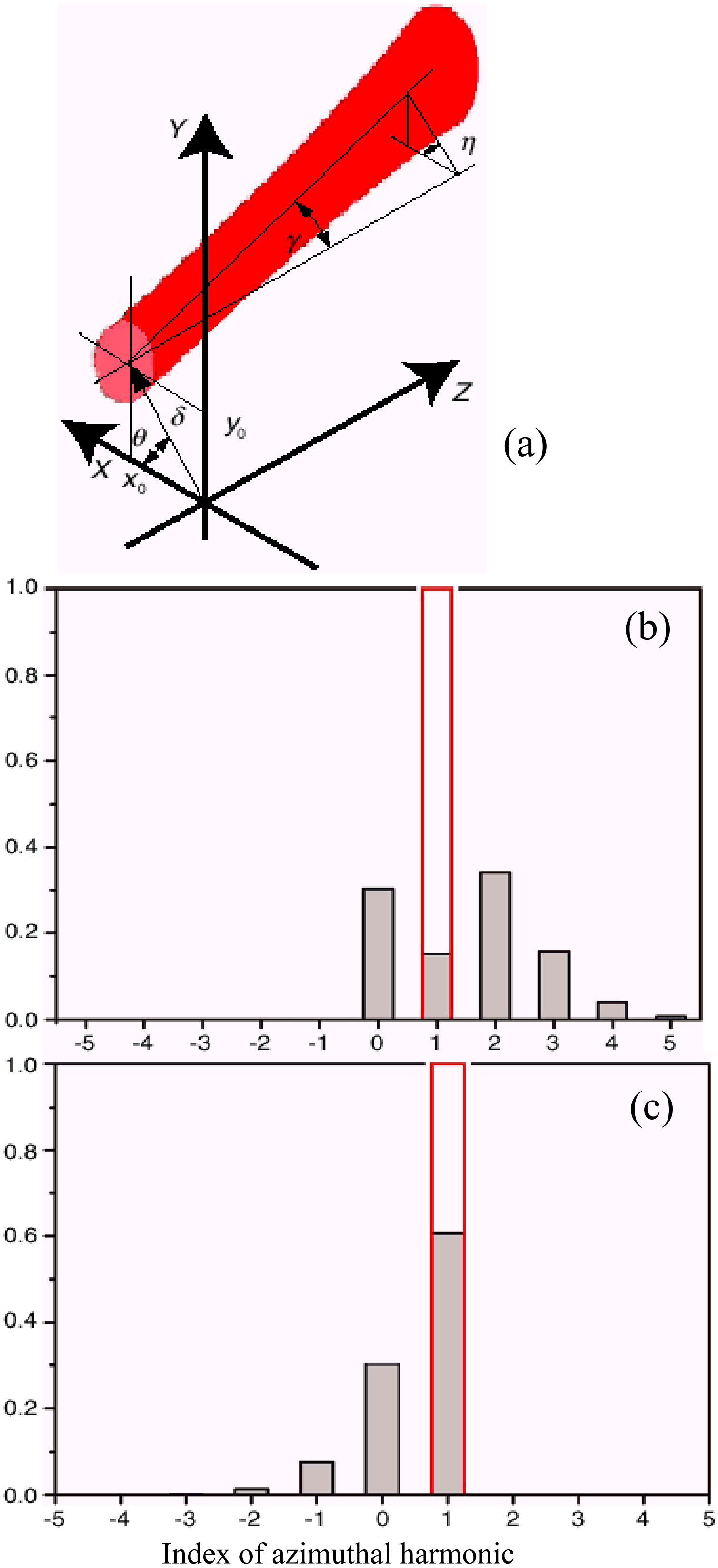

Fig. 30. (a) Schematic arrangement and (b, c) the OAM power spectra of the $LG_{01}$ beam shifted by $\delta = 0.7b_0$ and deflected by angle $\gamma = 10^{-4}$. Azimuthal parameters: (b) $\theta = 0$, $\eta = \pi/2$; (c) $\theta = \pi/2$, $\eta = 0$. "Empty" columns represent the beam OAM spectrum at $\delta = \gamma = 0$.

$$\mathcal{L}_O = \sum_l l \frac{\Phi_l}{c\omega} = \sum_l \mathcal{L}_{Ol} \tag{85}$$

(see Eq. (33)), where $\mathcal{L}_{Ol}$ is the "partial" OAM of the $l$-th component.

Notice an interesting detail. Expansion (80) establishes that the analyzed beam is absolutely equivalent to the set of helical harmonics; at the same time, the sum OAM of harmonics (85) is always directed along axis $z$, even if the vector OAM of the beam itself would possess another direction. This seeming contradiction is explained by the fact that when Eq. (85) was derived, axis $z$ as the AM reference axis was imposed "forcedly", that is, equality (85) relates only to $z$-component of the beam OAM. In other words, Eq. (85) represents the "total" beam OAM that includes also its "extrinsic" part (see the beginning of Sec. 4) connected with the specific disposition of the beam relatively to axis $z$; the beam "intrinsic" OAM, measured with respect to its own axis, can differ from result (84).

In situations presented in Figs. 28 and 29, rule (85) holds exactly despite that beams described by Fig. 29 deflect from the nominal axis $z$. This is not surprising because the "extrinsic" OAM in case of the "pure" angular deflection vanishes (corresponding modifications of the OAM appear only when the beam deflection is accompanied by the transverse shift, see Eq. (29)). Besides, in paraxial limits, the difference between the total OAM and its $z$-component is proportional to the square of the deflection angle and is thus negligibly small.

The situation becomes more complicated in case of simultaneously "shifted + deflected" beams (Fig. 30). Here a significant part of the beam OAM is "extrinsic" (see Sec. 4, Fig. 6) and depends on mutual disposition of the real and nominal axes, and so the sum OAM values, following from the spectra of Figs. 30b and 30c, differ. In this case, the spectra broaden asymmetrically and modify noticeably on the beam rotation, which, undoubtedly, should be taken into account in development of practical procedures for expansion of beams in the OAM spectrum.

To conclude this section, we mention another prospective approach to the OAM spectrum retrieval [173, 185, 205]. Its main idea consists in imparting the additional phase factor $\exp(im\phi)$ to an analyzed beam so that expansion (80) for its complex amplitude distribution takes on the form

$$\sum_l a_l(r) \exp\left[i(l+m)\phi\right].$$

Then the difference in spatial behavior of the vortex and non-vortex components is employed: only the mode with $l = -m$ has non-zero axial intensity, and it is this mode that possesses, generally, minimum divergence and can be focused into the smallest spot. Therefore, it can be selected by appropriate technical means (e.g., by the pin-hole diaphragm or by coupling into a single mode optical fiber etc.), so the weight of this mode can be measured. The necessary spiral phase ramp can be attributed to a beam by any of the known methods considered in Sec. 4.2. The use of holographic techniques [173, 205] is especially suitable because a single hologram can produce a set of output beams with different $m$ providing thus the means for simultaneous determination of helical harmonics with different $l$.

### 7.6. Non-collinear rotational Doppler effect

In Sec. 7.1, we emphasized that the condition for RDE observation is rotatory motion of the source (beam) or the observer (registering system) around the beam axis. However, there frequently occur more general situations where the beam axis does not coincide with the axis around which the rotation takes place. A characteristic and rather widespread example is supplied by the scheme of conical beam evolvement, or scanning [134, 174] (Fig. 31) when the output beam is formed by means of reflection from a tilted mirror that rotates near a fixed axis (other systems performing the same scanning are also possible and can be described analogously [174]). In this situation, a certain version of the RDE also takes place, and can be called "non-collinear" RDE.

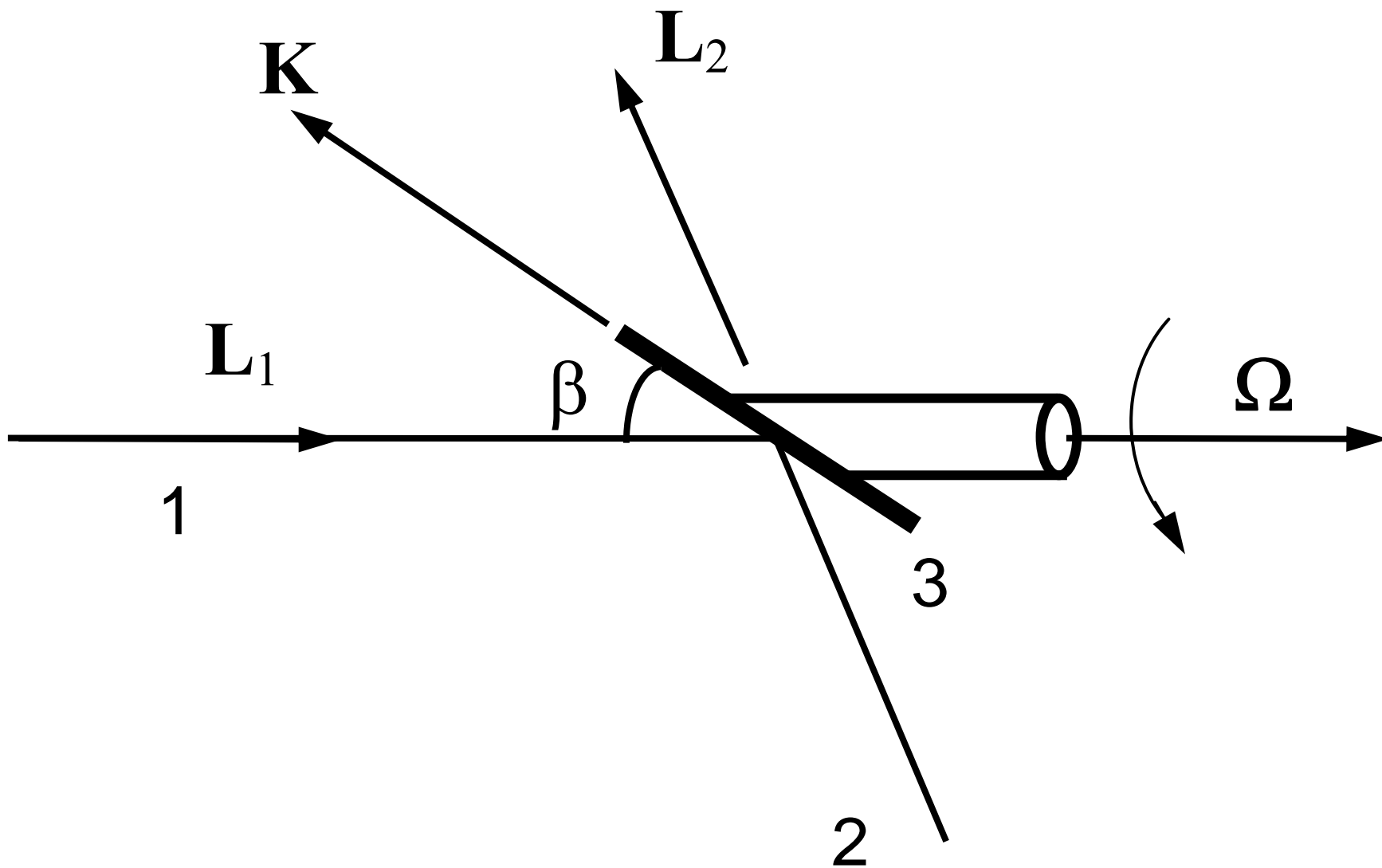

Fig. 31. Sketch of the non-collinear RDE observation upon the conical scanning of helical light beam. Input beam 1 is reflected from the tilted mirror 2 rotating around the input beam axis with angular velocity **Ω**. Upon reflection, the output beam 4 is formed whose axis "glades" over a conical surface.

We start its considering from the energy analysis similar to that made in Sec. 7.2 with the help of Fig. 25. Let the input beam be a CS beam with azimuthal index $l$, that is, every its photon bear the OAM of the value $l\hbar$. Operating as before, we find the torque applied to the beam (see Fig. 31)

$$|\mathbf{K}| = 2|\mathbf{L}|c \cos\beta = 2\mathcal{N} l\hbar \cos\beta$$

and corresponding change of the beam energy

$$\Delta\mathcal{E} = (\mathbf{K}\cdot\mathbf{\Omega}) = -\,2\mathcal{N} l\hbar\Omega \cos^2\beta, \text{ or } -2l\hbar\Omega \cos^2\beta \text{ "per photon"},$$

whence it follows that the output beam frequency must change by

$$\Delta\omega = -2l\,\Omega \cos^2\beta. \tag{86}$$

This result seems quite reasonable but it entails rather strange consequences. Owing to the frequency shift (86), the phase of the helical beam after the rotation through an angle $\theta = \Omega t$ must have changed by

$$\Delta(k\varphi) = \Delta\omega t = -2l\Omega t \cos^2\beta = -2l\theta \cos^2\beta, \tag{87}$$

i.e., after the full revolution ($\theta = 2\pi$), when the system of Fig. 31 restores its initial configuration, the phase increment is not multiple of $2\pi$!

In light of the mentioned connection of the phase and azimuthal orientation of a CS beam (Secs. 7.1, 7.3), this result looks even more paradoxical than it seems at first glance. For example, if the input beam contains a superposition of helical harmonics of the form (80), the phase increments (87) signify that after the mirror makes full revolution the output beam transverse profile does not reproduce but appears to be turned by the angle depending on the mirror tilt

$$\Delta\alpha_1 = -\,4\pi\cos^2\beta \tag{88}$$

– quite contradictory to the everyday experience.

There is another paradox linked to the conical scan (Fig. 32). If the input beam carry a certain image, then in the "forward" reflection taking place at the mirror tilt $\beta < 45°$, this image makes two complete revolutions during each full revolution of the mirror while in case of the "back" reflection ($\beta > 45°$), the output image seems not to rotate at all (one can very easily make sure of this just keeping the mirror in a hand). Changing the mirror tilt, it is possible to continuously transform the first situation into the second one; however, in any thinkable circumstances, only an integer number of the image revolutions can be seen during a single cycle of the mirror rotation, and no continuous transition from the "two revolutions" to "no revolution" can be imagined.

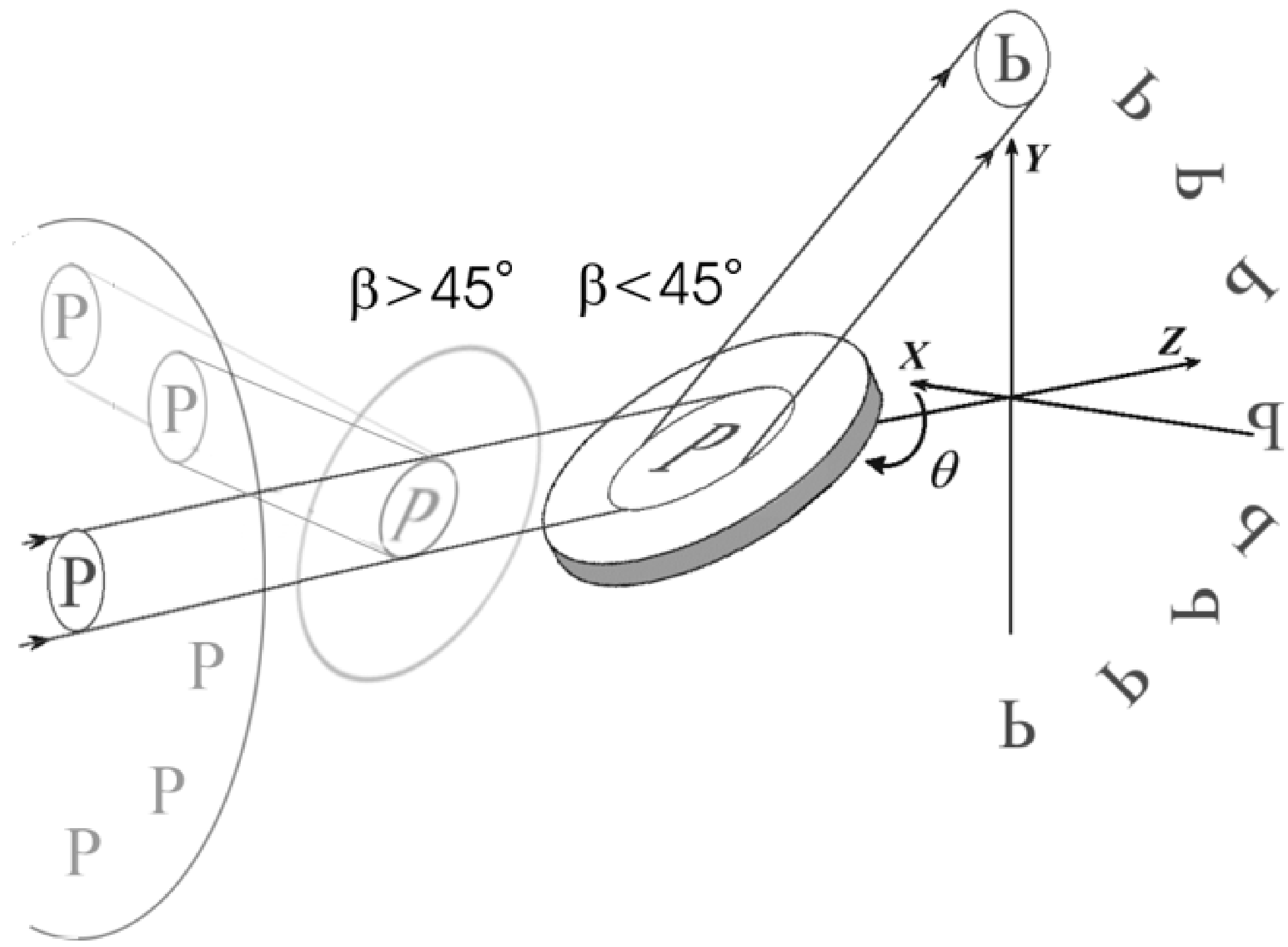

Fig. 32. Rotation of the beam transverse structure upon the conical scanning. During a single revolution of the mirror, the image makes two revolutions in case of the "forward" reflection and does not rotate in case of the "back" reflection.

Obviously, a key to solution of these "puzzles" lies in the careful analysis of conditions for observation of the mentioned frequency shifts and the image rotations. In this situation, the specific linkage between the observed phase (or frequency) of a helical beam and the relative disposition of the beam and registering system manifests itself especially clear. At the same time, not all reference systems are equally valid. This is seen already from the above-presented energy reasoning leading to Eqs. (86) and (87): there must exist a "preferential" reference system, in which the beam behavior at the conical scanning looks the most "natural".

Of course, the question on the reference system choice is implicitly present in the usual "collinear" situation as well, but there it is solved trivially: the beam axis is fixed, so it is natural to expect the observer to be fixed. In the non-collinear case, in agreement to remarks at the end of Sec.

7.1, the observer is "attached" to the plane orthogonal to the beam axis (Fig. 24), and is thus obliged to move. Therefore, the question on the character of its motion becomes highly important.

To study it, we address the "globe" model (Fig. 33). The input beam goes upwards along axis *Z*; the tilted mirror is situated in point M. When it rotates, the output beam axis (the unit vector of the axis direction $\mathbf{t}(\theta)$) moves along a "parallel" AB. The registering system is located in the plane tangent to the spherical surface; in every moment, its reference axis with unit vector $\mathbf{e}(\theta)$ is orthogonal to the current position of $\mathbf{t}(\theta)$. During the beam conical scanning, the reference axis remains constant (from a certain "intrinsic" point of view) but in respect to the laboratory system, can rotate around **t**. In Ref. [174] it is shown that "natural" motion of the reference system in the conical scanning process can be determined by the following equation for vector **e**

$$\frac{d\mathbf{e}}{d\theta} = -\left(\frac{d\mathbf{t}}{d\theta}\cdot\mathbf{e}\right)\mathbf{t}. \qquad (89)$$

Such motion of the reference system is similar to the hypothetical motion of a balanced arrow, freely (without friction) "pinned" to the output beam axis **t**. When the axis moves infinitely slowly, this arrow tries to preserve its position but the traveling axis "involve" it in the certain motion. Let its initial position coincides with a meridian (A in Fig. 33). After the axis travels a small path on the sphere, all points of the arrow travel the same paths; but meridians slightly "converge" to the pole, and therefore at the end B of this path, the arrow and a current meridian form an angle which can be easily calculated from Eq. (89) or immediately from Fig. 33:

$$\Delta\alpha = \Delta\theta\cos 2\beta.$$

During the mirror rotation, such changes in mutual orientation accumulate, so that after a full revolution the arrow will appear deflected by the angle $2\pi\cos 2\beta$ relatively to its initial position.

With allowance for such motion of the reference system, all the above-mentioned paradoxes of the non-collinear RDE find their explanations [174]. In this system, after the mirror rotates through angle $\Delta\theta$, the current angular displacement of the beam transverse profile amounts to $\Delta\theta(1 + \cos 2\beta) = 2\Delta\theta\cos^2\beta$. First, this implies that the beam profile rotates with angular rate $2\Omega\cos^2\beta$, which agrees with the frequency shift (86); second, this really allows to perform continuous transition from the "double" rotation at $\beta = 0$ ("forward" reflection, Fig. 32) to the "zero" rotation at $\beta = \pi/2$ ("backward" reflection). Apparent "jumps" in the character of the beam profile behavior (Fig. 32) occur because of attempts to fix the reference system or to change it stepwise.

The described phenomena possess a purely geometric nature and are conditioned by the peculiar rules of combination of rotations around non-collinear axes. Their analogs occur in various branches of physics beginning with the classical mechanics [175]. For example, Eq. (89) describes the parallel translation of vector **e** over the unit sphere surface [160]. Generally, it entails that after the parallel translation along a closed contour, the vector does not reproduce its initial orientation but appears to be turned by an angle measured by value of the solid angle subtended by the contour [161]. The same Eq. (89) expresses the Rytov law for the plane-of-polarization rotation of a twisted light ray [160]. The above-found motion of the "natural" reference system with respect to the "globe" meridians in Fig. 33 is identical to the oscillation plane motion of a famous Foucault pendulum [132, 175].

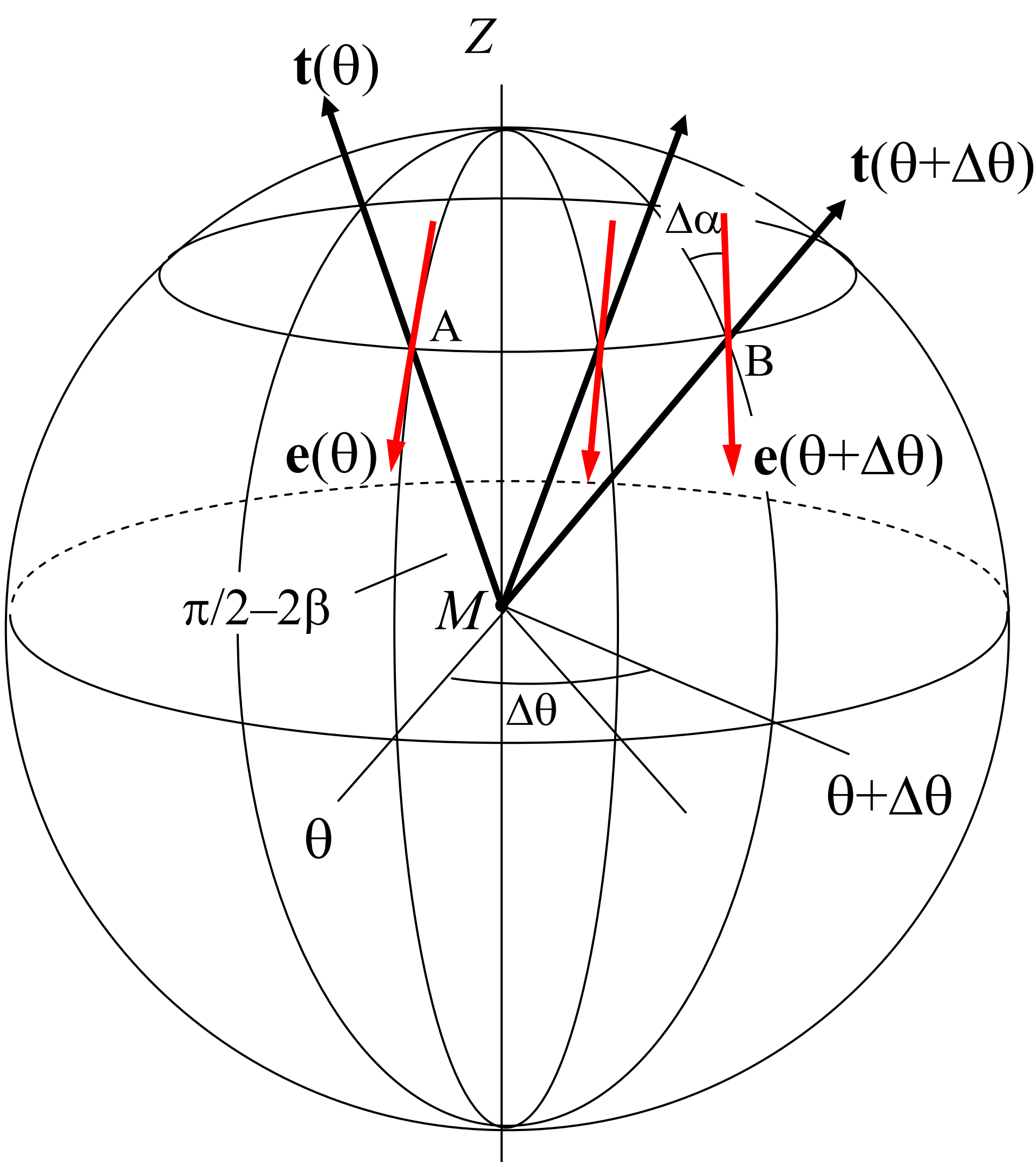

Fig. 33. The "globe model" for conical beam scanning: $Z$ is the input beam axis, $\theta$ is angle of rotation of the tilted mirror situated in point $M$, $\mathbf{t}(\theta)$ is the unit vector of the output beam axis. It moves so that it always crosses the "parallel" with "latitude" $\pi/2-2\beta$; orthogonal to $\mathbf{t}(\theta)$ unit vector

These phenomena are manifestations of the Hannay's geometric phase [161] (the special case of the topological Berry phase [162] emerging in the course of adiabatic evolution of non-holonomic systems). The topological nature of the effects makes them insensitive to the specific way of the beam scanning; the analogous non-integer phase shift and the reference system rotation will take place if the beam axis circumscribes an arbitrary closed contour on the unit sphere. The additional phase (75) and the beam rotation (76) are also manifestations of the geometric phase that, however, are visible only for the "natural" observer. In the laboratory frame, the geometric phases associated with motion of the reference system and with the beam motion in respect to this reference system, exactly compensate each other and that is why neither of them is observed. Nevertheless, their existence implicitly manifests itself in the peculiar picture of the beam transverse pattern rotation depending on the mirror tilt (Fig. 32).

## 8. PROPERTIES OF A BEAM FORCEDLY ROTATING AROUND ITS OWN AXIS

Throughout whole this review, we discussed various aspects of the rotational behavior in paraxial light beams; let us recall them. The first one concerns the instantaneous rotation of the electromagnetic field in CS beams (see Secs. 4.1 and 5.2). This sort of rotation produces no macroscopic energy transportation and has no direct dynamical meaning, although is a source of the vortex properties of CS beams. The second aspect is associated with the macroscopic energy circulation and manifests itself, for example, in the visible rotational transformation of the transverse intensity distribution in the course of the beam propagation. It is this class of the rotatory behavior that is responsible for the beam OAM; it can be treated in dynamical terms and even such mechanical notions as angular velocity and moment of inertia are applicable to it (see Sec. 5). This sort of the beam rotation occurs spontaneously, due to some intrinsic reasons conditioned by the beam own structure. At last, in the course of the RDE discussion (Sec. 7) we encountered beams whose rotation emerged due to certain external action, just like a mechanical body can be set in rotation by corresponding action of other bodies (for example, with the help of a rotating optical system, see Fig. 25).

Such "coercively rotating" beams (from now on, let us call them simply "rotating beams") demonstrate some interesting features that have become a subject of analysis quite recently [176, 177, 48]. The approach of Ref. [177] grounded on the Maxwell equations transformed to a rotating frame is more general but rather formal. Instead, in Refs. [176, 48] the problem is studied on the base of RDE, and this enables to disclose physical properties of rotating beams much better. Both approaches are equivalent and within the bounds of common validity give identical results.

Now let us dwell in more detail on the argumentation of Ref. [176] based on the idea that any beam rotation is equivalent to the combination of coordinated frequency shifts of its helical harmonic components (Sec. 7.5). The later work by Nienhuis [48] contains the substantial generalization and quantitative development of the same approach but its physical essence is better understood with simple examples. The most straightforward situation of this sort is supplied by superposition of the $LG_{0,+1}$ and $LG_{0,-1}$ modes of equal power Φ, which forms a Hermite-Gaussian beam with total power 2Φ and the well known "two-spot" intensity distribution [53] (Fig. 34). If the LG components with $l = \pm 1$ have frequencies $\omega_+$ and $\omega_-$, this pattern rotates with angular velocity $\Omega = (\omega_+ - \omega_-)/2$.

To highlight the "pure" effect of the transverse profile rotation, we assume the beam polarization to be plane and constant (such situation is quite possible in practice [165]). The electric and magnetic fields of such a beam can be presented similarly to Eq. (6)

$$\boldsymbol{E} = \boldsymbol{E}_+ + \boldsymbol{E}_-, \quad \boldsymbol{H} = \boldsymbol{H}_+ + \boldsymbol{H}_-,$$

where, in contrast to (6), both summands in the right-hand sides possess the same directions but different frequencies $\omega_+$ and $\omega_-$. The Poynting vector distribution is obtained from the corresponding analogs of Eq. (11) and (16); after averaging over the oscillation period, its transverse part consists of the constant

$$\mathbf{S}_= = \frac{ic}{16\pi}\left[\frac{1}{k_+}\left(u_+\nabla u_+^* - u_+^*\nabla u_+\right) + \frac{1}{k_-}\left(u_-\nabla u_-^* - u_-^*\nabla u_-\right)\right] \tag{90}$$

($k_\pm = \omega_\pm/c$) and slowly varying (with frequencies Ω and 2Ω) contributions. The latter ones when substituted into (17) give zero contributions, so the beam OAM is completely determined by the value (90) and equals to

$$\mathcal{L} = -\frac{2\Phi}{c\omega}\frac{\Omega}{\omega} = -\mathcal{L}_{2\Phi}\frac{\Omega}{\omega}, \tag{91}$$

where $\omega^2 = \omega_+\omega_-$, $\mathcal{L}_{2\Phi}$ is the linear OAM density for a CS beam with power 2Φ.

There is a transparent analogy between the result (91) and the case of a usual CS beam. In correspondence to the picture presented in Sec. 4.1, rotation of the instantaneous Hermite-Gaussian distribution of electromagnetic field with angular velocity ω gives rise to the OAM $\mathcal{L}_{2\Phi}$, so it is not surprising that $\omega/\Omega$ times slower rotation "produces" proportionally smaller OAM (though one should notice that now not instantaneous but oscillation-averaged optical field rotates). However, really unexpected is the sign in Eq. (91): OAM of the rotating beam appears to be directed oppositely to the rotation angular velocity. Then, if one assimilates the beam to a mechanical body, he should provide this body with a negative moment of inertia [48].

Nevertheless, this conclusion is strange only at first glance. The considered rotating beam is a special case of the superposition of coaxial helical harmonics (85), though with different frequencies, and its OAM can be calculated as an algebraic sum of the OAMs of $LG_{0,+1}$ and $LG_{0,-1}$ components:

$$\mathcal{L} = \mathcal{L}_+ + \mathcal{L}_- = \frac{\Phi}{c\omega_+} - \frac{\Phi}{c\omega_-} = \frac{\Phi}{c}\left(\frac{1}{\omega_+} - \frac{1}{\omega_-}\right),$$

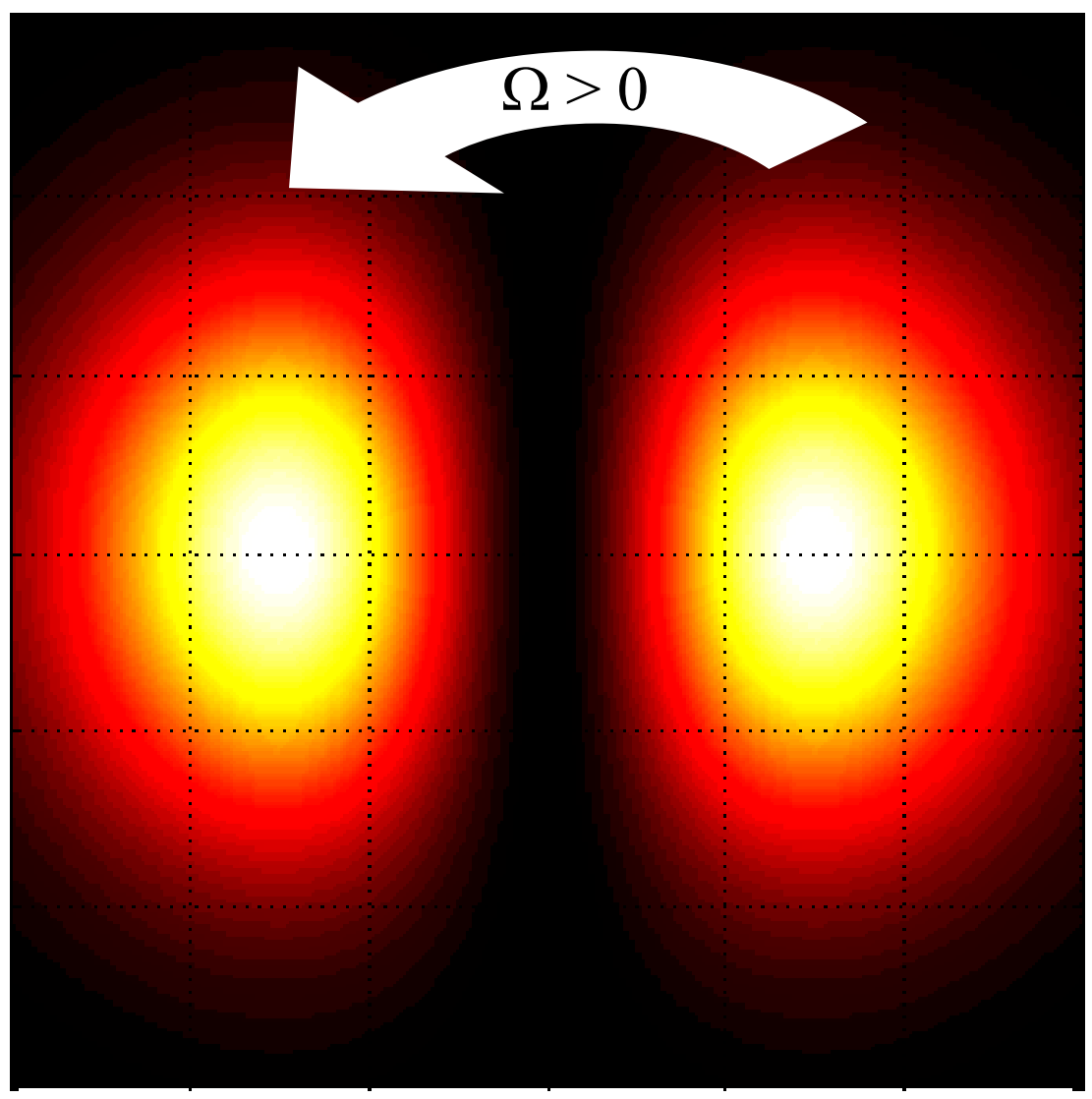

Fig. 34. Transverse intensity profile of the beam formed by superposition of the $LG_{0,+1}$ and $LG_{0,-1}$ modes. White arrow shows direction of the beam rotation at $\omega_+ > \omega_-$ (view "against" the beam propagation).

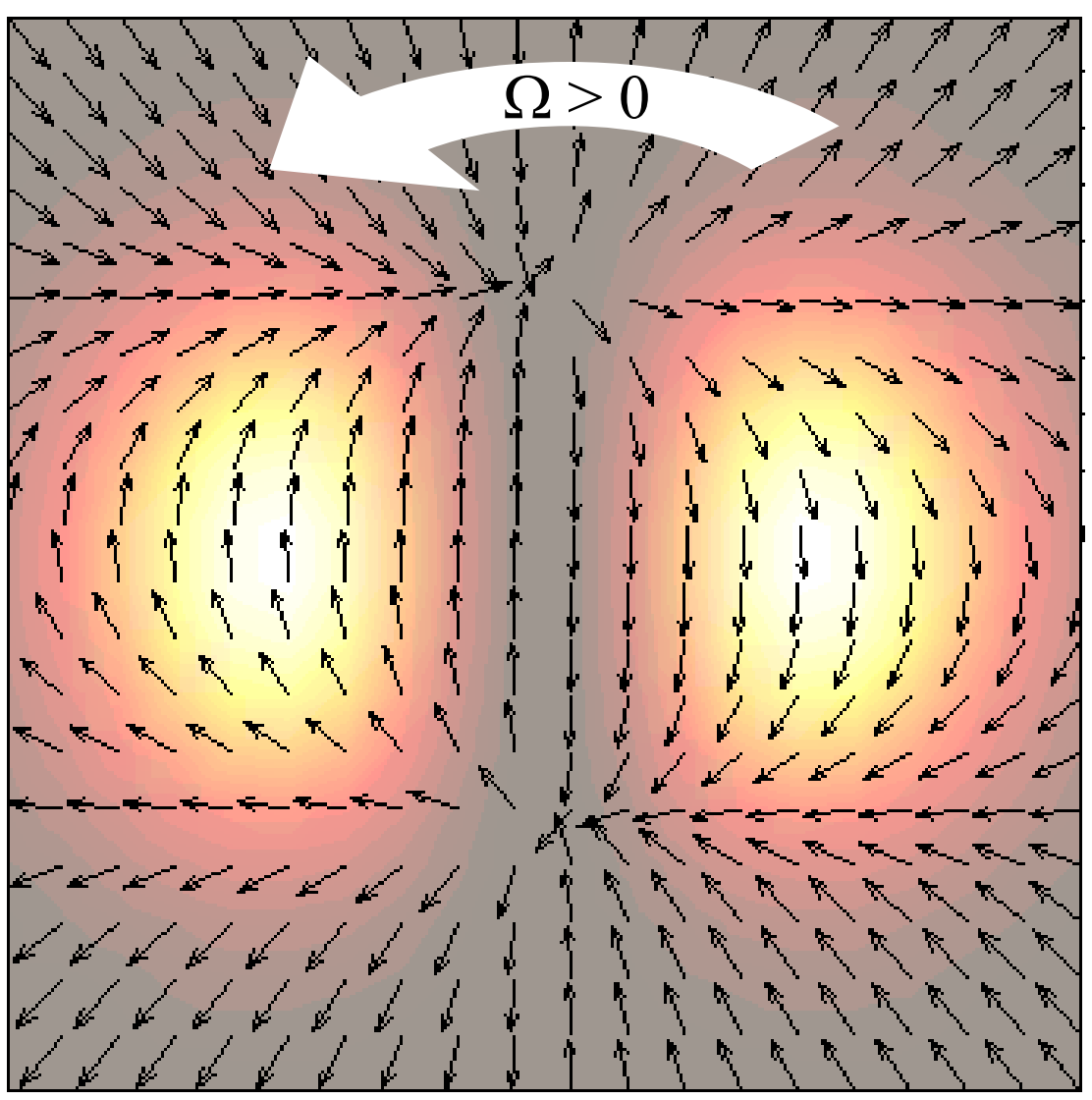

Fig. 35. The field of directions of the transverse energy transport in the waist cross section of the rotating beam at $\omega_+ > \omega_-$.

whence Eq. (91) immediately follows. Now the nature of this "negative" OAM is clearly seen: if two CS beams (31) with the same $\psi(r)$ and $l$ have equal powers and different frequencies, the mode with smaller frequency possesses higher absolute OAM value. The same can be formulated in "quantum language": the mode with higher frequency contains smaller number of photons (each of them carries the same OAM $\hbar$), and so yields lower contribution to the resulting OAM of the combined beam. Handedness of the beam rotation is determined by the energy circulation in the higher-frequency component while the OAM is directed in accord with OAM of the lower-frequency component. Fig. 35 provides an additional illustration of this situation: although the beam "in a whole" rotates counter-clockwise, the pattern of transverse energy flows in its waist cross section explicitly shows the overall energy circulation of the opposite sense.

To uncover physical manifestations of this circulation, look attentively at the three-dimensional behavior of a rotating beam. At moderate distances from the initial (waist) cross section, the beam has helical configuration (Fig. 36) determined by equation

$$2\phi + \Delta kz - \Delta\omega t = \mathrm{const}$$

where $\Delta k = k_+ - k_-$, $\Delta\omega = \omega_+ - \omega_-$. In particular, if $\Delta\omega(z/c - t) = 0$, azimuthal orientation of the current "two-spot" intensity distribution is conserved. This means that separate "portions" of the light energy (transverse "slices" of the rotating beam "body") propagate "as they are", experiencing only the usual spreading due to diffraction. Unlike the spiral beams considered in Sec. 5, in this case the "light matter" (the beam electromagnetic mass) is not involved in rotation.

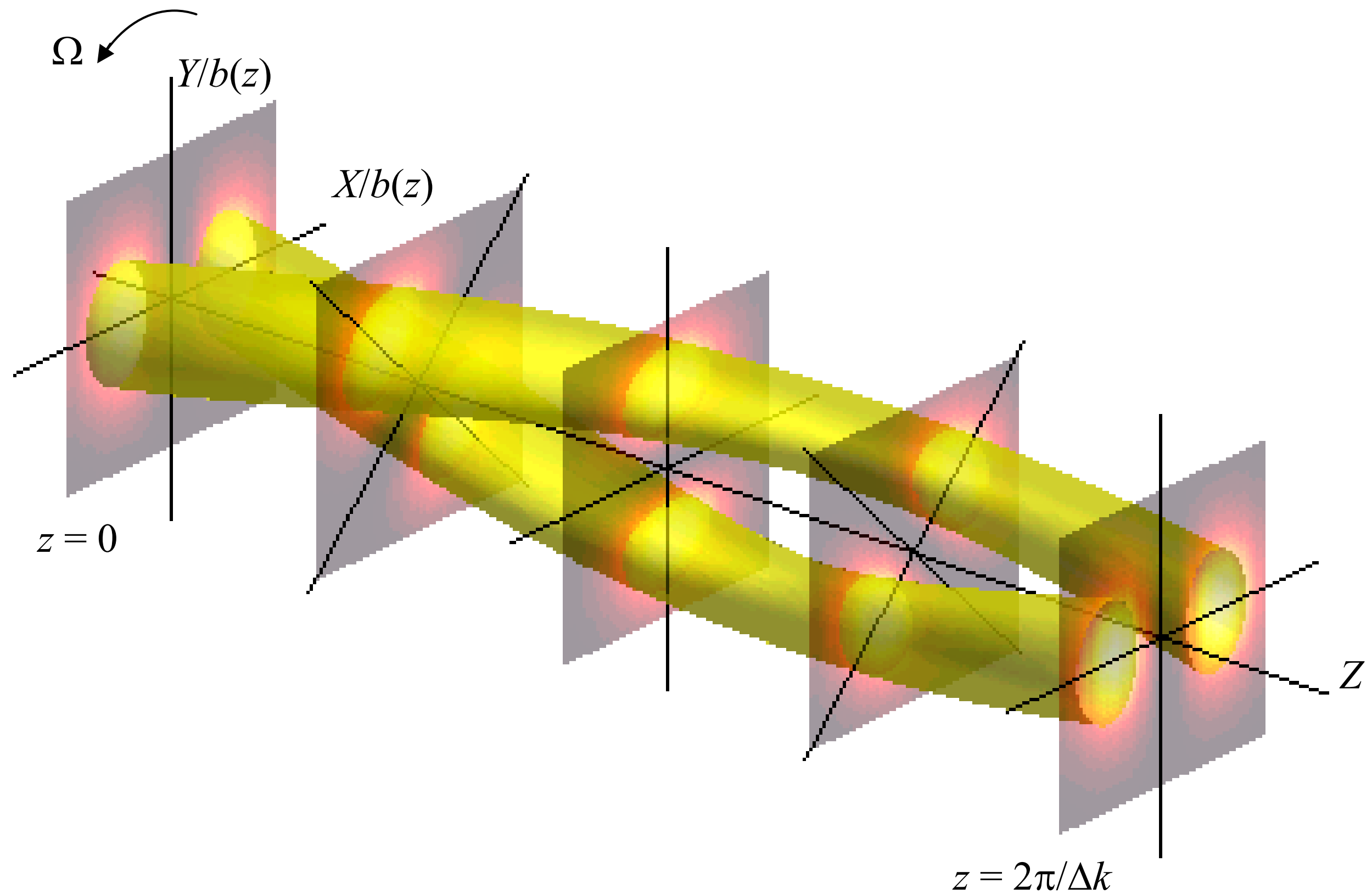

This feature also joins the rotating beam behavior with the instantaneous rotating pattern of a CS beam. As was discussed in Sec. 4.1, it implies, by the way, that the visible beam rotation is a sort of a "bright spot" motion, which has no mechanical meaning and no direct responsibility for the beam AM. The helicoid of Fig. 36, just like its analog in Fig. 7c, appears due to the limited speed of light in consequence of delays in "transmission" of the transverse structure details in the longitudinal direction. The main difference of Fig. 36 from Fig. 7c is the spatial scale, now determined not by the wavelength but by the helicoid pitch $z_{\mathrm{I}} = 4\pi/|\Delta k| = 2\pi c/|\Omega|$. Normally, on this distance a significant beam divergence occurs (not shown in Fig. 36); difference in the twist handedness in Figs. 7 and 36 is inessential and is caused by the different directions of the beam rotation in the "initial" cross sections (in Fig. 7, a case of $l = -1$ is shown).

However, Fig. 36 contains no explicit manifestations of the "negative" (with respect to rotation) AM (91). They are noticeable at very long distances from the initial cross section (Fig. 37) – of the order $z_{II} = (\omega/\Omega)z_R$ where $z_R$ is the Raleigh length (40) of the constituent LG modes. Then, really, a separate "flying slice" of the beam experiences a rotation-like deformation directed oppositely to the rotation of the beam "as a whole". This deformation is just the sought "material rotation" with which the beam OAM is associated and whose origin is connected with the transverse energy circulation in the initial cross section (Fig. 35). From the optical point of view, such deformation is caused by the fact that the constituent LG modes with different frequencies undergo slightly different transformations in the course of propagation; At distances of order $z_{II}$ the discrepancies of their spatial configurations become noticeable and a typical pattern of the interference between coaxial $LG_{0,+1}$ and $LG_{0,-1}$ beams with different WF curvatures appears [37, 85].

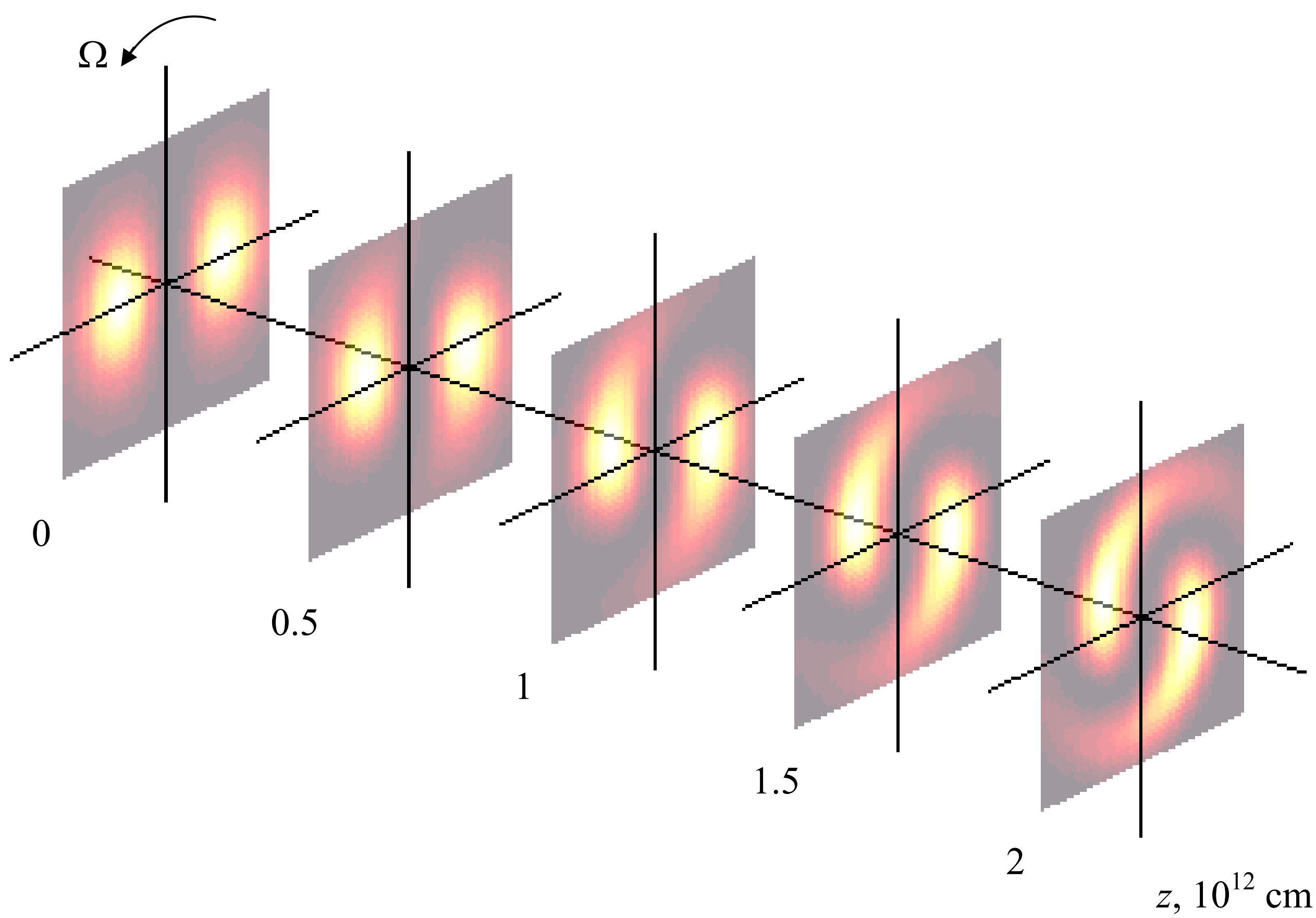

Fig. 37. Pattern of a transverse "flying slice" of the rotating beam at very high distances. Values of parameters accepted in calculations: $k_{+} = 10^{5}$ cm$^{-1}$, $\Delta k = 10^{-5}$ cm$^{-1}$ ($\Omega = 1.5\cdot 10^{5}$ c$^{-1}$, $z_{\mathrm{I}} = 1.3\cdot 10^{6}$ cm), $b_0 = 0.3$ mm ($z_R = 10^{3}$ cm, $z_{\mathrm{II}} = 2\cdot 10^{13}$ cm). Images are labeled by the current distances from the beam waist, corresponding time moments are $t = z/c$. The beam broadening is not shown (see capture to Fig. 36).

In this view, another peculiarity of the rotating beam becomes evident: strictly speaking, the usual Hermite-Gaussian mode pattern with the nodal line (an edge WF dislocation [26]) presented in Fig. 34, really exists only in the initial cross section ($z = 0$). At any non-zero $z$, the spatial parameters of the component LG beams differ and their superposition form, in fact, a beam with the screw WF dislocation, though strongly anisotropic (see Sec. 4.4, Fig. 14).

The simple example presented in this section reflects essential features of any beam forcedly rotated under some external action. The same physical behavior can be deduced by considering arbitrary rotating beam composed by however complicated composition of helical harmonics [176, 48]. In particular, in any such beam there takes place a certain rotation-like motion of the beam "matter" directed oppositely to the visible rotation. Besides, any rotating beam possesses the complex 3D configuration with the two longitudinal scales: $z_I$, defining the helical structure of the beam, and $z_{II}$, which characterizes the distance at which manifestations of its "negative" OAM become noticeable. Many of the presented arguments can be applied also to light waves with the slowly rotating uniform polarization [176, 48]. They can be represented by superpositions of two "opposite" CP waves with different frequencies and the similar arguments prove that such waves carry AM equal to algebraic sum of the AMs of separate CP components. This AM (now of the spin character) is also proportional but opposite to the rotation velocity.

On the whole, the description of a rotating light beam discussed in this section is logical and self-consistent; at the same time, attempts of its generalization lead to contradictions. First of all, the quantum-mechanical reasoning employed above essentially uses the multiphoton nature of "involved" beams. In cases of single- or few-photon states even the "preparation" of a rotating beam encounters fundamental difficulties because the photon energy is strictly "dictated" by its frequency. Single-photon azimuthal harmonics with different frequencies would inevitably have different amplitudes, so their superposition can never form a "true" Hermite-Gaussian beam.

A number of problems emerge when trying to consider the same phenomena in a frame rotating with the beam. In this frame, $\Omega = 0$ and frequencies of the helical components become identical. Their superposition thus produces a usual self-similar Hermite-Gaussian structure with no signs of helicity. Besides, if in the laboratory frame these components had identical amplitudes (and contained different numbers of photons), after transforming to the frame rotating together with the beam the photon numbers do not change, so amplitudes should modify. Again, this means that in a rotating frame the beam loses the "true" Hermite-Gaussian configuration (in the considered simplest case of the $LG_{0,+1}$ and $LG_{0,-1}$ superposition, the edge WF dislocation is replaced by the strongly anisotropic OV). Probably, all the mentioned (and some unmentioned) contradictions originate from the approximate character of simplified coordinate transformation (73), which in most situations is fairly appropriate quantitatively but leads to incorrect qualitative results in hypothetical limiting cases. The future consistent analysis of rotating beams must be based on general relativistic methods with taking into account the non-inertial nature of rotational frames [178, 179].

## 9. CONCLUSION

At the end of this review, we would like to stress on the exclusive position of light beams with AM among the objects of modern physics. By the very physical nature, they cannot be attributed to any traditional branch: they do not belong to the "pure" optics or mechanics; equally they cannot be understood on the base of only classical or only quantum notions. Problems appearing in their studies can be solved only with employment of the "whole" physics, and they clearly and demonstratively display the unity and universal character of the physical world. Beams with the "vortex" properties turn out to be the unique objects whose behavior allows to illustrate, in a very spectacular and impressive form, some of the most fundamental and, apparently, very abstract physical ideas. One can hardly find other objects which "unite" the topology of wave fields and the general relativistic space-time transformations, ponderomotive interactions of electromagnetic fields and the laws of symmetry, problems of non-holonomity and topological phase – and all this in easily conceivable and immediately observable situations (which, by the way, makes beams with AM very useful in the practice of physical education).

Of course, not all the mentioned aspects could be reflected within the scope of this review. From the multitude of concepts and facts associated with the AM of light beams, we have chosen only a modest part that seems to us the most interesting or instructive. A lot of important problems remained outside our attention; besides, many of them are far from the complete solution and are subjects of intensive research (for their general overview, it would be sufficient, for example, to look at contents of the recent topical collections [180, 181]).

We mention only several problems that are now subjects of great interest or, probably, will become at the focus of attention within the next few years. First of all, these are non-linear phenomena with participation of the AM-beams [123–125, 182–184]. They are interesting not only 'per se' or as illustrations of the AM conservation and transformation laws but also as a useful instrument of studying fundamental phenomena, in particular, the quantum "entanglement" [184, 185].

In our opinion, great prospects are associated with the usage of the beam AM and its constituents discussed in Secs. 3 and 4 for characterizing the beam spatial configuration. At the moment, this techniques is well developed for scalar models but the growing interest to vectorial and elliptic fields of paraxial beams [186, 187] stimulates to the search of new means for their description built on the base of the energy flow patterns inside the beam [188–193]. This opens impressive possibilities of "implantation" the AM ideas in the new, exclusively rich of phenomena and practically important field that includes the optics of crystals and inhomogeneous media, in particular, optical waveguides and fibers [194–196]. Another prospective area, only slightly touched in the last section, concerns the non-monochromatic singular beams [196, 197, 48].

At last, one can notice that almost all investigations of the light field AM relate to laser beams, i.e. the artificial objects, and this can create an expression as if its observation requires the special "laboratory" conditions. This impression is illusory, and besides the common example of speckle fields [17, 172] (which always contain singularities and, therefore, local "islets" of the transverse energy circulation), it is possible to point out the existence of beams with OAM in the cosmic radiation [198].

To finalize the paper, we make a brief mention of possible practical applications of beams with OAM. Here, the first place belongs to helical beams with isolated amplitude zeros that can serve, e.g., for the channeling or spatial localization of particles or atoms [37, 70]. The high sensitivity of the singular point position to the least perturbations enables to efficiently control the parameters of such "optical tweezers", to catch, to sort and to transport particles in a prescribed manner. This creates wide prospects for the study and diagnostics of isolated microobjects, for realization of selective physico-chemical interactions ("microreactors") and selective micro-influences. The unique spatial characteristics of helical beams provide also many other application possibilities. For example, they offer a controllable way to modify the autocorrelation function of an optical beam used for the irregular media diagnostics [199, 200] which allows, by comparison of the back scattering characteristics of CS beams with different topological charges, obtain valuable structural information on the probed medium [201, 202]. Even the "darkness" of an OV core can be utilized: If the optical system of a telescope is provided with a spiral phase plate (see Sec. 4.2), an "image" of even a very bright source will contain "absolutely dark" areas forming "windows" from which the radiation of adjacent weak sources, normally "masked" by intense glare, can be "peered" [203]. This method also improves angular resolution of sources within the diffraction limit and is proposed for observation of dim exoplanets close to the parent star [204].

Perhaps, the most promising applications of helical beams with AM are associated with the information processing. Unlike the usual information encoding methods using the physical objects with two well distinguishable states, the circular helical photons are able of carrying the "quantized" OAM equal to any integer number of $\hbar$. This enables to increase the information capacity of a physical carrier and to "write" in a single beam up to few tens bit of information [173]. Importantly, the information is encoded in the beam topological structure that is very stable to external disturbances, which improves the communication channel immunity. On the other hand, for the information readout one needs to know the whole beam cross section, and this provides the communication channel with the additional intercept security.